\documentclass{aastex}           
\usepackage{spr-astr-addons}     
\usepackage{float}
\usepackage{amsmath}
\usepackage{amssymb}
\usepackage{multirow}
\usepackage{array}

\begin{document}

\title{On the Connection of Gamma-Ray Bursts and X-Ray\\
Flashes in the BATSE and \textit{RHESSI} Databases}

\shorttitle{On the Connection of Gamma-Ray Bursts and X-Ray Flashes}
      
\author{J. \v{R}\'{\i}pa}
\affil{
National Taiwan University,\\
Leung Center for Cosmology and Particle Astrophysics,\\
No.1, Sec.4, Roosevelt Road, Taipei 10617, Taiwan (R.O.C)
}
\email{jripa@ntu.edu.tw}

\and
    
\author{A. M\'esz\'aros}
\affil{
Charles University in Prague,\\
Faculty of Mathematics and Physics, Astronomical Institute,\\
V Hole\v{s}ovi\v{c}k\'ach 2, CZ 180 00 Prague 8, Czech Republic
}
\email{meszaros@cesnet.cz}

\begin{abstract}
Classification of gamma-ray bursts (GRBs) into groups has been intensively studied
by various statistical tests in previous years. It has been suggested that there was
a distinct group of GRBs, beyond the long and short ones, with intermediate durations.
However, such a group is not securely confirmed yet.
Strangely, concerning the spectral hardness, the observations from the \textit{Swift} and \textit{RHESSI}
satellites give different results.
For the \textit{Swift}/BAT database it is found that the
intermediate-duration bursts might well be related to so-called X-ray flashes (XRFs).
On the other hand, for the \textit{RHESSI} dataset
the intermediate-duration bursts seem to be spectrally too hard to be given by XRFs.
The connection of the intermediate-duration bursts and XRFs for the BATSE database is not clear as well.
The purpose of this article is to check the relation between XRFs and GRBs for the BATSE
and \textit{RHESSI} databases, respectively. We use an empirical definition
of XRFs introduced by other authors earlier.
For the \textit{RHESSI} database we also use a transformation between the detected counts and the
fluences based on the simulated detector response function.
The purpose is to compare the hardnesses of GRBs with the definition of XRFs.
There is a $1.3-4.2$\,\% fraction of XRFs in the whole BATSE database.
The vast majority of the BATSE short bursts are not XRFs because only $0.7-5.7$\,\%
of the short bursts can be given by XRFs. However, there is a large uncertainty in the
fraction of XRFs among the intermediate-duration bursts. The fraction of $1-85$\,\% of
the BATSE intermediate-duration bursts can be related to XRFs. For the long bursts this fraction is
between 1.0\,\% and 3.4\,\%. The uncertainties in these fractions are large, however
it can be claimed that all BATSE intermediate-duration bursts cannot be given by XRFs.
At least $79$\,\% of \textit{RHESSI} short bursts,
at least $53$\,\% of \textit{RHESSI} intermediate-duration bursts,
and at least 45\,\% of \textit{RHESSI} long bursts should not be given by XRFs.
A simulation of XRFs observed by \textit{HETE-2} and \textit{Swift} 
has shown that \textit{RHESSI} would detect, and in fact detected, only one long-duration XRF
out of 26 ones observed by those two satellites.
We arrive at the conclusion that the intermediate-duration bursts in the BATSE database
can be partly populated by XRFs, but the \textit{RHESSI} intermediate-duration bursts are
most likely not given by XRFs. The results, claiming that the \textit{Swift}/BAT
intermediate-duration bursts are closely related to XRFs do not hold for the BATSE
and \textit{RHESSI} databases. 
\end{abstract}

\keywords{gamma-ray burst: general $-$\\
X-rays: bursts}

\section{Introduction}
\label{sec:intro}

Gamma-ray bursts (GRBs) are diverse objects.
The existence of two astrophysically different groups of GRBs, denoted as ``short-''
and ``long-duration'' bursts,
is now well established \citep{maz81,kou93,nor01,bal03,bor04,me06b,zha09}.
However, there is a non-negligible overlap in durations between these two groups \citep{kan10, kan11, brom13, tar15b}.

In addition, the occurrence of a group of in\-ter\-me\-dia\-te-du\-ra\-tion GRBs in data samples of several satellites
has been intensively studied using various statistical methods.
It was suggested that there could be such a group
\citep{ho98,muk98,balas01,ho02,var05,ho08,va08,ho09,hoi09,rip09,huj09,ho10,zit15}.
On the other hand, several works doubt its existence \citep{hak00,hak03,raj02,koen12,tar15a,tar16,bha16}.
Moreover, different statistical tests applied on different datasets of different satellites give varying
significance claiming its existence. Even though several statistical tests claimed occurrence of this group,
its astrophysical meaning is not well established yet and it remains unclear.
For example, the anti-correlation between the hardness and duration in the
\textit{CGRO}/BATSE\footnote{http://www.batse.msfc.nasa.gov/batse/grb/catalog} \citep{fish94} database is
fully unclear \citep{ho06}.

Recently, two essential steps were taken in the clarification of the physical meaning of these intermediate-duration bursts.
First, a detailed statistical analysis of data from the
\textit{Swift}/BAT instrument\footnote{http://swift.gsfc.nasa.gov/docs/swift/swiftsc.html} \citep{geh04} arrived
at the conclusion that they are related to so-called X-ray flashes (XRFs)
\citep{ver10,kob13}. Second, a similar detailed statistical analysis of the
\textit{RHESSI}\footnote{http://hesperia.gsfc.nasa.gov/hessi/index.html} \citep{lin02,smi03,haj04,wig04}
database showed that the intermediate-duration bursts in this database were similar to the short ones \citep{rip12}.
This means that in this database the intermediate-duration bursts are spectrally as hard as the short ones,
and thus they hardly can be identified with the spectrally soft XRFs.
Hence, it is clear that the instrumental effects are important concerning the GRB classification.

The purpose of this article is to study the connection of GRBs and XRFs both for the BATSE and \textit{RHESSI} datasets.
The main aim is to estimate the fraction of XRFs among the intermediate-duration GRBs separately for both databases.

The paper is organized as follows.
In Section~\ref{sec:xrf_definition} we present a review of the definitions of XRFs.
Sections~\ref{sec:samples} and \ref{sec:methods} define the used samples and methods.
In Sections~\ref{sec:batse_results} and \ref{sec:rhessi_results} we study the BATSE and the \textit{RHESSI} databases, respectively.
Section~\ref{sec:instrum_effects} discusses the instrumental effects of the BATSE, \textit{RHESSI}, \textit{Swift}, and
\textit{HETE-2} instruments. Section~\ref{sec:conclusions} summarizes the results.

\section{Two definitions of XRFs}
\label{sec:xrf_definition}

The notion of XRF was introduced by \citet{hei01}
for the bursts detected in the Wide Field Cameras (WFC) on \textit{BeppoSAX}\footnote{http://www.asdc.asi.it/bepposax/} satellite in the energy range $2-25$\,keV,
but not detected in the Gamma-Ray Bursts Monitor (GRBM) on the same satellite in the energy range $40-700$\,keV.
Hence, XRFs are soft long events, which emit mainly in the X-ray band at $^{<}_{\sim}\, 25$\,keV \citep{hei01,ved09}.
\citet{kip03} studied the peak energy and peak flux distributions of XRFs observed by the \textit{BeppoSAX}/WFC instrument and
compared them to the spectral properties of the bright BATSE GRBs \citep{pre00}.
The authors claimed that XRFs have significantly lower values of peak energy than
the bright BATSE GRBs with the Kolmogorov-Smirnov (K-S) probability of $P_{KS}=1.5\times10^{-8}$.
They also found that XRFs are inconsistent even with the weakest 5\,\% of the bright BATSE GRBs with the K-S probability of $P_{KS}\approx10^{-5}$.
However, from their Figure~3 it seems that XRFs are soft and weak events on the tail of the GRB distribution.
\citet{kip03} stated in their conclusion that XRFs could be a low-energy extension of the long GRB population.
Similarly, also \citet{sak05} concluded that XRFs and long GRBs can arise from the same phenomenon.

Without going into details of the astrophysical models of XRFs we briefly review
the three scenarios that have been suggested to explain their origin:
\begin{enumerate}
\item In the high-redshift scenario a normal long GRB, placed at a
high redshift, would be seen as XRF due to the shift of the peak energy to the X-ray band \citep{hei03}.
\item The off-axis model claims that XRFs are ordinary long GRBs viewed off-axis of the relativistic outflow jet.
Different jet structures have been proposed to explain the properties of XRFs \citep{yam02,dad04,eich04,la05,to05,zha04}.
\item It has also been suggested that the soft spectrum of XRFs could be due to the intrinsic properties of the long GRBs,
e.g., sub-energetic or an inefficient fireball \citep{der99,zha02,rr02,mo04}.
\end{enumerate}

The origin of XRFs still remains unclear \citep{grb12}. This paper does not focus on specific models of XRFs and thus
the question of the correctness of the models will be omitted in this work.

\citet{lamb03,lamb04} and \citet{sak05} defined XRFs in the sample of the \textit{HETE-2} satellite\footnote{http://space.mit.edu/HETE/} \citep{ric03}
as those events for which the ratio $\log[S_{2-30}/S_{30-400}] > 0$,
where $S_{E_i-E_j}$ defines the energy fluence in the energy range $E_i < E < E_j$ (in keV).
Hereafter we will call this \textit{HETE-2} XRF definition as ``Def1''.
Def1 implies that for XRFs the fluence in the range of $30-400$\,keV should be smaller compared with the fluence in the range of $2-30$\,keV.
Since the ratio of the fluences is called ``hardness'' (denoted by $H$) for the given energy ranges, it is always necessary to precise which
energy ranges are used in the definition. For example, it can be written: $H_{\frac{50-100}{20-50}} = S_{50-100}/S_{20-50}$.
Note that the used fluences should have dimension erg\,cm$^{-2}$.

\citet{sak08} introduced a different definition of XRFs for the sample of the \textit{Swift} satellite
which better suits the energy range of the BAT instrument ($15-150$\,keV).
They defined an event as XRF if $0.76 > S_{50-100}/S_{25-50} = H_{\frac{50-100}{25-50}}$.
Hereafter we will refer to this definition as ``Def2''.
Hence, Def2 means that for XRFs the fluence in the range $50-100$\,keV should be smaller
than 76\,\% of the fluence in the range $25-50$\,keV.

The two definitions are not identical because they use different energy bands.
To compare them one needs the form of the time integrated spectra.

We choose three spectral models to be considered in our analysis:
power law (PL); power law with exponential cutoff (CPL); and Band function (shortly ``Band'', \citet{band93}).
The Band function with general parameters takes the form

\begin{eqnarray} \label{eq:Band_general_1}
N_\mathrm{E}(E) = \left\{
\begin{array}{ll}
f_1(E) & \quad \mathrm{for~} E\leq E_{\mathrm{break}} \\
f_2(E) & \quad \mathrm{for~} E\geq E_{\mathrm{break}}, \\
\end{array} \right.
\end{eqnarray}
where
\begin{equation} \label{eq:Band_general_2}
f_1(E) = K_1\left(\frac{E}{E_\mathrm{piv}}\right)^{\alpha}\exp\left(-\frac{E}{E_{0}}\right),
\end{equation}
and
\begin{equation} \label{eq:Band_general_3}
f_2(E) = K_2\left(\frac{E}{E_\mathrm{piv}}\right)^{\beta}.
\end{equation}
The amplitudes $K_1$ and $K_2$ are chosen so that\\
$f_1(E_{\mathrm{break}}) = f_2(E_{\mathrm{break}})$, i.e.
\begin{equation} \label{eq:Band_general_4}
K_2 = K_1\left[(\alpha-\beta)\frac{E_0}{E_\mathrm{piv}}\right]^{(\alpha-\beta)}\exp(\beta-\alpha),
\end{equation}
where $E_\mathrm{piv}$ is a fixed value usually at 100\,keV. Hence, only one amplitude is independent.
For $E_0$ it holds
\begin{equation} \label{eq:Band_general_5}
E_0=\frac{E_\mathrm{peak}}{2+\alpha}=\frac{E_{\mathrm{break}}}{\alpha-\beta},
\end{equation}
where $E_\mathrm{peak}$ is called peak energy and it must be $\alpha > \beta$.
The differential photon spectrum $N_E(E)$ is in units ph\,cm$^{-2}$\,s$^{-1}$\,keV$^{-1}$ (``ph'' refers to photon).
In these formulas $E$ is the energy of photons in keV. In the CPL model the whole spectrum is described by $f_1(E)$,
where the amplitude $K_1$ is an independent free parameter. For the PL model the whole spectrum is described by $f_2(E)$,
where the amplitude $K_2$ is an independent free parameter. The Band function has four independent parameters,
CPL (PL) has three (two) ones. In the computation of hardnesses the amplitude $K_1$ or $K_2$ is always cancelled, and hence we
should consider only three parameters in Eqs.~(\ref{eq:Band_general_1}-\ref{eq:Band_general_3}) ($\alpha$, $E_0$, $\beta$).
Instead of $E_0$ either $E_\mathrm{peak}$ or $E_\mathrm{break}$ can be used, too, due to Eq.~(\ref{eq:Band_general_5}).
For details of the GRB spectral models see, e.g. \citet{band93,llo00,meszp06,wig08,gold13}.
Add also that detailed statistical studies of the spectra show that the best option for the 
three needed parameters ($\alpha$, $E_0$, $\beta$) is to consider them as independent variables
\citep{bag98,ryd05,bor06,bag09,axe15}.

For XRFs the typical peak energy in the Band function is $E_{\mathrm{peak}}=30$\,keV \citep{pre00,hei01,kip03,sak05,DAle06,sak08,ved09}.
\citet{sak08} derived their definition of XRFs for the \textit{Swift}/BAT sample based on the typical spectral parameters of XRFs:
$\alpha=-1$ for the low-energy spectral index and $\beta=-2.5$ for the high-energy spectral index with $E_{\mathrm{peak}}=30$\,keV,
i.e. $E_{\mathrm{break}} = 45$\,keV.

If we adopt this definition, we can extrapolate the limiting hardness of XRFs for another energy band.
Since the energy fluence is
\begin{equation}
S_{E_1-E_2}=\int_{E_1}^{E_2} EN_E(E) dE,
\end{equation}
one obtains, e.g. for $S_{50-100}/S_{20-50}$, where the energy bands are partly below $45$\,keV,
\begin{equation} \label{eq:H_XRF_def2_BATSE}
S_{50-100}/S_{20-50} =
\frac{\int_{50}^{100} E f_2 dE}{\int_{20}^{45} E f_1 dE + \int_{45}^{50} E f_2 dE} \approx 0.6.
\end{equation}

For $S_{120-1500}/S_{25-120}$ the XRF limiting hardness is given as
\begin{equation} \label{eq:H_XRF_def2_RHESSI}
S_{120-1500}/S_{25-120} =
\frac{\int_{120}^{1500} E f_2 dE}{\int_{25}^{45} E f_1 dE + \int_{45}^{120} E f_2 dE} \approx 0.6.
\end{equation}

\begin{figure}[t]
\centering
\includegraphics[width=0.48\textwidth]{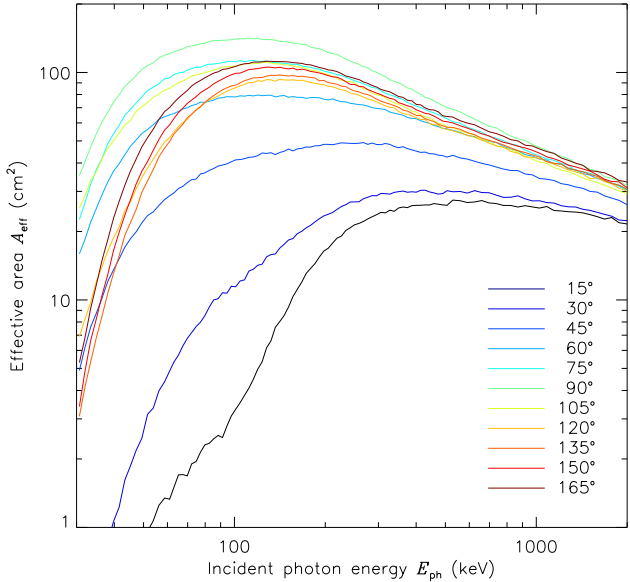}
\caption{
The photopeak effective area of rear segments of the \textit{RHESSI} detectors in the range $30-2000$\,keV.
The overall effective area is summed over all the nine detectors except for malfunctioning No.2 and averaged over
the six intervals of the azimuth angle: $0^\circ - 60^\circ$, $60^\circ - 120^\circ$, $120^\circ - 180^\circ$,
$180^\circ - 240^\circ$, $240^\circ - 300^\circ$, $300^\circ - 360^\circ$.
The dependency is shown for eleven different off-axis angles of incoming radiation.
}
\label{fig:rhessi_eff_area}
\end{figure}

Comparing the two definitions one may say that there are several problems
and hence their use is never straightforward.

First of all, the two different definitions are valid for different energy bands.
Of course, it is possible to use Eq.~(\ref{eq:Band_general_1}-\ref{eq:H_XRF_def2_RHESSI})
for the calculation of a hardness for any other energy bands if one assumes a GRB spectrum.
However, this means that, e.g., the limiting values in Eqs.~(\ref{eq:H_XRF_def2_BATSE}) and (\ref{eq:H_XRF_def2_RHESSI})
themselves can be changed due to the change of the spectral parameters.

The second problem concerns Def1 itself, because for the BATSE instrument and for the \textit{RHESSI} satellite
the efficiency of the detection of the photons with energies smaller than $\sim(25-30)\,$keV drops down rapidly.
Thus Def1 uses the energy bands which are - in essence - not detected observationally.
Concerning Def2 there is no such problem as the energies $\sim(25-100)\,$keV are observed both by
the BATSE instrument and by the \textit{RHESSI} satellite. In fact, this is the principle argument why
\citet{sak08} introduced Def2 instead of Def1 \citep{sak05} for the \textit{Swift} database.

A third problem arises from the dimension erg\,cm$^{-2}$ of the fluences.
Since the hardness itself is a ratio of two fluences, it is a dimensionless number.
Then it seems that any dimension of the fluence can be used.
For example for the \textit{RHESSI} satellite the fluence is available in instrumental counts for the vast majority of bursts and
then a ratio of two such total counts from different energy bands can also define a ``pseudo-hardness''.
However, these two hardnesses, even for the same object, in general need not be identical due to a specific properties of
the detector's response.

Following mainly the second argument, which strong\-ly prefers
the use of Def2 instead of Def1, we restrict ourselves to Def2 in the following chapters.
This means that the definition of the XRF is following: An event is classified as an XRF
if $0.76 > S_{50-100}/S_{25-50}$ and the fluences $S_{50-100}$ and $S_{25-50}$ have dimensions erg\,cm$^{-2}$.

\begin{figure}[t]
\centering
\includegraphics[width=0.48\textwidth]{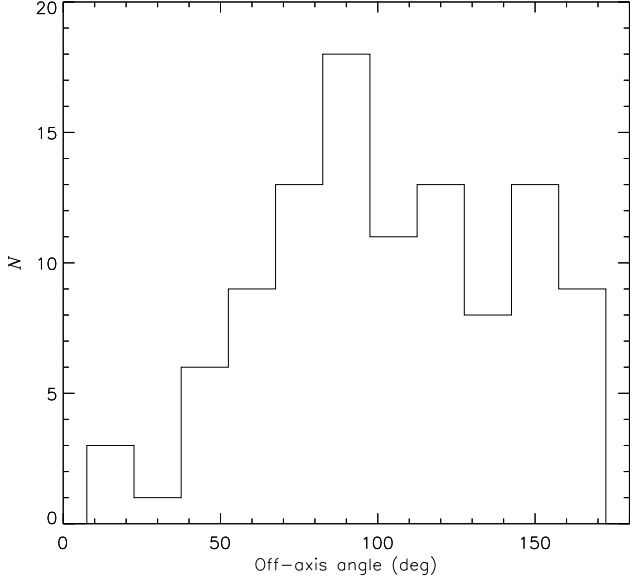}
\caption{
The distribution of the off-axis angles of 104 localized GRBs from the total sample of 427 GRBs observed by the \textit{RHESSI} satellite.
}
\label{fig:rhessi_off-axis_distribution}
\end{figure}

Add still that the application of this Def2 on the BATSE and mainly \textit{RHESSI} databases is never a simple task
due to the instrumental effects. In addition, the fluence measured by \textit{RHESSI} is not given in units erg/cm$^2$.
These effects will be discussed in more details in the next sections. Here we note only the following.
Fig.~\ref{fig:rhessi_eff_area} shows the photopeak effective area of
the rear segments of the \textit{RHESSI} detectors. It is summed over all nine detectors except for malfunctioning No.2.
The response function as well as the effective area used in this work were provided by E. Bellm (private communication).
The simulated response functions were based on the satellite's mass model in the Monte Carlo suite MGEANT
(for details see \citet{bel08a, bel10, bel11}). Most of the localized GRBs detected by \textit{RHESSI} were seen under
the off-axis angle $>50^\circ$ (see Fig.~\ref{fig:rhessi_off-axis_distribution}).
Although the energy range $25-50$\,keV used in Def2 is near the edge of the \textit{RHESSI}'s sensitivity,
the effective area for the off-axis angles $>50^\circ$ is still sizeable.
This suggests that one can still detect fluence in that energy range.
For the effective area of the BATSE Large Area Detector (LAD) see, e.g. Fig.~3 of \citet{fish85} or Fig.~1 of \citet{pen99}.
Concerning BATSE, the measurements only from LAD are used in this work.
Similarly to \textit{RHESSI}, although the energy range $25-50$\,keV is near the edge of the sensitivity of
BATSE LAD, the effective area is still sizeable and fluence in that energy range is still detectable.
The above arguments strongly suggest the use of Def2 with the XRF limiting hardness given by
Eq.~(\ref{eq:H_XRF_def2_BATSE}) and Eq.~(\ref{eq:H_XRF_def2_RHESSI})
instead of Def1 for the BATSE instrument and \textit{RHESSI} satellite, respectively.

In what follows we will use abbreviation ``Def.'' instead of Def2.

\section{The samples}
\label{sec:samples}

\subsection{The BATSE samples}
\label{sec:BATSE_samples}

In the paper we will employ two types of BATSE Catalogs.
The first one is BATSE Current Catalog\footnote{http://gammaray.msfc.nasa.gov/batse/grb/catalog/current/}
(shortly ``Current Catalog''). It contains 2702 events and uses 4-energy channel data
in the energy range from 20\,keV to $>300$\,keV.
For our purpose the important information (besides other records) included in this Catalog is:
the BATSE trigger numbers; $T_{90}$\,[s] durations; uncertainties in $T_{90}$\,[s];
fluences $S1$ in channel 1 ($20-50$\,keV); fluences $S2$ in channel 2 ($50-100$\,keV);
fluences $S3$ in channel 3 ($100-300$\,keV); and $1\,\sigma$ statistical uncertainties in all fluences.
All fluences and their uncertainties have units of erg\,cm$^{-2}$.
The Current Catalog does not contain the hardness ratios and they must be calculated from the measured fluences.
From 2702 events 1927 GRBs have simultaneously measured $T_{90}$ with uncertainties and three fluences
($S1$, $S2$ and $S3$) with uncertainties.

The second one is the BATSE Complete Spectral Catalog (shortly ``Spectral Catalog'') released in the electronic 
form\footnote{http://www.batse.msfc.nasa.gov/$\sim$goldstein/}. For more details see also \citet{gold13}.
This Spectral Catalog contains time integrated spectral fits (fluence spectra) for 2106 events.
It employs CONT data type and the medium energy resolution (MER) data type. 
It uses 14 channels out of 16 energy channels. This corresponds to energies between $\sim 25$\,keV and $\sim 1.8$\,MeV.
For our purpose the important information (besides other records) included in this Catalog is:
the BATSE trigger number; the best-fit spectral parameters for PL, CPL, and Band functions (fluence spectral information);
the uncertainties in the best-fit spectral parameters; the $\chi^2$ of each fit;
and the number of degrees of freedom of each fit.
The Spectral Catalog does not contain the hardness ratios and they must be calculated from the measured
spectral parameters assuming a particular spectral model.
All 2106 events from the Spectral Catalog are present in the Current Catalog as well.

\subsection{The RHESSI sample}
\label{sec:RHESSI_sample}

In the case of the \textit{RHESSI} satellite we will use the sample, which was used already by \citet{rip09,rip12},
and which contains 427 events. However, oppositely to the BATSE datasets,
the accurate spectral fits are possible only for about $67$ GRBs of all 427 events.
The reason is that the \textit{RHESSI}'s detector \citep{smi02} response function depends on the incident angle of incoming photons and thus
the knowledge of the GRB sky position is essential to obtain a precise spectral fit. However, there are only 104 localized GRBs
in the \textit{RHESSI} database used. Also the measured flux from a GRB should have a high signal-to-noise ratio (S/N\,$\gtrsim 10$)
to allow a reliable spectral fit. This effect further decreases the suitable GRBs
because 37 localized GRBs had $\mathrm{S/N}<10$.
Hence, there are only 67 GRBs ($67/427\approx16$\,\%)  
in the \textit{RHESSI} database for which reliable spectral fits can be carried out.
This means that we cannot apply ProcI here and we will provide ProcII only for the sample of 427 GRBs.

In addition, unlike in the BATSE sample, the fluence in the \textit{RHESSI} GRB database is not given in units of erg\,cm$^{-2}$.
Instead, it is available in the instrumental counts.
A ratio of two such numbers of counts in different energy bands define a ``pseudo-hardness''.
The database analyzed in works \citet{rip09,rip12} used the energy ranges of $25-120$\,keV and $120-1500$\,keV.
Hence, the ``pseudo-hardness'' $\widetilde{H}_{\frac{120-1500}{25-120}} = C_{120-1500}/C_{25-120}$\,(cnt\,cnt$^{-1}$) can be defined,
where $C$ are the numbers of the detected counts in the given energy ranges.
The conversion between the ``pseudo-hardness'' $\widetilde{H}_{\frac{120-1500}{25-120}}$\,(cnt\,cnt$^{-1}$) and the 
hardness $H_{\frac{120-1500}{25-120}}$ (erg\,cm$^{-2}$\,erg$^{-1}$\,cm$^{2}$) will be described in detail in
Subsection~\ref{sec:rhessi_cnt_erg_conversion}.

\section{The methods}
\label{sec:methods}

In the determination of the fraction of XRFs in the BATSE database we use the hardness
$H_{21}\equiv H_{\frac{50-100}{20-50}}=S_{50-100}/S_{20-50}$ because its energy range
ensures sensitivity at 30\,keV, which is the typical peak energy in the Band function of XRFs.
For the separation of bursts into three groups we use the classification published by
\citet{ho06} in their Fig.~1 on the hardness $H_{32}\equiv H_{\frac{100-300}{50-100}}=S_{100-300}/$
$S_{50-100}$. The group-membership list itself was provided by
I. Horv\'ath (private communication).

\begin{figure}[h]
\centering
\includegraphics[width=0.48\textwidth]{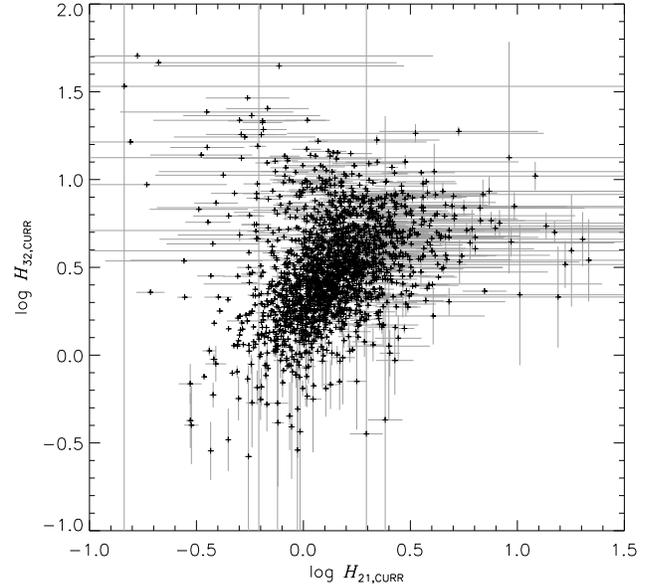}
\caption{The hardnesses $H_{21\rm,CURR}$ vs $H_{32\rm,CURR}$ from the BATSE Current Catalog for 1927 are shown.}
\label{fig:batse_H21_curr_H32_curr}
\end{figure}

There are 5 events with measured $T_{90}$ with uncertainties and simultaneously with
measured fluence $S1$ (with uncertainties), but without measured fluence $S3$ (with uncertainties).
On the other hand, there are 28 events with measured $T_{90}$ (with uncertainties) and simultaneously with
measured fluence $S3$ (with uncertainties), but without measured fluence $S1$ (with uncertainties).
Fig.~\ref{fig:batse_H21_curr_H32_curr} shows the hardnesses $H_{21\rm,CURR}$ and $H_{32\rm,CURR}$
calculated from the fluences $S1$, $S2$, and $S3$ from the Current Catalog
plotted against each other.
The Pearson correlation coefficient
between $H_{21\rm,CURR}$ and $H_{32\rm,CURR}$ is 0.1 and between
log $H_{21\rm,CURR}$ and log $H_{32\rm,CURR}$ is 0.3.
In reality, $H_{21\rm,CURR}$ and $H_{32\rm,CURR}$ should be correlated since both values are
related to the spectral parameters. Perhaps the Pearson correlation coefficients are low, because of the
large error bars of the hardnesses and thus large scatter of the data points.
The reason for the low correlation coefficients can also be given by some unknown systematic errors of the
fluences in the BATSE Catalog.
In any case, this fact does not disfavor our method of deriving fractions of XRFs in the BATSE datasets.

Since the energy ranges used in hardness $H_{21}$ are not exactly identical to the ranges of the hardnesses
used in Def., one can use two different procedures to decide if there are XRFs in the BATSE database.
First, one can either calculate the hardnesses for the energy ranges used 
in Def. for any burst, and then compare these calculated hardnesses with the Def. limit, or, second,
one can calculate the limiting hardness $H_{21}$ for XRFs,
and then compare it with the measured $H_{21}$ values. 
The limiting $H_{21}$ for XRFs is given by Eq.~(\ref{eq:H_XRF_def2_BATSE})
and hence the second procedure appears to be simpler.
Nevertheless, if there are measured spectral parameters for any burst, then both procedures are
possible, and it is obviously better option to provide both procedures and compare them.
Concerning the BATSE dataset we will use both procedures.

The methods used for the \textit{RHESSI} sample cannot be identical to that of the BATSE case,
because for \textit{RHESSI} there is no equivalent of the Spectral Catalog. This means that for
\textit{RHESSI} only the method, which is analogous to the method based on the BATSE Current Catalog, can be used.
In addition, a care is needed because the fluences are not given in units erg\,cm$^{-2}$.

\section{Fraction of XRFs in the BATSE database}
\label{sec:batse_results}

\begin{table*}
\caption{\label{tab:batse_H_50-100/25-50}
The numbers of events classified as XRFs ($H_{\frac{50-100}{25-50}} \le 0.76$) or GRBs ($H_{\frac{50-100}{25-50}} > 0.76$) by Def. in ProcI.
in the BATSE sample. ``Inter.'' means the intermediate group; ``No-group'' means that no group-membership has been assigned by \citet{ho06}.
The values not written in parentheses were obtained directly from the measured data.
The values written in parentheses are median and 90\,\% CL obtained by method described in Subsection~\ref{sec:batse_uncertainties_in_fractions}.}
\centering
\begin{tabular}{@{}cccccccccc@{}}
\tableline \\[-9pt]
GRBs/XRFs             & Total         & PL                & CPL               & Band              & Short               & Inter.             & Long                 & No-group    \\  
\tableline \\[-9pt]  
\multirow{2}{*}{GRBs} & 1605          & 154               & 904               & 547               & 424                 &  76                & 1055                 & 50          \\ [1pt]
                      & $(1593\pm 6)$ & $(152^{+4}_{-3})$ & $(898^{+4}_{-3})$ & $(542^{+4}_{-3})$ & $(427^{+19}_{-35})$ & $(31^{+62}_{-14})$ & $(1081^{+18}_{-30})$ & $(50\pm 1)$ \\ [7pt]
\multirow{2}{*}{XRFs} & 21            & 17                &  2                & 2                 & 3                   &  5                 & 11                   & 2           \\ [1pt]
                      & $(33\pm 6)$   & $(19^{+3}_{-4})$  & $(8^{+3}_{-4})$   & $(7^{+3}_{-4})$   & $(7\pm 3)$          & $(3^{+4}_{-2})$    & $(21^{+6}_{-5})$     & $(2\pm 1)$  \\ [7pt]
\multirow{2}{*}{Sum}  & 1626          & 171               & 906               & 549               & 427                 & 81                 & 1066                 & 52          \\ [1pt]
                      &               &                   &                   &                   & $(434^{+21}_{-36})$ & $(34^{+65}_{-15})$ & $(1103^{+18}_{-32})$ &             \\
\tableline
\end{tabular}
\end{table*}

\begin{table*}
\caption{\label{tab:batse_H_50-100/20-50_procII}
The numbers of events classified as XRFs ($H_{21\rm,CURR} \leq 0.6$) or GRBs ($H_{21\rm,CURR} > 0.6$) by Def. in ProcII,
i.e. with hardness below or above the limit given by Eq.~(\ref{eq:H_XRF_def2_BATSE}) in the BATSE sample.
``Inter.'' means the intermediate group; ``No-group'' means that no group-membership has been assigned by \citet{ho06}.
The values not written in parentheses were obtained directly from the measured data.
The values written in parentheses are median and 90\,\% CL obtained by method described in Subsection~\ref{sec:batse_uncertainties_in_fractions}.}
\centering
\begin{tabular}{@{}ccccccc@{}}
\tableline \\[-9pt]
GRBs/XRFs                   & Total              & $\;\;$ & Short               & Inter.             & Long                 & No-group        \\
\tableline \\[-9pt]
\multirow{2}{*}{GRBs}       & 1861               &        & 471                 & 84                 & 1302                 & 4               \\[1pt]
                            & $(1857^{+7}_{-6})$ &        & $(481^{+23}_{-39})$ & $(32^{+67}_{-16})$ & $(1338^{+20}_{-34})$ & $(4^{+0}_{-1})$ \\[7pt]
\multirow{2}{*}{XRFs}       &   71               &        &  23                 & 15                 & 32                   & 1               \\[1pt]
                            & $(75^{+6}_{-7})$   &        & $(19^{+7}_{-6})$    & $(14^{+8}_{-6})$   & $(40^{+5}_{-7})$     & $(1^{+1}_{-0})$ \\[7pt]
\multirow{2}{*}{Sum}        & 1932               &        & 494                 & 99                 & 1334                 & 5               \\[1pt]
                            &                    &        & $(501^{+23}_{-43})$ & $(45^{+77}_{-19})$ & $(1378^{+22}_{-38})$ &                 \\
\tableline
\end{tabular}
\end{table*}

\subsection{Procedure I}
\label{subsec:batse_procI}

In the first procedure (shortly ``ProcI'') we take the measured spectral parameters of the observed 
bursts, then calculate the hardness ratios for the energy ranges used in
Def., and using the obtained hardnesses we determine whether an object
is classified as an XRF.

We use the Spectral Catalog. There are all bursts fitted by the PL model.
If there is any other fit using the other model(s), then we take the best-fit one,
i.e., the model with the highest goodness-of-fit (GOF). 
We make a restriction that we accept a fit only if ${\rm GOF} >5$\,\%. 
A further restriction is that a burst must also have a measured $T_{90}$ duration with uncertainties in the
Current Catalog. After these two restrictions the number of remaining events is 
1626, which is the sample used in ProcI. From these events PL, CPL, and Band function
were the best-fit models in 171, 906, and 549 cases, respectively.
From 1626 events 427 ones belong to the short group,
81 objects belong to the intermediate one, 1066 objects belong to the long group,
and 52 events are not separated into the groups by \citet{ho06}.

To provide ProcI for the 1626 events we calculate the following hardnesses for any burst: 
$H_{\frac{50-100}{25-50}}$, which is the hardness ratio used in Def.;
$H_{\frac{50-100}{20-50}} \equiv H_{21}$, which is the hardness ratio calculable from the Current Catalog.
The second hardness $H_{21}$ is calculated for checking because then we have two
$H_{21}$ hardness values for a given object: one from the Spectral Catalog (in what follows
denoted as $H_{21\rm,SPEC}$) and one from the Current Catalog (in what follows denoted as $H_{21\rm,CURR}$). 
Comparison of these two values allows us to test the precision of ProcI.

The numbers of events classified as XRFs or GRBs by Def. in ProcI
are presented in Table~\ref{tab:batse_H_50-100/25-50}. We find that there are
1605 objects (98.7\,\% of total 1626 events) with hardness $H_{\frac{50-100}{25-50}} > 0.76$ and 
21 (1.3\,\%) objects with hardness $H_{\frac{50-100}{25-50}} \le 0.76$.
This means that there are 1605 objects classified as GRBs and 21 objects classified as XRFs.
The numbers for the three spectral models separately are provided, too. 
The CPL model and Band function give similar fractions, being comparable with the fractions of the
whole sample, but for the PL models the fraction of XRFs is remarkably high (17 [9.9\,\%] from 171 events).
Concerning the groups there are only few XRFs for the short bursts (3 [0.7\,\%] from 427 objects);
for the intermediate group the fraction of XRFs is the highest (5 [6\,\%] from 81 objects);
for the long bursts the fraction of XRFs is practically identical to that of the whole sample
(11 [1.0\,\%] by Def. from 1066 objects); and for the bursts with unknown group-membership
the fraction of XRFs is (2 [4\,\%] from 52 objects).

\subsection{Procedure II}
\label{subsec:batse_procII}

In the second procedure (shortly ``ProcII'') we take the measured hardness ratios $H_{21\rm,CURR}$
of the events from the Current Catalog and compare them with the limiting hardness for XRFs
given by Eq.~(\ref{eq:H_XRF_def2_BATSE}).

\begin{figure}[t]
\centering
\includegraphics[width=0.48\textwidth]{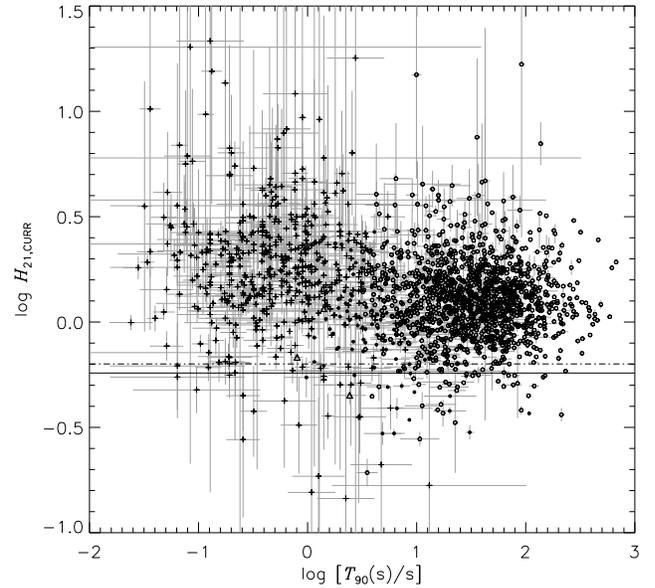}
\caption{
The hardness ratio $H_{21\rm,CURR}$ 
vs. $T_{90}$ durations of 1932 BATSE bursts with identified group of short (crosses),
intermediate (full circles), long bursts (open circles), and ones without assigned group-membership (triangles).
The horizontal solid line is the XRF limit from Eq.~(\ref{eq:H_XRF_def2_BATSE}).
The objects above this line are not classified as XRFs; the objects below this line are classified as XRFs using ProcII.
The horizontal dash-and-dot line marks the confidence interval (CI) meaning that any object lying above has a probability of $<10^{-5}$
to be an XRF by Def. This CI was obtained from the simulation shown in Fig.~\ref{fig:batse_H_50-100/25-50_H21_simulated}.
The plotted error bars were calculated from the uncertainties given in the BATSE Current Catalog and
using the error propagation theory.
}
\label{fig:batse_H21_curr_T90}
\end{figure}

Fig.~\ref{fig:batse_H21_curr_T90} shows the hardness ratio $H_{21\rm,CURR}$ against the $T_{90}$ duration.
The sample here contains 1932 events because in this figure all BATSE GRBs
having defined hardness ratios $H_{21\rm,CURR}$ and durations $T_{90}$, both with defined uncertainties,
in the Current Catalog are plotted. 
The horizontal solid line is the limit for XRFs given by Eq.~(\ref{eq:H_XRF_def2_BATSE}).
The objects below (above) this line are (are not) XRFs using ProcII.
The number of GRBs with hardness lower or equal to the XRF limit is 71, i.e., 3.7\,\% of the whole sample.
For the three groups separately one obtains:
from the 494 short bursts 23 objects (4.7\,\% of the short ones) are below the limit;
from the 99 intermediate GRBs 15 objects are below the limit (15\,\% of the intermediate ones);
from the 1334 long ones 32 objects are below the limit (2.4\,\% of the long ones).
The numbers are also summarized in Table~\ref{tab:batse_H_50-100/20-50_procII}.

The two procedures gave similar - but not exactly identical - results.
The second procedure gave higher fraction of XRFs in total sample as well as in the individual groups.
Also the samples, used in ProcI and ProcII, respectively, were not the same.
Hence, a discussion of the two procedures with their uncertainties is clearly required.

\subsection{Precision of Procedure I}
\label{subsec:batse_precis_procI}

The first checking of ProcI can be done as follows.
One may expect that the two $H_{21}$ hardnesses are identical. Hence, statistically, $H_{21\rm,CURR}$ and
$H_{21\rm,SPEC}$ should give a linear relation $H_{21\rm,SPEC} = H_{21\rm,CURR}$ with an acceptable GOF.

\begin{figure}[H]
\centering
\includegraphics[width=0.365\textwidth]{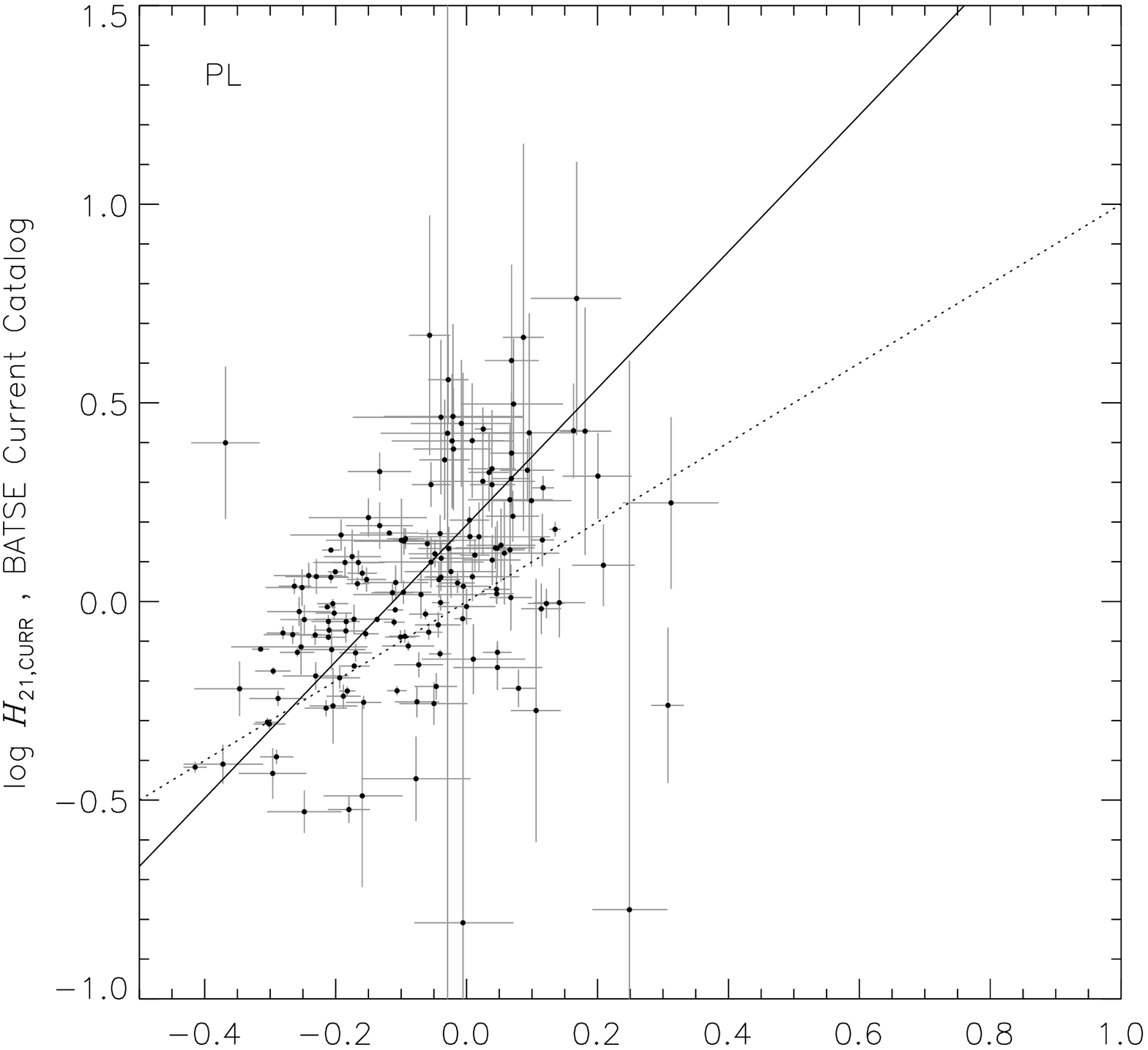}
\includegraphics[width=0.365\textwidth]{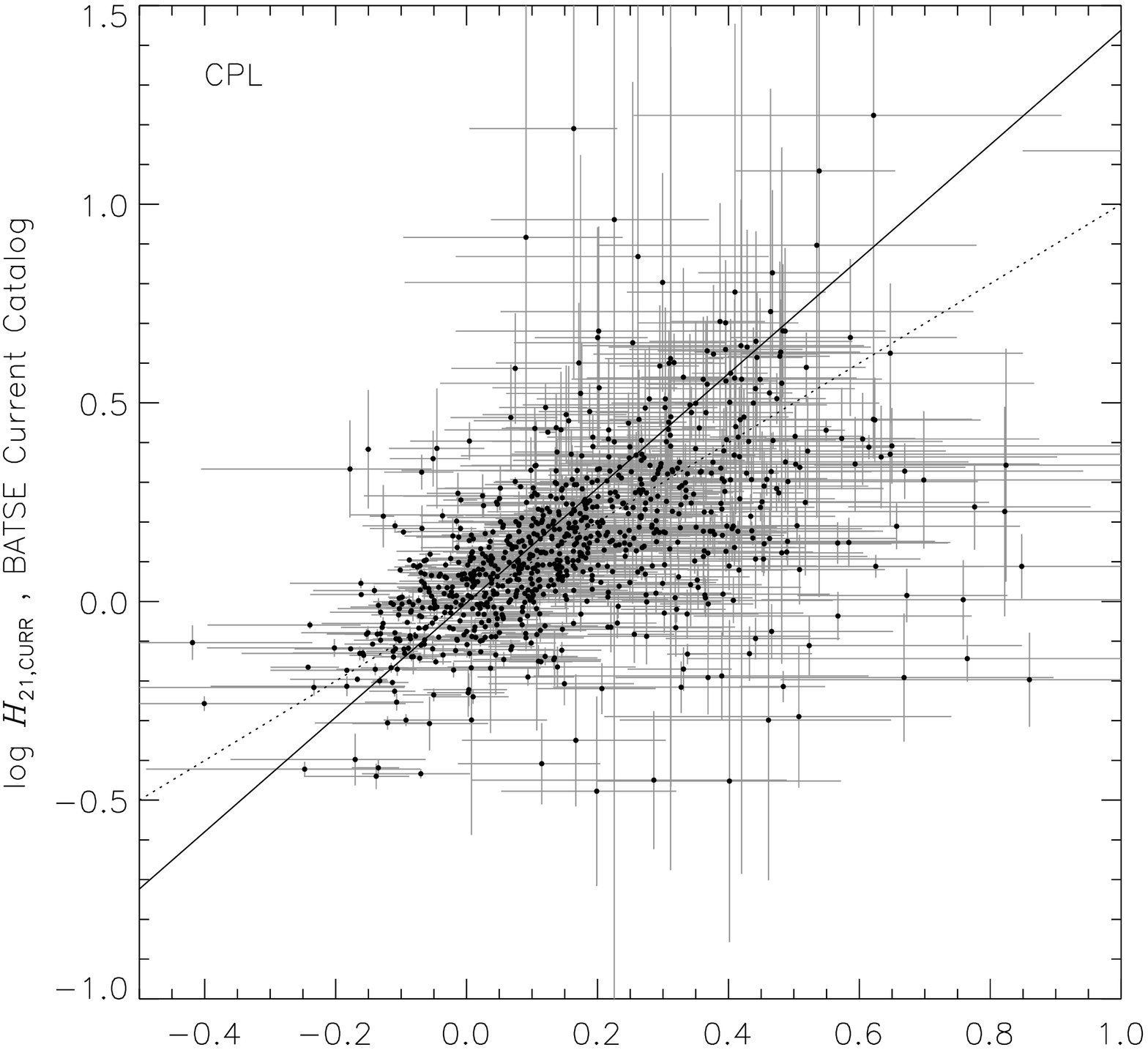}
\includegraphics[width=0.365\textwidth]{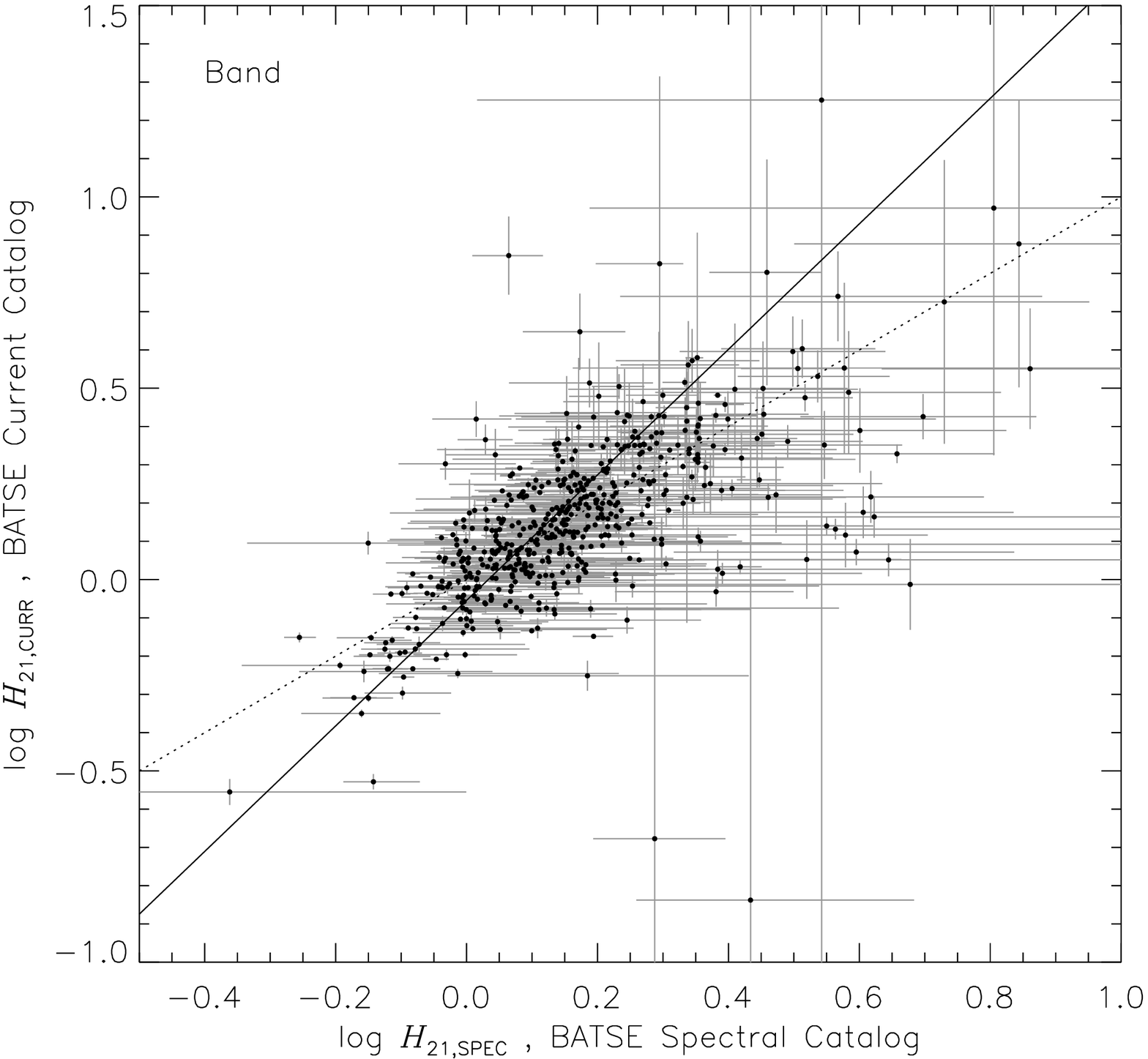}
\caption{
A comparison of the two kinds of hardnesses $H_{21}$ for 1560 bursts in the BATSE
sample are shown separately for three spectral models.
The error bars of $\log H_{21\rm,SPEC}$ delimit the 68\,\% mid quantile interval and were calculated using MC simulations
as described in Section~\ref{subsec:batse_precis_procI}.
The error bars of $\log H_{21\rm,CURR}$ are $1 \sigma$ statistical uncertainties and were calculated from
the uncertainties given in the BATSE Current Catalog and using the error propagation theory.
The solid line is the best fit. The dotted line denotes relation $\log H_{21\rm,CURR} = \log H_{21\rm,SPEC}$.
}
\label{fig:batse_H21_curr_H21_spec}
\end{figure}

A comparison of the two kinds of $H_{21}$ is shown in Fig.~\ref{fig:batse_H21_curr_H21_spec}.
There are 1560 bursts having both hardnesses derived. This number follows from the fact that in the sample,
containing 1626 GRBs and used in ProcI, 66 objects did not have measured fluences needed for the comparison with
$H_{21\rm,CURR}$. We fitted the values $\log H_{21\rm,CURR}$ and $\log H_{21\rm,SPEC}$
using the linear least square fitting \citep{press07} assuming a linear relation
$\log H_{21\rm,CURR} = a + b \log H_{21\rm,SPEC}$.
The two parameters should have the values $a = 0$ and $b=1$. The fitting was done for the three spectral models separately.

To obtain the uncertainties in $H_{21\rm,SPEC}$, needed to obtain the GOF, we proceeded as follows. 
Using the measured spectral parameters and their uncertainties from the Spectral 
Catalog we generated 10\,000 Monte Carlo (MC) simulated spectra for each event.
We assumed that the given uncertainties in the spectral parameters in the Spectral Catalog
are one standard deviations of the Gaussian distribution.
Concerning the Band function we required $\alpha>-2.0$, $\beta<\alpha$ and $E_{\rm peak}>0$.
Sometimes the simulated parameter $\beta$ was lower than -20.
This occurred when there was a relatively high uncertainty of this particular parameter.
If the energy $E<100$\,keV then the integrated spectrum can reach an extremely high values
and lead to an overflow of the program during the numerical integration.
Therefore, we also introduced a condition: if $\beta < -20$ then we set $\beta = -20$.
Concerning CPL we required $E_{\rm peak}>0$.
If these criteria were not met, the spectral parameters were generated again.

The simulated spectra were numerically integrated to obtain fluences and
then the logarithms of hardnesses. Thus, for each event, a distribution of 10\,000 
logarithmic hardnesses was obtained. In such a distribution the quantiles $Q(0.3173/2)$ and $Q(1-0.3173/2)$ 
delimit the 68\,\% mid quantile interval. Doing so one derives the Gaussian-equivalent 1-$\sigma$ interval delimited by
$-\sigma_{\log H_{21\rm,SPEC}}$ and $+\sigma_{\log H_{21\rm,SPEC}}$ asymmetric uncertainties, respectively.
We used routine ``FITEXY'' of the IDL\footnote{http://www.harrisgeospatial.com/ProductsandSolutions/\\
GeospatialProducts/IDL.aspx}
Astronomy Users Library\footnote{http://idlastro.gsfc.nasa.gov/} for fitting because it accounts for the uncertainties
along both axes. This routine is based on the procedure described by \citet{press07}.
It requires symmetric uncertainties. Therefore we used the mean value
$\sigma_{\log H_{21\rm,SPEC}}$ of the asymmetric uncertainties
$-\sigma_{\log H_{21\rm,SPEC}}$ and $+\sigma_{\log H_{21\rm,SPEC}}$.
The symmetric uncertainties in $\log H_{21\rm,CURR}$ were calculated from the error propagation theory
of the uncertainties in the fluences mentioned in the BATSE Current Catalog.

The best fit for PL spectral model (158 bursts) is following: 
$a = 0.193 \pm 0.006$, $b = 1.719 \pm 0.044$ with $\chi^2 = 1815$;
the best fit for CPL spectral model (868 bursts) is following:
$a = -0.003 \pm 0.003$, $b = 1.441 \pm 0.015$ with $\chi^2 = 8863$;
and the best fit for the Band function (534 bursts) is following:
$a = -0.054 \pm 0.004$, $b = 1.640 \pm 0.020$ with $\chi^2 = 4596$.
In all three cases it holds that GOF is in essence zero, and hence the fits are {\it not} acceptable from the statistical point of view. 
From the three figures it follows that fitting with a straight line is incorrect, because for some objects the one sigma error bars 
are far away from the best fit line. The essentially zero GOF value claims that the scattering is too large.
In addition, both $a$ and $b$ lie away from the expected $a=0$; $b=1$ values.
For all three spectral models the slope $b$ is higher than $b=1$ which means that the relation between
$\log H_{21\rm,CURR}$ and $\log H_{21\rm,SPEC}$ is systematically steeper than it should be
in case of $H_{21\rm,CURR} = H_{21\rm,SPEC}$. This means there is a systematic error between
$H_{21\rm,CURR}$ and $H_{21\rm,SPEC}$.
This systematic error can be probably explained by different time intervals used in the fitting of the time-averaged fluence
spectra in the Spectral Catalog and in the calculation of the fluences in the Current Catalog.
Due to the well known time evolution of the GRB spectra over the different periods of their light curves
(see e.g. \citet{kane06}) such a systematic discrepancy could occur. In the Spectral Catalog the time-integrated fluence spectra
were estimated over the duration of the observed emission, where the observed emission is defined as $3.5\sigma$
over the estimated background in the $20-2000$\,keV energy range \citep{gold13}.
In the Current Catalog the fluences were obtained by the integration over the $T_{90}$ durations
(G. J. Fishman, private communication).

\begin{figure}[t]
\centering
\includegraphics[width=0.48\textwidth]{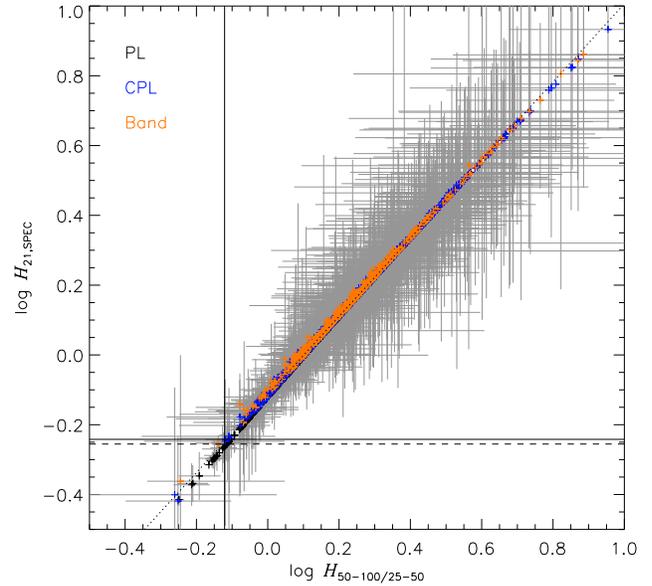}
\caption{The hardness ratio $H_{21\rm,SPEC}$ vs. $H_{\frac{50-100}{25-50}\rm}$,
both calculated from the Spectral Catalog,
for 1626 BATSE bursts. The solid vertical line denotes the limit from Def., 
the horizontal solid line denotes the limit from Eq.~(\ref{eq:H_XRF_def2_BATSE}).
The dashed horizontal line denotes the highest hardness 
$\log H_{21\rm,SPEC} = -0.26$ in the left placed quadrants.
This value can also serve as an empirical XRF upper limit for $H_{21\rm,SPEC}$ on
Fig.~\ref{fig:batse_H21_spec_T90}.
The dotted line is the best linear fit between the logarithmic values.}
\label{fig:batse_H_50-100/25-50_H21_spec}
\end{figure}

A second check of ProcI can be done as follows. One may verify the correctness of the values $H_{21\rm,SPEC}$
if they are compared with $H_{\frac{50-100}{25-50}}$.
We note that hardnesses $H_{\frac{50-100}{25-50}}$ and $H_{21\rm,SPEC}$ are both calculated from
the Spectral Catalog and thus they can be compared regardless the systematic difference between the $H_{21\rm,SPEC}$
and $H_{21\rm,CURR}$ values.
This comparison is done in Fig.~\ref{fig:batse_H_50-100/25-50_H21_spec}.
It shows a strong correlation between both hardnesses. Pearson correlation coefficient is $\rho = 0.998$.
Moreover, the XRF limit from Eq.~(\ref{eq:H_XRF_def2_BATSE}) and the highest value in the left placed quadrants 
are different only by $0.02$.
This can also serve as an empirical upper value of $H_{21\rm,SPEC}$ for XRFs obtained from Def. The number of objects are shown in 
Table~\ref{tab:batse_H_50-100/25-50_H21_spec}. They can also serve as a test of ProcI with Def.
The number of XRFs following from the limit given by Eq.~(\ref{eq:H_XRF_def2_BATSE}) is 29 (1.8\,\% of the whole sample).
For the three groups separately one has:
from the 427 short bursts 4 objects (0.9\,\%) are XRFs;
from the 81 intermediate-duration bursts 8 objects (10\,\%) are XRFs;
from the 1066 long ones 15 objects (1.4\,\%) are XRFs.
In addition, from the 52 events without group-membership 2 objects (4\,\%) are XRFs
(again the group-membership is taken from \citet{ho06}).
It can be said that the numbers and their trends are similar to the values of Table~\ref{tab:batse_H_50-100/25-50},
i.e. similar to the results of ProcI.

\begin{table*}
\caption{\label{tab:batse_H_50-100/25-50_H21_spec}
Numbers of BATSE events in four segments separated by the solid-lines in Fig.~\ref{fig:batse_H_50-100/25-50_H21_spec}.; 
``Inter.'' means the intermediate group; ``No-group'' means that no group-membership is assigned by \citet{ho06}.
The values not written in parentheses were obtained directly from the measured data.
The values written in parentheses are median and 90\,\% CL obtained by method described in Subsection~\ref{sec:batse_uncertainties_in_fractions}.}
\centering
\begin{tabular}{@{}ccccccccc@{}}
\tableline \\[-9pt]
Quadrant                      & Total         & PL                &  CPL              & Band              & Short               & Inter.             & Long                 & No-group        \\
\tableline \\[-9pt]
\multirow{2}{*}{Top-left}     & 0             &  0                & 0                 & 0                 & 0                   & 0                  & 0                    & 0               \\[1pt]
                              & (0)           &  (0)              & (0)               & (0)               & (0)                 & (0)                & (0)                  & (0)             \\[7pt]
\multirow{2}{*}{Top-right}    & 1597          &  148              & 902               & 547               & 423                 & 73                 & 1051                 & 50              \\[1pt]
                              & $(1590\pm 6)$ & $(151^{+3}_{-4})$ & $(898^{+3}_{-4})$ & $(542^{+3}_{-4})$ & $(427^{+19}_{-36})$ & $(31^{+62}_{-14})$ & $(1078^{+20}_{-28})$ & $(50\pm 1)$     \\[7pt]
\multirow{2}{*}{Bottom-left}  & 21            &  17               & 2                 & 2                 & 3                   & 5                  & 11                   & 2               \\[1pt]
                              & $(33\pm 6)$   & $(19^{+3}_{-4})$  & $(8^{+3}_{-4})$   & $(7^{+3}_{-4})$   & $(7\pm 3)$          & $(3^{+4}_{-2})$    & $(21^{+6}_{-5})$     & $(2\pm 1)$      \\[7pt]
\multirow{2}{*}{Bottom-right} & 8             &  6                & 2                 & 0                 & 1                   & 3                  & 4                    & 0               \\[1pt]
                              & $(3\pm 3)$    & $(1^{+3}_{-1})$   & $(0^{+2}_{-0})$   & $(0^{+2}_{-0})$   & $(0^{+1}_{-0})$     & $(0^{+1}_{-0})$    & $(2^{+3}_{-2})$      & $(0^{+1}_{-0})$ \\[7pt]
Sum                           & 1626          &  171              & 906               & 549               & 427                 & 81                 & 1066                 & 52              \\
\tableline
\end{tabular}
\end{table*}

\begin{figure*}
\centering
\includegraphics[height=0.31\textheight]{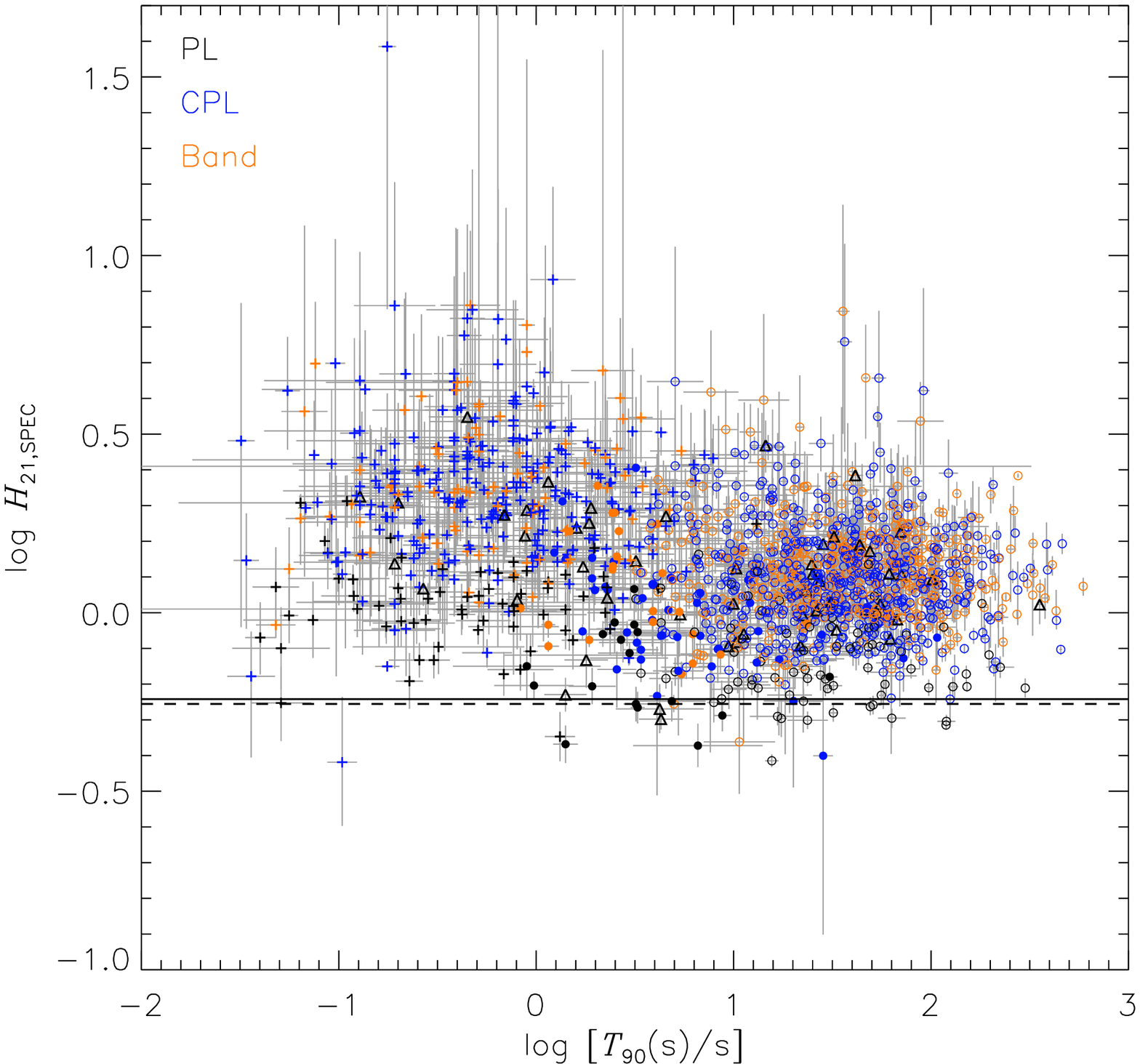}
\includegraphics[height=0.31\textheight]{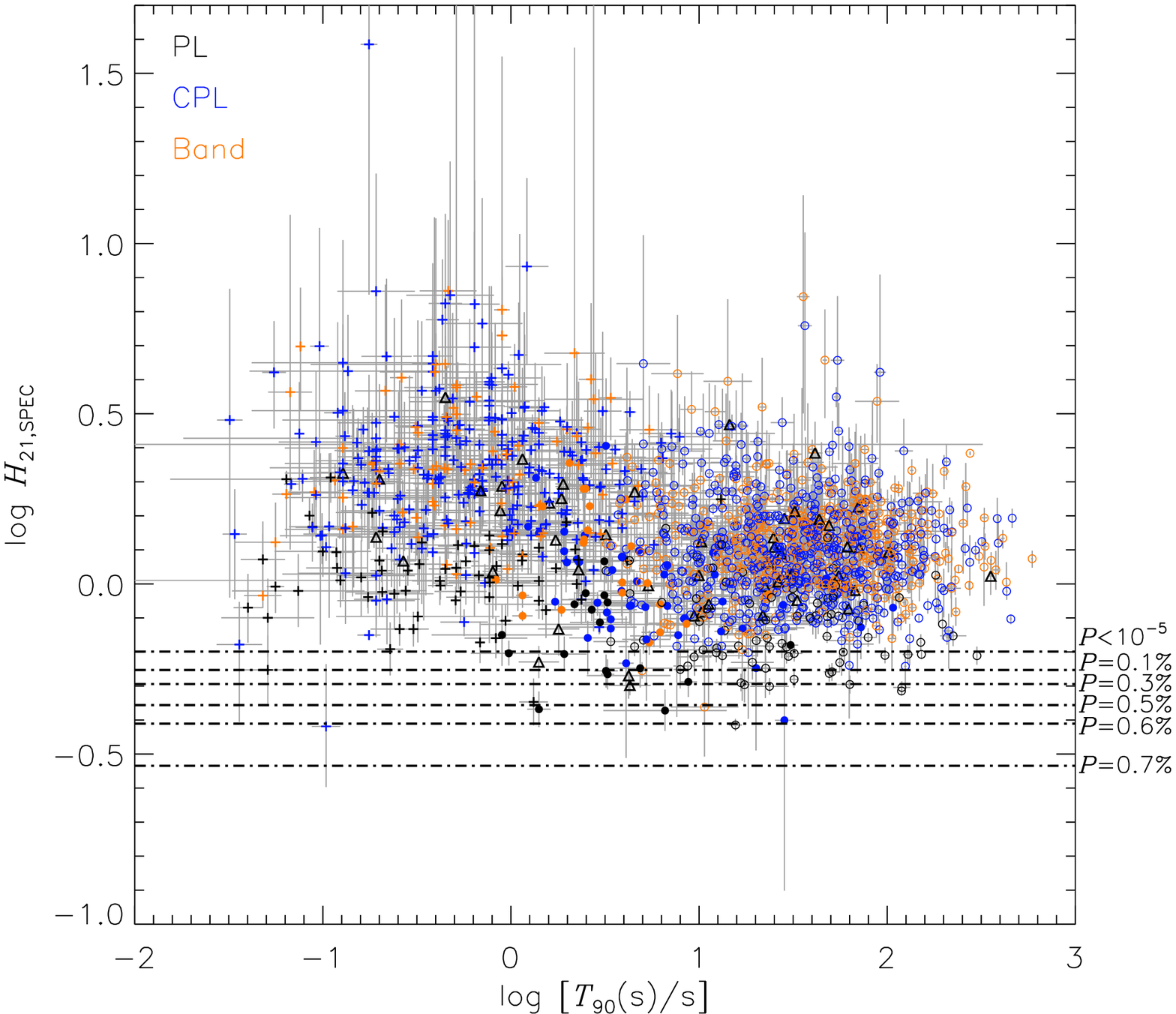} 
\caption{
Left: The hardness ratios $H_{21\rm,SPEC}$ vs. $T_{90}$ durations of the BATSE GRB sample containing
1626 GRBs. The crosses, full and open circles have the same meaning as the objects on Fig.~\ref{fig:batse_H21_curr_T90};
the open triangles denote GRBs with no group-membership, because they are not mentioned by \citet{ho06}.
The dashed line denotes the highest hardness of an observed event still classified as XRF using Def.
(see Fig.~\ref{fig:batse_H_50-100/25-50_H21_spec}).
It can serve as an empirical XRF upper limit.
The solid line defines the limit from Eq.~(\ref{eq:H_XRF_def2_BATSE}).
Right: A similar plot, but with displayed dash-and-dot lines marking the confidence intervals (CIs) and meaning
that any object above the given line has a probability of $<10^{-5}$, 0.1\,\%, 0.3\,\%, 0.5\,\%, 0.6\,\%, and 0.7\,\%,
respectively, to be an XRF by Def. These CIs were obtained from the simulation shown in Fig.~\ref{fig:batse_H_50-100/25-50_H21_simulated}.
The error bars of $\log H_{21\rm,SPEC}$ delimit the 68\,\% mid quantile interval
and were calculated using MC simulations as described in Section~\ref{subsec:batse_precis_procI}.
Error bars of $T_{90}$ were obtained from the uncertainties given in the BATSE Current Catalog.
Different colors of data points correspond to different best-fit spectral models.
}
\label{fig:batse_H21_spec_T90}
\end{figure*}

Fig.~\ref{fig:batse_H21_spec_T90} illustrates bursts with determined $H_{21\rm,SPEC}$ divided into the short,
intermediate and long groups according to the group-membership published by \citet{ho06}.
This figure is similar to Fig.~\ref{fig:batse_H21_curr_T90}, but it uses the sample containing
1626 events with the $H_{21\rm,SPEC}$ values. The horizontal solid line is again 
the limit for XRFs given by Eq.~(\ref{eq:H_XRF_def2_BATSE}). The empirical limit from
Fig.~\ref{fig:batse_H_50-100/25-50_H21_spec} (i.e., from Def.) is also shown.
Note that in Fig.~\ref{fig:batse_H21_spec_T90} there is not seen any clear separation between XRFs and GRBs.
It suggests that there might be no astrophysical difference between XRFs and GRBs.

Reviewing the uncertainties of ProcI, one can see that the values $H_{21\rm,SPEC}$
and $H_{21\rm,CURR}$ do not give an acceptable one-to-one correspondence for a given object.  
From the spectral models used the PL models gave the highest difference between the fit and the expected
$\log H_{21\rm,SPEC} = \log H_{21\rm,CURR}$, however one cannot claim any significant differences
among the spectral models.

\subsection{Precision of Procedure II}
\label{sec:batse_precision_of_procII}

In order to check the precision of ProcII one has to check the precision of the limit from Eq.~(\ref{eq:H_XRF_def2_BATSE}).
To provide this checking, we performed numerical simulations as follows.
We took the best fit spectral parameters of the Band function of GRBs in the BATSE Spectral Catalog
for the cases where the GOF was greater than 5\,\%.
From the 2106 bursts in the Catalog 1548 events had determined parameters of the Band function,
then from the 1548 events 1178 ones had GOF $>5\,\%$.
The parameters were taken from the ranges: $-2 < \alpha \leq2$, $\beta \ge -5$,
and $E_{\rm peak} \ge 7$\,keV, which further limited the size of the sample to 1077 events.
The obtained distributions of these spectral parameters are shown in Fig.~\ref{fig:batse_band_spec_params}.
Note that similar distributions were published already in \citet{pre00,bis11, gold13}.
Using these distributions we simulated 1~million mock GRB spectra and then
calculated the hardnesses $H_{\frac{50-100}{25-50}}$, and $H_{21}$.

The result of the simulation is presented in Fig.~\ref{fig:batse_H_50-100/25-50_H21_simulated}.
The figure shows the relation between the hardness $H_{\frac{50-100}{25-50}}$ used by Def. and the hardness $H_{21}$.
In what follows we mark the number of XRFs in the simulation due to Def. ($H_{\frac{50-100}{25-50}}<0.76$) as $n_{\rm XRF}$.
Oppositely we mark the number of GRBs in the simulation due to Def. ($H_{\frac{50-100}{25-50}}>0.76$) as $n_{\rm GRB}$.
It holds that if above an arbitrarily chosen horizontal line defined by $H_{\rm 21,i}$ there are $n_{\rm XRF,i}$ and $n_{\rm GRB,i}$ events then
for any object above this line the probability that it is an XRF is $P=n_{\rm XRF,i}/(n_{\rm XRF,i} + n_{\rm GRB,i})$.
On the other hand, the probability that an object below this line is not an XRF is
$P^{\ast}=n^{\ast}_{\rm GRB,i}/(n^{\ast}_{\rm XRF,i} + n^{\ast}_{\rm GRB,i})$, where $n^{\ast}_{\rm GRB,i}$ and $n^{\ast}_{\rm XRF,i}$
are the numbers of the events below this line.
The probability $P$ above the solid horizontal line $H_{21}\approx0.6$ given by Eq.~(\ref{eq:H_XRF_def2_BATSE}) is
$P=5\times10^{-2}\,\%$. This means the probability of an event above this line to be an XRF due to Def.
The probability $P^{\ast}$ below this solid horizontal line is $P^{\ast}=0.967$.
This means that above the line given by Eq.~(\ref{eq:H_XRF_def2_BATSE}) there is only
$P=5\times10^{-2}\,\%$ probability for an XRF to be wrongly identified, but below this line an ordinary GRB can well be
wrongly identified as an XRF by a high 96.7\,\% probability.
In Fig.~\ref{fig:batse_H_50-100/25-50_H21_simulated} there are marked following confidence intervals (CIs):
$P<10^{-5}$, $P=0.1\,\%$, $P=0.3\,\%$, $P=0.5\,\%$, $P=0.6\,\%$, and $P=0.7\,\%$, respectively.
These confidence intervals are also shown in Fig.~\ref{fig:batse_H21_spec_T90} illustrating their effects on the real data.
Table~\ref{tab:batse_H_50-100/25-50_H21_simulated} summarizes the values from the simulation.

It is interesting to compare the value $\log H_{21}=$ $-0.20$ for the confidence interval $P<10^{-5}$
from Fig.~\ref{fig:batse_H_50-100/25-50_H21_simulated}, i.e. the highest hardness of an XRF by Def. obtained from the simulations,
with the limiting XRF hardness obtained by Eq.~(\ref{eq:H_XRF_def2_BATSE}), $\log H_{21} \approx -0.24$.
Both values are very similar, which supports the correctness of the limit following from Eq.~(\ref{eq:H_XRF_def2_BATSE}).

All this means that the fractions of XRFs - summed in Table~\ref{tab:batse_H_50-100/20-50_procII} - are confirmed.  

\begin{table}[h]
\caption{\label{tab:batse_H_50-100/25-50_H21_simulated}
The table summarizes the confidence intervals $P$ marked at given hardnesses
$H_{21,\rm{i}}$ in Fig.~\ref{fig:batse_H_50-100/25-50_H21_simulated}.
The quantities $n_{\rm XRF,i}$ ($H_{21} \ge H_{21,\rm{i}}$ and $H_{\frac{50-100}{25-50}} \le 0.76$)
and $n_{\rm GRB,i}$ ($H_{21} \ge H_{21,\rm{i}}$ and $H_{\frac{50-100}{25-50}} > 0.76$)
are the numbers of XRFs and GRBs by Def. with hardness $H_{21} \ge H_{21,\rm{i}}$ in the simulation, respectively.
The $P$ is the probability that any object with $H_{21} \ge H_{21,\rm{i}}$ is an XRF by Def.
}
\centering
\begin{tabular}{@{}cccc@{}}
\tableline \\[-9pt]
$\log H_{21,\rm{i}}$  &   $n_{\rm XRF,i}$  &  $n_{\rm GRB,i}$  &  $P$\,(\%)  \\
\tableline \\[-9pt]
-0.20                          &    1               & 989370            &  $<10^{-3}$ \\
-0.25                          &  994               & 992630            &  0.1        \\
-0.29                          & 2987               & 992678            &  0.3        \\
-0.36                          & 4989               & 992678            &  0.5        \\
-0.41                          & 5993               & 992678            &  0.6        \\
-0.53                          & 6998               & 992678            &  0.7        \\
\tableline
\end{tabular}
\end{table}

\subsection{Uncertainties in the numbers of GRBs/XRFs}
\label{sec:batse_uncertainties_in_fractions}

In order to determine the uncertainties in the numbers of GRBs and XRFs mentioned in
Tables~\ref{tab:batse_H_50-100/25-50}-\ref{tab:batse_H_50-100/25-50_H21_spec} we proceed as follows.
\begin{enumerate}
\item The uncertainties in the $T_{90}$ duration effect the classification of the bursts into the
short, intermediate, and long groups. To account for that we take BATSE Current Catalog and
calculate hardnesses $H_{32,\rm{CURR}}$, which is the hardness used by \citet{ho06} in the GRB classification.
We take only the events for which the hardness ratios $H_{32,\rm{CURR}}$
and durations $T_{90}$ with measured uncertainties are defined. This gives us a sample of 1954 events.
\newline
\item We calculate the uncertainties in log \break $H_{32,\rm{CURR}}$ from the standard error propagation theory.
\newline
\item Next, for each event, we add the random Gaussian noise to $\log H_{32,\rm{CURR}}$ with the mean value the same as the
measured value and the standard deviation the same as $1\sigma$ uncertainty in $\log H_{32,\rm{CURR}}$.
Thus we obtain shifted values $\log H'_{32,\rm{CURR}}$.
\newline
\item Then, for each event, independently on log \break $H_{32,\rm{CURR}}$, we add the random Gaussian noise to $\log T_{90}$
with the mean value the same as the measured value and the standard deviation the same as $1\sigma$
uncertainty in $\log T_{90}$. Thus we obtain shifted values $\log T'_{90}$.

\item We follow the method described by \citet{ho06} to determine a new group-membership assigned to each event
in this ``shifted'' sample. This means we follow their Eqs.~(1-2) to fit the sum of three bivariate
Gaussian distributions on the \{$\log T'_{90}; \log$ $H'_{32,\rm{CURR}}$\} plane, then maximize the Likelihood
function $L$ (see their Eq.~(3)), calculate the membership probabilities (see their Eq.~(5)), and finally
from the maxima of the membership probabilities a group-membership is assigned to each event.
\newline
\item We repeat steps $3.-5.$ $1000\times$ which gives us 1000 unique lists each containing 1954 bursts with
assigned group-memberships.
\end{enumerate}

In the steps 3. and 4. we restrict the ``shifted'' hardness and duration to be in the range
$-3 \le \log H'_{32,\rm{CURR}} \le 4$ and $-3 \le \log T'_{90}$ $\le 4$.
If the ``shifted'' $\log H'_{32,\rm{CURR}}$ or $\log T'_{90}$ were not in this range we generated the
random Gaussian noise again. The reason is that if an event has relatively high uncertainties
it can be often shifted too far from the overall distribution and the likelihood $L$ would go to $-\infty$.
There were two such events with very high uncertainties in hardness which made fitting impossible, that is
why we introduced these restrictions. These restrictions are wide enough and they do not effect the vast majority of the events.

We used the ``AMOEBA'' function under the IDL programming language for finding the best fit of the sum of three bivariate Gaussian distributions
by maximizing the likelihood $L$. The ``AMOEBA'' function is based on the routine with the same name described in \citet{press07}.

Concerning the estimation of the uncertainties in the numbers of XRFs/GRBs in Table~\ref{tab:batse_H_50-100/20-50_procII} we
proceed as follows.

\begin{enumerate}
\item For each event of the 1932 bursts displayed in Fig.~\ref{fig:batse_H21_curr_T90},
we add the random Gaussian noise to $\log H_{21,\rm{CURR}}$ with the mean value the same as the
measured value and the standard deviation the same as $1\sigma$ uncertainty in $\log H_{21,\rm{CURR}}$.
Thus we obtain ``shifted'' values $\log H'_{21,\rm{CURR}}$.

\begin{figure}[H]
\centering
\includegraphics[width=0.45\textwidth]{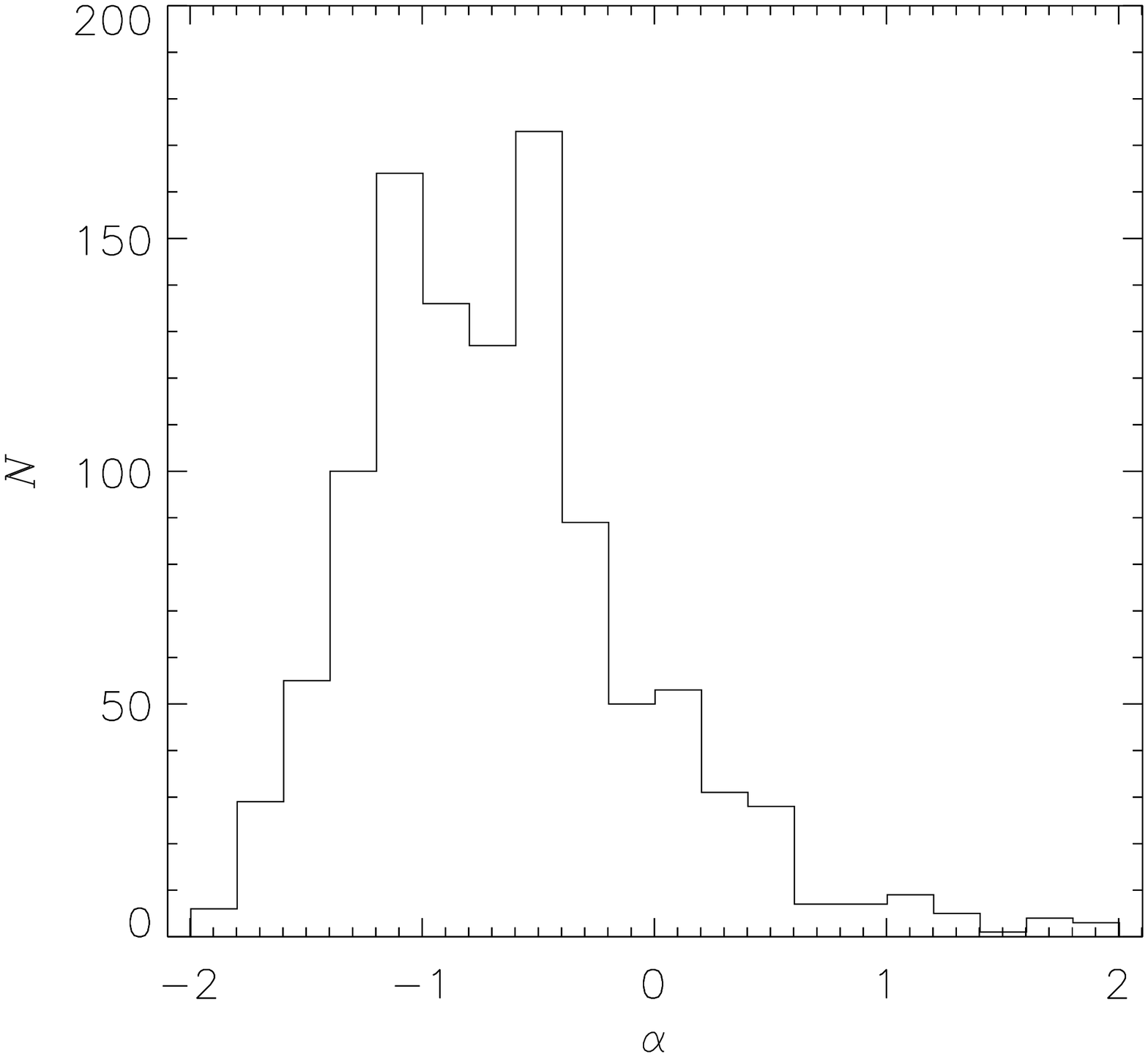}
\includegraphics[width=0.45\textwidth]{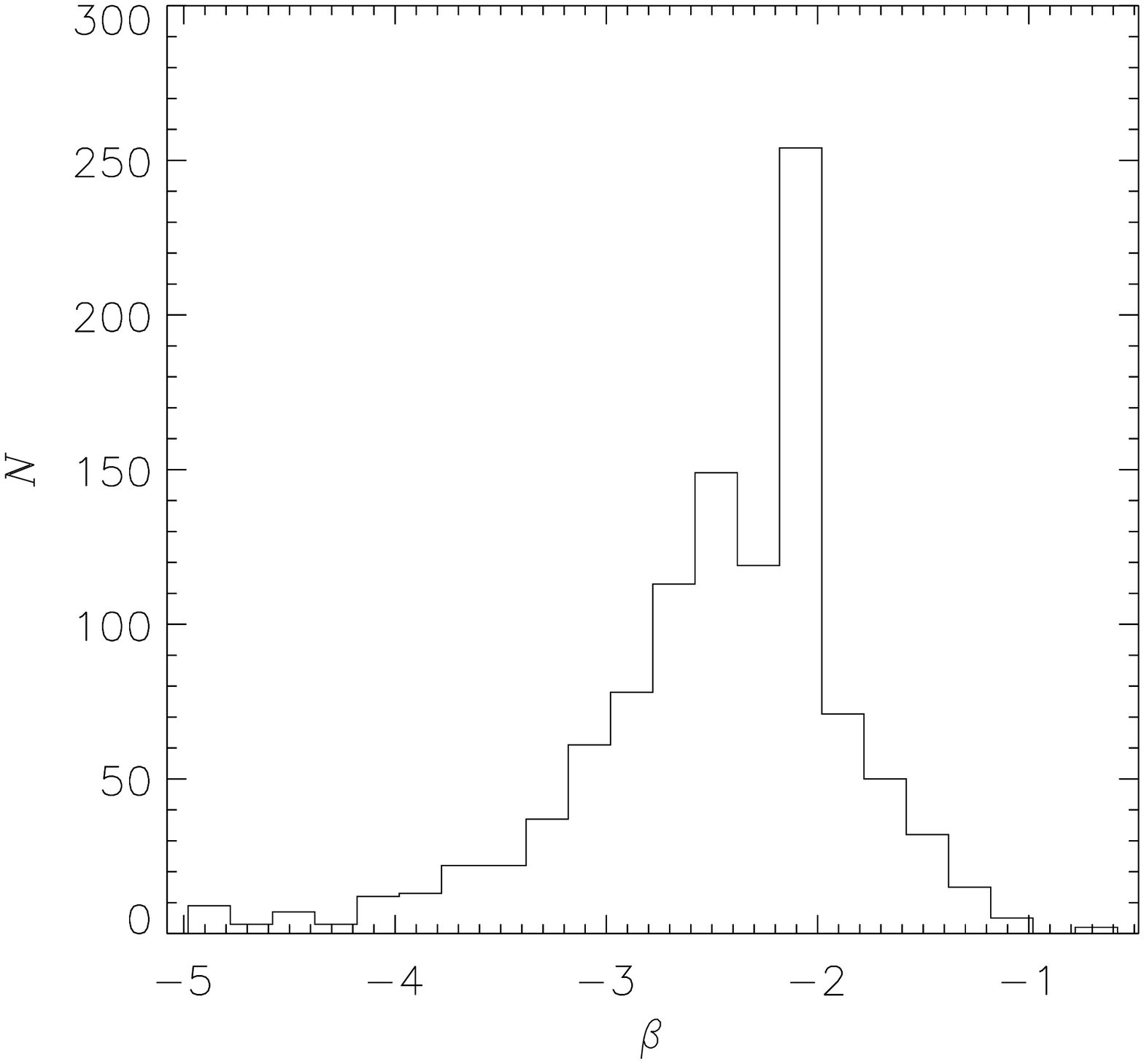}
\includegraphics[width=0.45\textwidth]{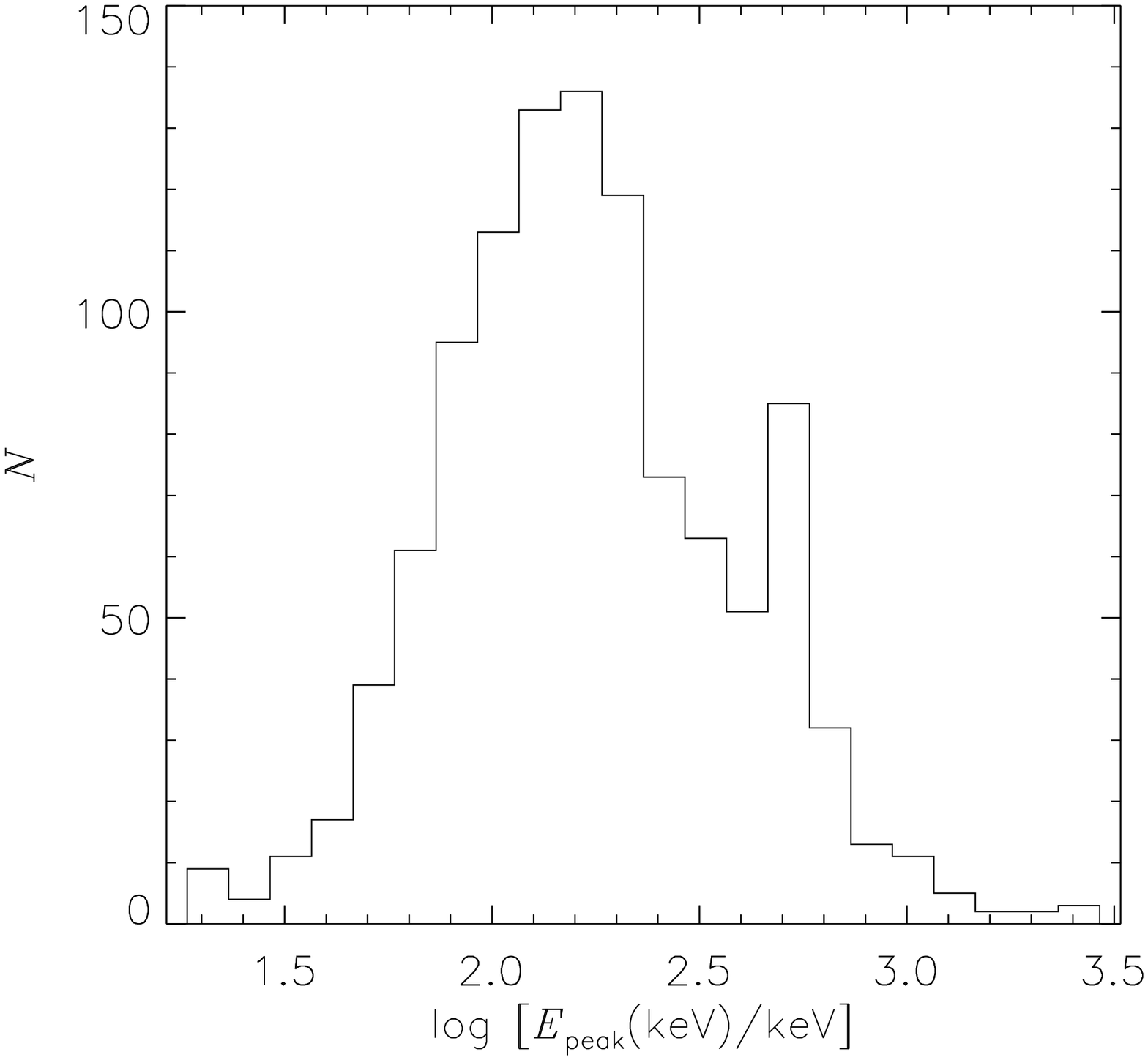}
\caption{
The distributions of the spectral parameters of the Band function of 1077 GRBs from the BATSE Spectral Catalog.
}
\label{fig:batse_band_spec_params}
\end{figure}

\item Next, using the ``shifted'' values of log \break $H'_{21,\rm{CURR}}$ we calculate the numbers of GRBs and XRFs
as described in Subsection~\ref{subsec:batse_procII}.
We use one of the 1000 lists with assigned group members of bursts as described above in point 6.  
\newline
\item We repeat these points $1.-2.$ $1000\times$ and each time we use a different list with
the assigned group members of bursts.
\newline
\item This gives us distribution of the fractions of XRFs/ GRBs for each group separately and in total. 
From these distributions we calculate the medians and 90\,\% confidence levels (CL).
These results are written in parenthesis in Table~\ref{tab:batse_H_50-100/20-50_procII}. 
\end{enumerate}

Concerning the estimation of the uncertainties in the numbers of XRFs/GRBs in Tables~\ref{tab:batse_H_50-100/25-50}
and \ref{tab:batse_H_50-100/25-50_H21_spec} we proceed as follows. The values in these tables were obtained by
comparing $H_{21\rm,SPEC}$ and $H_{\frac{50-100}{25-50}}$, both calculated using the BATSE Spectral Catalog.
The sample contains 1626 events and there is a strong correlation between both hardnesses
(see Fig.~\ref{fig:batse_H_50-100/25-50_H21_spec}). Pearson correlation coefficient is $\rho = 0.998$.
Therefore, one cannot treat them as independent variables.
The linear least square fitting \citep{press07} assuming a linear relation
$\log H_{21\rm,SPEC}=a+b\log H_{\frac{50-100}{25-50}}$ gives
$a=-0.113 \pm 0.001$, $b=1.120 \pm 0.003$ with $\chi^2$ = $34.3$ and GOF = 1.0.
The linear fit is perfectly acceptable.

\begin{figure}[h]
\centering
\includegraphics[width=0.48\textwidth]{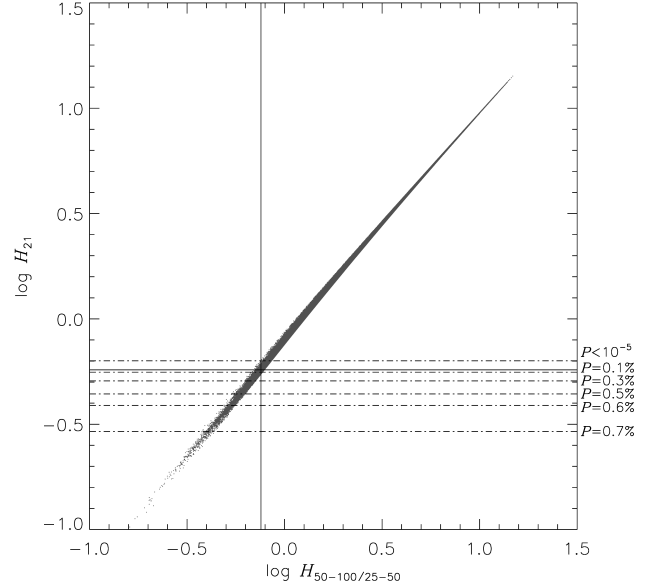}
\caption{The hardness ratio $H_{21}$ vs. $H_{\frac{50-100}{25-50}}$ of 1~million simulated GRB spectra.
The vertical (horizontal) solid line is the XRF limit from Def. (Eq.~(\ref{eq:H_XRF_def2_BATSE})).
The events which lie on the left from the vertical solid line are classified as XRFs by Def.
The events which lie below the horizontal solid line are classified as XRFs using the limit given by Eq.~(\ref{eq:H_XRF_def2_BATSE}).
The dash-and-dot lines mark the CIs meaning that any object above that line has
a probability of $<10^{-5}$, 0.3\,\%, 0.5\,\%, 0.6\,\%, and 0.7\,\%, respectively, to be an XRF by Def.
}
\label{fig:batse_H_50-100/25-50_H21_simulated}
\end{figure}

Same as in Subsection~\ref{subsec:batse_precis_procI} we used routine ``FITEXY'' for fitting
because it accounts for the uncertainties along both axes.
It requires symmetric uncertainties. Therefore we used the mean values
$\sigma_{\log H_{21\rm,SPEC}}=\langle -\sigma_{\log H_{21\rm,SPEC}};$
$+\sigma_{\log H_{21\rm,SPEC}}\rangle$
and
$\sigma_{\log H_{\frac{50-100}{25-50}}}=\langle -\sigma_{\log H_{\frac{50-100}{25-50}}};$
$+\sigma_{\log H_{\frac{50-100}{25-50}}}\rangle$
of the asymmetric uncertainties. Therefore we proceed as follows.
\begin{enumerate}
\item For each event of the 1626 bursts displayed in Fig.~\ref{fig:batse_H_50-100/25-50_H21_simulated},
we add the random Gaussian noise to $\log H_{\frac{50-100}{25-50}}$ with the mean value the same as the
measured value and the standard deviation the same as $\sigma_{\log H_{\frac{50-100}{25-50}}}$.
Thus we obtain ``shifted'' values $\log H'_{\frac{50-100}{25-50}}$.
\newline
\item Next, due to the strong correlation between \break $H_{21\rm,SPEC}$ and $H_{\frac{50-100}{25-50}}$
we do not add a random noise to hardnesses $\log H_{21\rm,SPEC}$, but we calculate ``shifted'' values
$\log H'_{21\rm,SPEC}$ from the best linear fit mentioned above, i.e.
$\log H'_{21\rm,SPEC}=-0.113+1.12\log H'_{\frac{50-100}{25-50}}$.
\newline
\item Then, using the ``shifted'' values of the $\log H'_{\frac{50-100}{25-50}}$ and $\log H'_{21\rm,SPEC}$,
we calculate the numbers of GRBs and XRFs as described in Subsection~\ref{subsec:batse_procI}.
We also calculate the number of events in each of four quadrants in a similar plot as shown in
Fig.~\ref{fig:batse_H_50-100/25-50_H21_simulated},
but this time with $\log H'_{\frac{50-100}{25-50}}$ and $\log H'_{21\rm,SPEC}$.
We use one of the 1000 lists with assigned group members of bursts as described above in point 6.
\newline
\item We repeat these points $1.-3.$ $1000\times$ and each time we use a different list with
the assigned group members of bursts.
\newline
\item This gives us distribution of the fractions of XRFs / GRBs for each group separately and in total
as well as the distributions of the number of the events in the four quadrants of
Fig.~\ref{fig:batse_H_50-100/25-50_H21_simulated}.
From these distributions we calculate the medians and 90\,\% CL uncertainties.
These results are written in parenthesis in Tables~\ref{tab:batse_H_50-100/25-50}
and \ref{tab:batse_H_50-100/25-50_H21_spec}. 
\end{enumerate}

\subsection{Summary of the fraction of XRFs}
\label{sec:batse_summary_of_fractions}

The fractions of XRFs in the BATSE samples by ProcI (Table~\ref{tab:batse_H_50-100/25-50}),
from checking of ProcI (Subsection~\ref{subsec:batse_precis_procI}), and by ProcII (Table~\ref{tab:batse_H_50-100/20-50_procII})
are summarized in Table~\ref{tab:batse_summary_of_fractions}.

The fractions of XRFs obtained by different procedures have the same trend. The largest fraction of XRFs is in the group
of intermediate-duration bursts. ProcII gives systematically larger fractions of XRFs.
The reason can be because the uncertainties in $\log H_{21\rm,CURR}$ (used in ProcII) for the groups of
short and intermediate-duration bursts are larger than the uncertainties in $\log H_{21\rm,SPEC}$
(used in ProcI). This larger scatter in $\log H_{21\rm,CURR}$
can contribute to the larger fraction of XRFs when ProcII is used
(compare Fig.~\ref{fig:batse_H21_curr_T90} and  Fig.~\ref{fig:batse_H21_spec_T90}).
This suggests that ProcI might be more reliable than ProcII in case of the BATSE datasets.
In any case, taking into account all these uncertainties, and hence the large scatters in the fractions,
it can be claimed that the whole entire intermediate-duration group cannot be given by XRFs only.

\begin{table*}
\caption{\label{tab:batse_summary_of_fractions}
A summary of the fractions of XRFs [\%] in the BATSE samples by different procedures.
The values summarizes Tables~\ref{tab:batse_H_50-100/25-50}, \ref{tab:batse_H_50-100/20-50_procII} and Subsection~\ref{subsec:batse_precis_procI}.
``Total'' means fraction from a whole sample. ``Short'', ``Inter.'', and ``Long'' mean fractions in the individual groups.
``No-group'' means fraction of the events without assigned group-membership.
``N/A'' means not applicable because in ProcII all events in the sample had assigned group-membership.
The fractions written in parentheses were obtained from the median and 90\,\% CL uncertainties in the numbers
described in Subsection~\ref{sec:batse_uncertainties_in_fractions}.
}
\centering
\begin{tabular}{@{}cccccc@{}}
\tableline                                                                                                                       \\[-9pt]
Procedure               & Total                 & Short                 & Inter.             & Long                  & No-group  \\
\tableline                                                                                                                       \\[-9pt]
\multirow{2}{*}{ProcI}  & 1.3                   & 0.7                   & 6                  & 1                     & 4         \\[1pt]
                        & $(2.0\pm0.4)$         & $(1.6^{+0.9}_{-0.7})$ & $(9^{+28}_{-8})$   & $(1.9^{+0.6}_{-0.5})$ & $(4\pm2)$ \\[7pt]
Checking of ProcI       & 1.8                   & 0.9                   & 10                 & 1.4                   & 4         \\[7pt]
\multirow{2}{*}{ProcII} & 3.7                   & 4.7                   & 15                 & 2.4                   & N/A       \\[1pt]
                        & $(3.9^{+0.3}_{-0.4})$ & $(3.8^{+1.9}_{-1.3})$ & $(31^{+54}_{-25})$ & $(2.9\pm0.5)$         & N/A       \\
\tableline
\end{tabular}
\end{table*}

\section{Fraction of XRFs in the \textit{RHESSI} database}
\label{sec:rhessi_results}

\subsection{Determination of the fraction}
\label{sec:rhessi_results_determ_frac}

For ProcII and the \textit{RHESSI} sample the limiting hardness of XRFs
$\log H_{\frac{120-1500}{25-120}} \approx -0.2$ is given by Eq.~(\ref{eq:H_XRF_def2_RHESSI}).
However, this value of hardness should be converted into the ``pseudo-hard\-ness'' $\widetilde{H}_{\frac{120-1500}{25-120}}$ in order to 
provide ProcII, because we have the measured ``pseudo-hardness'' for 427 objects.
The details of this conversion will be presented in Subsection~\ref{sec:rhessi_cnt_erg_conversion} and in Fig.~\ref{fig:rhessi_Hcnt_Herg}.
Here we use only the results from Fig.~\ref{fig:rhessi_Hcnt_Herg} with marked CIs meaning
that any object above the given ``pseudo-hardness'' $\widetilde{H}_{\frac{120-1500}{25-120},\rm{i}}$ has a probability of
$<10^{-5}$, 0.1\,\%, 0.2\,\%, 0.5\,\%, 1.0\,\%, 2.0\,\%, and 3.0\,\%, respectively, to be an XRF
by using the limit from Eq.~(\ref{eq:H_XRF_def2_RHESSI}).

\begin{figure}[h]
\centering
\includegraphics[width=0.48\textwidth]{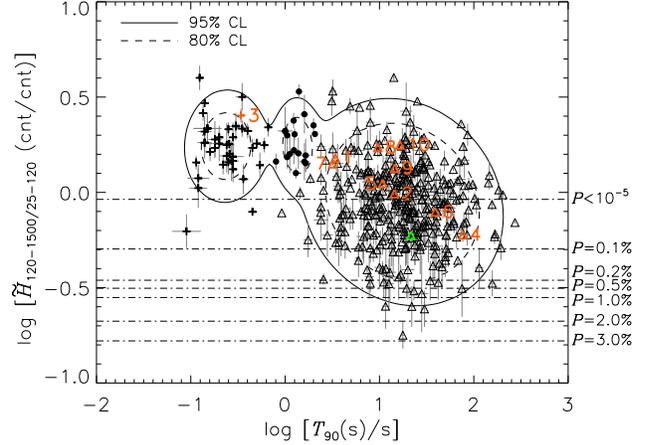}
\caption{
The ``pseudo-hardness'' $\widetilde{H}_{\frac{120-1500}{25-120}}$\,(cnt\,cnt$^{-1}$) vs. $T_{90}$ durations
with identified short, intermediate, and long groups of the 427 \textit{RHESSI} GRBs as published by \citet{rip12}.
The CL marks the confidence levels of the best maximum likelihood fit with three bivariate lognormal functions
as calculated and described in paper \citet{rip12}.
Here we add the dash-and-dot lines which mark the confidence intervals meaning that any object above the given line has
a probability of $<10^{-5}$, 0.1\,\%, 0.2\,\%, 0.5\,\%, 1.0\,\%, 2.0\,\%, and 3.0\,\%, respectively,
to be an XRF by using the limit given by Eq.~(\ref{eq:H_XRF_def2_RHESSI}).
These confidence intervals were obtained from the simulation shown in Fig.~\ref{fig:rhessi_Hcnt_Herg}.
Objects colored orange are the ten GRBs (see Tables~\ref{tab:spec_fits} and \ref{tab:spec_fluences}) selected to check the conversion between $H_{\frac{120-1500}{25-120}}$
(erg\,cm$^{-2}$\,erg$^{-1}$\,cm$^{2}$) and $\widetilde{H}_{\frac{120-1500}{25-120}}$ (cnt\,cnt$^{-1}$) employing the
simulated response function of the detector (see Fig.~\ref{fig:rhessi_Hcnt_Herg}).
The numbers refer to the measurements in Table~\ref{tab:spec_fluences}.
The error bars were taken from \citet{rip09,rip12}.
The green triangle marks GRB 030528 observed both by \textit{RHESSI} and \textit{HETE-2} and classified as XRF by \citet{sak05}.
}
\label{fig:rhessi_Hcnt_T90}
\end{figure}

Fig.~\ref{fig:rhessi_Hcnt_T90} shows the ``pseudo-hardness'' $\widetilde{H}_{\frac{120-1500}{25-120}}$ against the $T_{90}$ duration.
That figure is similar to Figure~1 of \citet{rip12}, but here we add the CIs for the XRF limit.
The objects above the most upper horizontal dash-and-dot line,
i.e., the $<10^{-5}$ CI ($\log \widetilde{H}_{\frac{120-1500}{25-120}} > -0.04$), are not XRFs.
The objects below this line can already be XRFs by using the limit given by Eq.~(\ref{eq:H_XRF_def2_RHESSI}).
The numbers of events above and below the $<10^{-5}$ CI are summarized in Table~\ref{tab:rhessi_pseudo-hardness_T90}.
From that table it follows that there are 232 objects with $\log \widetilde{H}_{\frac{120-1500}{25-120}} > -0.04$, i.e., 
54.3\,\% of the whole sample, which are not XRFs. For the groups separately one obtains:
from the 42 short bursts 40 objects (95\,\% of the short ones) are above the limit;
all 18 intermediate-duration bursts are above the limit;
from the 367 long ones 174 objects (47.4\,\% of the long ones) are above the limit.
Here the classification into the groups is the one used by \citet{rip12}.
These results imply that for the \textit{RHESSI} sample at least 95\,\% of the short burst,
all intermediate-duration bursts, and at least 47.4\,\% of the long ones are not given by XRFs.
The two short bursts, which are below the $<10^{-5}$ CI limiting line (see Fig.~\ref{fig:rhessi_Hcnt_T90}),
could potentially be outliers.

\begin{table*}
\caption{\label{tab:rhessi_pseudo-hardness_T90}
The numbers of bursts with the ``pseudo-hardness'' $\widetilde{H}_{\frac{120-1500}{25-120}}$
above and below the $<10^{-5}$ CI ($\log \widetilde{H}_{\frac{120-1500}{25-120}} = -0.04$)
for the XRF limit following from Eq.~(\ref{eq:H_XRF_def2_RHESSI}) for the \textit{RHESSI} sample and for the three groups separately
classified by \citet{rip12}.
The values not written in parentheses were obtained directly from the measured data.
The values written in parentheses are median and 90\,\% CL obtained by method described in Subsection~\ref{sec:rhessi_uncertainties_in_fractions}.
}
\centering
\begin{tabular}{@{}ccccc@{}}
\tableline                                                                                                                                               \\[-9pt]
                                                                           & Total             & Short            & Intermediate     & Long              \\
\tableline                                                                                                                                               \\[-9pt]
\multirow{2}{*}{$\log \widetilde{H}_{\frac{120-1500}{25-120}} > -0.04$}    & 232               & 40               & 18               & 174               \\[1pt]
                                                                           & $(234^{+8}_{-6})$ & $(39^{+2}_{-5})$ & $(22\pm6)$       & $(175^{+8}_{-9})$ \\[7pt]
\multirow{2}{*}{$\log \widetilde{H}_{\frac{120-1500}{25-120}} \leq -0.04$} & 195               & 2                & 0                & 193               \\[1pt]
                                                                           & $(193^{+6}_{-8})$ & $(2\pm1)$        & $(1\pm1)$        & $(189\pm7)$       \\[7pt]
\multirow{2}{*}{Sum}                                                       & 427               & 42               & 18               & 367               \\[1pt]
                                                                           &                   & $(41^{+2}_{-5})$ & $(22^{+8}_{-6})$ & $(364\pm5)$       \\
\tableline
\end{tabular}
\end{table*}

\subsection{Precision of the fraction}
\label{sec:rhessi_precision_of_procII}

To check the precision of ProcII we proceed similarly to Subsection~\ref{sec:batse_precision_of_procII}
and check the precision of the limit from Eq.~(\ref{eq:H_XRF_def2_RHESSI}) itself.
A numerical simulation was carried out as follows.
We used the distributions of the GRB spectral parameters of Band function from Fig.~\ref{fig:batse_band_spec_params} and simulated 1~million mock
events. Then we calculated the hardnesses $H_{\frac{50-100}{25-50}}$ and $H_{\frac{120-1500}{25-120}}$.

\begin{figure}[h]
\centering
\includegraphics[width=0.48\textwidth]{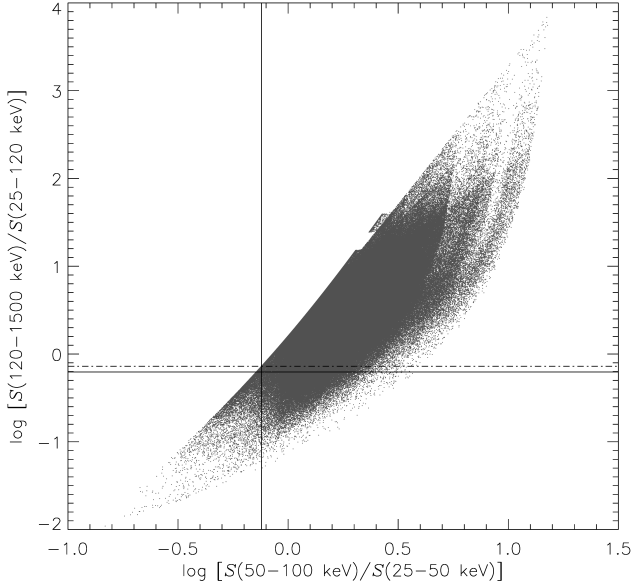}
\caption{
The hardness ratio $H_{\frac{120-1500}{25-120}}$ vs. $H_{\frac{50-100}{25-50}}$ of 1~million simulated GRB spectra.
The vertical (horizontal) solid line is the XRF limit from Def. (Eq.~(\ref{eq:H_XRF_def2_RHESSI})).
The events which lie on the left from the vertical solid line are classified as XRFs by Def.
The events which lie below the horizontal solid line are classified as XRFs
by using the limit from Eq.~(\ref{eq:H_XRF_def2_RHESSI}).
The dash-and-dot line marks the highest hardness $H_{\frac{120-1500}{25-120}}$
of an XRF by Def. obtained in the simulation.
}
\label{fig:rhessi_H_50-100/25-50_H_120-1500/25-120_simulated}
\end{figure}

The result of the simulation is presented in Fig.~\ref{fig:rhessi_H_50-100/25-50_H_120-1500/25-120_simulated}.
It shows the relation between the hardness $H_{\frac{50-100}{25-50}}$ used by Def. and the hardness $H_{\frac{120-1500}{25-120}}$.
Similarly to Subsection~\ref{sec:batse_precision_of_procII} we mark
the number of XRFs in this simulation due to Def. ($H_{\frac{50-100}{25-50}}<0.76$) as $n_{\rm XRF}$.
Oppositely we mark the number of GRBs due to Def. ($H_{\frac{50-100}{25-50}}>0.76$) as $n_{\rm GRB}$.
It holds that if above any horizontal line defined by $H_{\rm\frac{120-1500}{25-120},i}$ there are
$n_{\rm XRF,i}$ and $n_{\rm GRB,i}$ events then for any event above this line
the probability that it is an XRF is $P=n_{\rm XRF,i}/(n_{\rm XRF,i} + n_{\rm GRB,i})$.
The probability $P$ above the solid horizontal line $H_{\frac{120-1500}{25-120}}\approx0.6$ given by Eq.~(\ref{eq:H_XRF_def2_RHESSI}) is $P=0.0005$.
This means that an event above this line has only 0.05\,\% probability to be an XRF.
For the numbers of the events below the horizontal line the simulation gives
$P^{\ast}=n^{\ast}_{\rm GRB,i}/(n^{\ast}_{\rm XRF,i} + n^{\ast}_{\rm GRB,i}) = 0.836$.
This means that above the line given by Eq.~(\ref{eq:H_XRF_def2_RHESSI}) there is only
0.05\,\% probability for an XRF to be wrongly identified as not an XRF, but below this line an ordinary GRB can be
wrongly identified as an XRF by a high 83.6\,\% probability.
This also means that the limit from Eq.~(\ref{eq:H_XRF_def2_RHESSI}) is in fact a strong upper limit.

It is interesting to compare the value $\log$ $H_{\frac{120-1500}{25-120}}=-0.14$ from Fig.~\ref{fig:rhessi_H_50-100/25-50_H_120-1500/25-120_simulated},
i.e., the highest possible hardness $H_{\frac{120-1500}{25-120}}$ of an XRF by Def. (obtained in the simulation) with the limiting XRF hardness
obtained by Eq.~(\ref{eq:H_XRF_def2_RHESSI}), $\log H_{\frac{120-1500}{25-120}}=-0.20$.
Both values are very similar, which also supports the correctness of the limit following from Eq.~(\ref{eq:H_XRF_def2_RHESSI}).

\subsection{Uncertainties in the numbers}
\label{sec:rhessi_uncertainties_in_fractions}
In order to determine the uncertainties in the numbers mentioned in
Table~\ref{tab:rhessi_pseudo-hardness_T90} we proceed similarly to Subsection~\ref{sec:batse_uncertainties_in_fractions}.
The uncertainties in the $T_{90}$ duration effect the classification of the bursts into the short, intermediate, and long groups.
The uncertainties in the ``pseudo-hardness'' $\widetilde{H}_{\frac{120-1500}{25-120}}$ effect the number of bursts above or below the
limiting value. To account for that we proceed as follows.
\begin{enumerate}
\item For each event in the RHESSI sample we add the random Gaussian noise to $\log \widetilde{H}_{\frac{120-1500}{25-120}}$
with the mean value the same as the measured value and the standard deviation the same as $1\sigma$ uncertainty in
$\log \widetilde{H}_{\frac{120-1500}{25-120}}$.
Thus we obtain shifted values of $\log \widetilde{H'}_{\frac{120-1500}{25-120}}$.
\newline
\item Next, for each event, independently on log $\widetilde{H}_{\frac{120-1500}{25-120}}$,
we add the random Gaussian noise to $\log T_{90}$
with the mean value the same as the measured value and the standard deviation the same as $1\sigma$
uncertainty in $\log T_{90}$. Thus we obtain shifted values of $\log T'_{90}$.
\newline
\item We follow the method described by \citet{ho06} to determine the group-member\-ship assigned to each event
in this ``shifted'' sample.
\newline
\item Then, using the ``shifted'' values of log $\widetilde{H'}_{\frac{120-1500}{25-120}}$
we count the numbers of events with log $\widetilde{H'}_{\frac{120-1500}{25-120}}$ above or below -0.04
for each group separately and in total as described in Subsection~\ref{sec:rhessi_results_determ_frac}.
\newline
\item We repeat steps $1.-4.$ $1000\times$ which gives us distributions of the
numbers of bursts above and below the limiting ``pseudo-hardness'' as well as the distributions of
the numbers of short-, intermediate-, and long- duration bursts.
\newline
\item
From these distributions we calculate the medians and 90\,\% CL uncertainties.
These results are written in parenthesis in Table~\ref{tab:rhessi_pseudo-hardness_T90}. 
\end{enumerate}

\subsection{Conversion of the two types of hardnesses}
\label{sec:rhessi_cnt_erg_conversion}

\begin{figure}[t]
\centering
\includegraphics[width=0.48\textwidth]{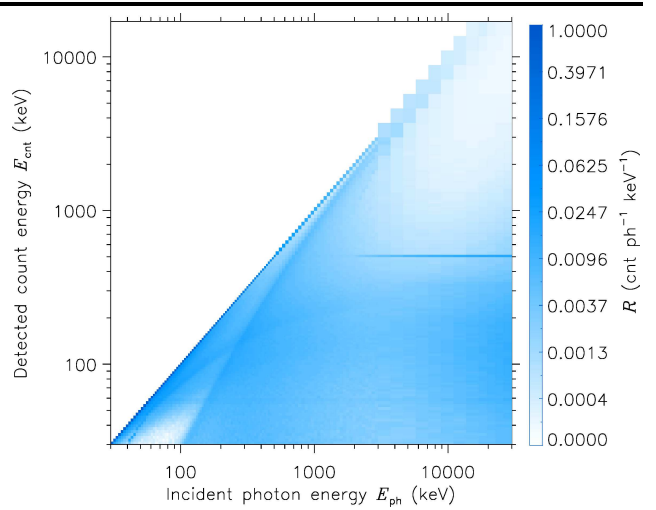}
\caption{
An example of the used \textit{RHESSI} off-axis detector response function provided by E. Bellm (private communication)
for the off-axis angle $90^\circ$, averaged over all azimuth angles, and normalized by the sizes of the count energy bins.
We used only the rear segments of all nine detectors expect for the malfunctioning No. 2.
The displayed response function $R$ is normalized by its maximum value, so it ranges from 0 to 1.
}
\label{fig:rhessi_drm}
\end{figure}

In order to convert the hardness $H_{\frac{120-1500}{25-120}}$ (erg cm$^{-2}$ erg$^{-1}$ cm$^{2}$) 
defined as a ratio of fluences into the ``pseudo-hardness'' $\widetilde{H}_{\frac{120-1500}{25-120}}$ (cnt\,cnt$^{-1}$)
defined as a ratio of detected counts, and vice versa, we proceed as follows:
\newpage
\begin{enumerate}
\item We begin with the distribution of the measured GRB spectral parameters
$\alpha$, $\beta$ and $E_\mathrm{peak}$ of Band function from the BATSE Spectral Catalog
as shown in Fig.~\ref{fig:batse_band_spec_params}.
\newline
\item From this distribution we randomly create an ensemble of 50\,000 sets of $\alpha, \beta$ and $E_\mathrm{peak}$.
This makes a basis of our sample of simulated GRB spectra.
\newline
\item We integrate each simulated spectrum in the energy ranges $25 - 120$\,keV and $120 - 1500$\,keV; then
we calculate hardness $H_{\frac{120-1500}{25-120}}$\,(erg\,cm$^{-2}$\,erg$^{-1}$\,cm$^{2}$).
\newline
\item After that we use the \textit{RHESSI} response function to obtain the number of counts in the same
energy ranges. We calculate the ``pseudo-hardness'' $\widetilde{H}_{\frac{120-1500}{25-120}}$\,(cnt\,cnt$^{-1}$) as follows:
\begin{equation}
\widetilde H_\mathrm{\frac{E_{cnt,1}-E_{cnt,2}}{E_{cnt,3}-E_{cnt,4}}}=
\frac{C_\mathrm{_{E_{cnt,1}-E_{cnt,2}}} }{C_\mathrm{_{E_{cnt,3}-E_{cnt,4}}} },
\end{equation}
where
\begin{equation}
\begin{split}
C_{ _{E_\mathrm{cnt,i}-E_\mathrm{cnt,j}}}
= \int_{t_{1}}^{t_{2}}\int_{E_\mathrm{ph,1}}^{E_\mathrm{ph,2}}\int_{E_\mathrm{cnt,i}}^{E_\mathrm{cnt,j}}
S\,N_\mathrm{E}(E_\mathrm{ph}) \\ R(E_\mathrm{ph},E_\mathrm{cnt})\,\mathrm{d}E_\mathrm{ph}\,\mathrm{d}E_\mathrm{cnt}\,\mathrm{d}t.
\end{split}
\end{equation}
Here $N_\mathrm{E}$ is the differential photon spectrum of the incident radiation 
(see Eqs.~(\ref{eq:Band_general_1}-\ref{eq:Band_general_3})) in units ph\,cm$^{-2}$\,s$^{-1}$\,keV$^{-1}$;
$S$ is the detector area in cm$^2$;
$R$ is the detector response function (differential, i.e., normalized by the count energy)
in units cnt\,ph$^{-1}$\,keV$^{-1}$ (``ph'' and ``cnt'' refer to photon and count, respectively);
$E_\mathrm{ph}$ is the incident photon energy;
$E_\mathrm{cnt}$ is the measured count energy.
We use $E_\mathrm{cnt,1}=120$\,keV, $E_\mathrm{cnt,2}=1500$\,keV, 
$E_\mathrm{cnt,3}=25$\,keV, $E_\mathrm{cnt,4}=120$\,keV,
$E_\mathrm{ph,1}=25$\,keV and $E_\mathrm{ph,2}=30000$\,keV to determine the sensitivity range of the detector.
$t_{1}$ and $t_{2}$ are the start and the end time of the analyzed flux, respectively.
An example of the \textit{RHESSI} detector response function is shown in Fig.~\ref{fig:rhessi_drm}.
\newline
\item Since the \textit{RHESSI} detector response depends on the off-axis angle of the incoming X-ray photons,
we calculate the steps II.$-$IV. using the response functions for
the eleven different off-axis angles, particularly for $15^\circ$, $30^\circ$, $45^\circ$, $60^\circ$, $75^\circ$, $90^\circ$,
$105^\circ$, $120^\circ$, $135^\circ$, $150^\circ$, and $165^\circ$.
The off-axis angles 0$^{\circ}$ and 180$^{\circ}$ mean the front direction (direction from the Sun) and the rear direction, respectively.
The overall response functions were also averaged over the responses in the six intervals of the azimuth angle:
$0^\circ - 60^\circ$, $60^\circ - 120^\circ$, $120^\circ - 180^\circ$, $180^\circ - 240^\circ$, $240^\circ - 300^\circ$, $300^\circ - 360^\circ$.
The reason is that the satellite spins along its axis with period $\approx 4$\,s.
\newline
\item As the number of the observed GRBs by the rear segments of the \textit{RHESSI} detector depends on the off-axis angle, the number of the
simulated GRB spectra for different off-axis angles must be accordingly weighted. There were 104 localized GRBs in our \textit{RHESSI} sample and
thus there were 104 known off-axis angles of these GRBs. Fig.~\ref{fig:rhessi_off-axis_distribution} shows their distribution. We weighted the number
of the simulated GRB spectra for each off-axis angle proportionally to the frequency of the GRB observations for the given off-axis angle bin.
Table~\ref{tab:rhessi_off-axis_distribution} shows the concrete numbers of these simulated spectra.
\end{enumerate}

\begin{figure}[t]
\centering
\includegraphics[width=0.48\textwidth]{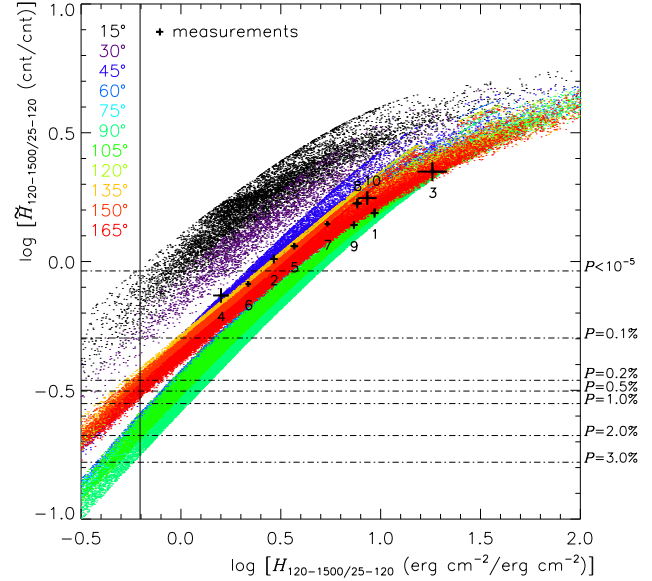}
\caption{
Conversion between the hardness $H_{\frac{120-1500}{25-120}}$ 
(erg\,cm$^{-2}$\,erg$^{-1}$\,cm$^{2}$) and the ``pseudo-hardness'' $\widetilde{H}_{\frac{120-1500}{25-120}}$\,(cnt\,cnt$^{-1}$) for \textit{RHESSI} GRBs.
Different color marks the hardness of the simulated GRB spectra with photons
coming into the detector under different off-axis angles running from 15$^{\circ}$ to 165$^{\circ}$
(displayed is a cut-out of the whole distribution).
Crosses denote the real measured GRBs with numbers referring to the values in Table~\ref{tab:spec_fluences}.
The vertical line marks the XRF limit from Eq.~(\ref{eq:H_XRF_def2_RHESSI}).
The dash-and-dot lines mark the confidence intervals meaning that any object above the given line has
a probability of $<10^{-5}$, 0.1\,\%, 0.2\,\%, 0.5\,\%, 1.0\,\%, 2.0\,\%, and 3.0\,\%, respectively,
to be an XRF by using the limit given by Eq.~(\ref{eq:H_XRF_def2_RHESSI}).
}
\label{fig:rhessi_Hcnt_Herg}
\end{figure}

Since the energy bins of the provided detector response functions started from 30\,keV and did not exactly match
the edges 25\,keV, 120\,keV, 1500\,keV of the needed energy ranges used in 
the ``pseudo-hardness'' $\widetilde{H}_{\frac{120-1500}{25-120}}$,
we linearly extrapolated and interpolated them in order to match the desired energy binning.
We use the responses for the rear segments of the detectors and we use all the nine detectors except for No.2
which has been shown to be malfunctioning.
This means that we use the same detectors and segments as they were used in the work by \citet{rip09}.
Fig.~\ref{fig:rhessi_Hcnt_Herg} shows the dependence between
$\widetilde{H}_{\frac{120-1500}{25-120}}$ and $H_{\frac{120-1500}{25-120}}$
together with CIs of the XRF limit following from Eq.~(\ref{eq:H_XRF_def2_RHESSI}).

\begin{table}[h]
\caption{\label{tab:rhessi_off-axis_distribution}
A summary of the distribution of the 104 localized GRBs in the \textit{RHESSI} sample, together with the weighted number of the mock
GRB spectra for each off-axis angle.}
\centering
\begin{tabular}{@{}cccc@{}}
\tableline \\[-9pt]
$\Delta \theta$\tablenotemark{a}            &  $\theta_{\rm{sim}}$\tablenotemark{b}            &  $N_{\rm{obs}}$\tablenotemark{c}            &  $N_{\rm{sim}}$\tablenotemark{d}  \\
(deg)                                       &  (deg)                                           &                                             &                                   \\
\tableline \\[-9pt]
$\langle 7.5\,; 22.5)$                      &  15.0                                            &   3                                         &  8333                             \\
$\langle 22.5\,; 37.5)$                     &  30.0                                            &   1                                         &  2778                             \\
$\langle 37.5\,; 52.5)$                     &  45.0                                            &   6                                         & 16667                             \\
$\langle 52.5\,; 67.5)$                     &  60.0                                            &   9                                         & 25000                             \\
$\langle 67.5\,; 82.5)$                     &  75.0                                            &  13                                         & 36111                             \\
$\langle 82.5\,; 97.5)$                     &  90.0                                            &  18                                         & 50000                             \\
$\langle 97.5\,; 112.5)$                    &  105.0                                           &  11                                         & 30556                             \\
$\langle 112.5\,; 127.5)$                   &  120.0                                           &  13                                         & 36111                             \\
$\langle 127.5\,; 142.5)$                   &  135.0                                           &   8                                         & 22222                             \\
$\langle 142.5\,; 157.5)$                   &  150.0                                           &  13                                         & 36111                             \\
$\langle 157.5\,; 172.5)$                   &  165.0                                           &   9                                         & 25000                             \\
\tableline
\end{tabular}
\tablenotetext{a}{The boundaries of the off-axis angle bins of the \\
measured distribution for the localized GRBs.}
\tablenotetext{b}{The off-axis angles for which the \textit{RHESSI} detector \\
response function were used.}
\tablenotetext{c}{The frequency of the observed GRBs for the given \\
off-axis angle bin.}
\tablenotetext{d}{The weighted number of the simulated GRB spectra \\
for the given off-axis angle.}
\end{table}

The confidence intervals CIs were calculated as follows.
Assume that there are $\widetilde{n}_{\rm XRF}$ events above an arbitrarily chosen horizontal line
marked by $\widetilde{H}_{\frac{120-1500}{25-120},\rm{i}}$, i.e.
$\widetilde{H}_{\frac{120-1500}{25-120}} \ge \widetilde{H}_{\frac{120-1500}{25-120},\rm{i}}$,
with $H_{\frac{120-1500}{25-120}} \le 0.6$.
Assume also that there are $\widetilde{n}_{\rm GRB}$ non-XRF events above the same horizontal line
($\widetilde{H}_{\frac{120-1500}{25-120}} \ge \widetilde{H}_{\frac{120-1500}{25-120},\rm{i}}$),
with $H_{\frac{120-1500}{25-120}} > 0.6$.
The limiting value $\approx0.6$ (logarithmic value $\approx-0.2$) follows from Eq.~(\ref{eq:H_XRF_def2_RHESSI})
and is denoted in Fig.~\ref{fig:rhessi_Hcnt_Herg} by a solid vertical line.
Then the probability $P$ that any object above $\widetilde{H}_{\frac{120-1500}{25-120},\rm{i}}$
is an XRF is $P=\widetilde{n}_{\rm XRF}/(\widetilde{n}_{\rm XRF} + \widetilde{n}_{\rm GRB})$.
Fig.~\ref{fig:rhessi_Hcnt_Herg} shows the $<10^{-5}$, 0.1\,\%, 0.2\,\%, 0.5\,\%, 1.0\,\%, 2.0\,\%, and 3.0\,\% CIs, respectively.
Table~\ref{tab:rhessi_Hcnt_Herg} summarizes the ``pseudo-hardnesses'' for the given CIs,
the numbers of XRFs and GRBs following from the limit given by Eq.~(\ref{eq:H_XRF_def2_RHESSI}), which are above the given
$\widetilde{H}_{\frac{120-1500}{25-120},\rm{i}}$, and the probabilities $P$.

Additionally, to check further the conversion between $H_{\frac{120-1500}{25-120}}$ (erg\,cm$^{-2}$\,erg$^{-1}$\,cm$^{2}$) and
$\widetilde{H}_{\frac{120-1500}{25-120}}$ (cnt\,cnt$^{-1}$), we select ten localized GRBs with diffe\-rent durations and ``pseudo-hardnesses''
$\widetilde{H}_{\frac{120-1500}{25-120}}$ from the database used by \citet{rip09}.
Hence, we verify the numerical simulations by a further analysis of real measured data.
Since we need a fitting of the spectrum for this verification, there are only 67 suitable objects that can be used.
To have a good fitting from these 67 GRBs we took only the objects, which had S/N ratios even higher,
namely above 26. With this higher cut 39 objects remained. From them we have chosen randomly 10 GRBs in order to check the conversion.
The durations $T_{90}$ of these bursts spread between $0.34$\,s and $77$\,s and the logarithmic ``pseudo-hardnesses''
$\log \widetilde{H}_{\frac{120-1500}{25-120}}$ spread between -0.13 and 0.35.
These ten objects are also shown in Fig.~\ref{fig:rhessi_Hcnt_T90}.

We fit the spectra and derive the spectral parameters $\alpha, \beta$ and $E_\mathrm{peak}$.
In four cases we fitted the Band function to the spectrum, because it was possible to 
determine well the spectral parameters.
Nevertheless, in six cases the spectrum was better described by the CPL (see Eq.~(\ref{eq:Band_general_2})).
The fitted spectra were time averaged over the intervals $\Delta t_{\mathrm{spec}}$   
similar to the $T_{90}$ durations of the bursts as published by \citet{rip09}.
The background count spectrum, averaged over the certain time intervals before and after the given GRB,
was subtracted from the total observed count spectrum during the fitting procedure.

\begin{table}[h]
\caption{\label{tab:rhessi_Hcnt_Herg}
The table summarizes the confidence intervals $P$ marked at given ``pseudo-hardness''
$\widetilde{H}_{\frac{120-1500}{25-120},\rm{i}}$ in Fig.~\ref{fig:rhessi_Hcnt_Herg}.
The quantities $\widetilde{n}_{\rm XRF,i}$
($\widetilde{H}_{\frac{120-1500}{25-120}} \ge \widetilde{H}_{\frac{120-1500}{25-120},\rm{i}}$ and $H_{\frac{120-1500}{25-120}} \le 0.6$)
and $\widetilde{n}_{\rm GRB,i}$
($\widetilde{H}_{\frac{120-1500}{25-120}} \ge \widetilde{H}_{\frac{120-1500}{25-120},\rm{i}}$ and $H_{\frac{120-1500}{25-120}} > 0.6$)
are the numbers of XRFs and GRBs (in the simulation) classified by Def. with the ``pseudo-hardness'' $\widetilde{H}_{\frac{120-1500}{25-120}}$
higher or equal to $\widetilde{H}_{\frac{120-1500}{25-120},\rm{i}}$ corresponding to the given confidence level $P$, respectively.
The limiting value 0.6 is given by Eq.~(\ref{eq:H_XRF_def2_RHESSI}).
The $P$ is the probability that any object with $\widetilde{H}_{\frac{120-1500}{25-120}}$ higher or equal to
$\widetilde{H}_{\frac{120-1500}{25-120},\rm{i}}$ is an XRF following the limit in Eq.~(\ref{eq:H_XRF_def2_RHESSI}).
}
\centering
\begin{tabular}{@{}cccc@{}}
\tableline \\[-9pt]
$\log \widetilde{H}_{\frac{120-1500}{25-120},\rm{i}}$   &   $\widetilde{n}_{\rm XRF,i}$ &   $\widetilde{n}_{\rm GRB,i}$ &  $P$\,(\%) \\[5pt]
\tableline \\[-9pt]
-0.04                 				   	&     1              		&   109130           		& $<10^{-3}$ \\
-0.30                 				   	&   221              		&   220441           		&   0.1      \\
-0.46                				   	&   525              		&   261856           		&   0.2      \\
-0.50                 				   	&   1342            		&   266931           		&   0.5      \\
-0.55                 				   	&   2735             		&   270724           		&   1.0      \\
-0.68                 				   	&   5636             		&   276124           		&   2.0      \\
-0.78                 				   	&   8553             		&   276528           		&   3.0      \\
\tableline
\end{tabular}
\end{table}

Next, we integrated the fitted spectra, derived the fluences
$S_{120-1500}$\,(erg\,cm$^{-2}$) and $S_{25-120}$\,(erg\,cm$^{-2}$) 
in the energy ranges $25-120$\,keV and $120-1500$\,keV, respectively,
and calculated the hardnesses $H_{\frac{120-1500}{25-120}}$.
Importantly, we also derived the detected counts 
and calculated the ``pseudo-hardness'' $\widetilde{H}_{\frac{120-1500}{25-120}}$\,(cnt\,cnt$^{-1}$)
for the same time intervals $\Delta t_{\mathrm{spec}}$ from the background-subtracted count light curves
(background was subtracted by fitting a linear or quadratic function as a background model to the 
certain time intervals before and after the given GRB).

The results of the spectral fits are summarized in Table~\ref{tab:spec_fits}.
The obtained fluences, detected counts and both hardnesses are written in Table~\ref{tab:spec_fluences}.
The comparison of these real data with the simulated values indicate that the trend is the same (see Fig.~\ref{fig:rhessi_Hcnt_Herg}).                  

\begin{table*}[t]
\caption{\label{tab:spec_fits}
Spectral fits of the \textit{RHESSI} GRBs used to check the conversion between
$H_{\frac{120-1500}{25-120}}$\,(erg\,cm$^{-2}$\,erg$^{-1}$\,cm$^{2}$)
and $\widetilde{H}_{\frac{120-1500}{25-120}}$\,(cnt\,cnt$^{-1}$).
}
\centering
\begin{tabular}{@{}lccccccccc@{}}
\tableline \\[-9pt]
GRB\tablenotemark{a}           & Reference                       & Model\tablenotemark{c}           & $\alpha$\tablenotemark{d}           & $\beta$\tablenotemark{e}           & $E_\mathrm{peak}$\tablenotemark{f}           & Range\tablenotemark{g}           & $\Delta t_{\mathrm{spec}}$\tablenotemark{h}           & dof\tablenotemark{i}           & $\chi^2_r$\tablenotemark{j} \\
                               & Time\tablenotemark{b}           &                                  &                                     &                                    & (keV)                                        & (keV)                            & (s)                                                   &                                &                             \\[3pt]
\tableline \\[-9pt]
020418                         & 17:43:08.4                      & CPL                              & $-0.652^{+0.071}_{-0.072}$          & \textemdash                        & $718_{-51}^{+59}$                            & 24-1700                          & 2.77                                                  & 89                             & 1.19                        \\[3pt]
021020                         & 20:12:51.3                      & Band                             & $-0.98^{+0.16}_{-0.18}$             & $-2.34^{+0.15}_{-0.30}$            & $271_{-41}^{+57}$                            & 24-2000                          & 14.4                                                  & 89                             & 0.78                        \\[3pt]
021201                         & 05:30:04.0                      & CPL                              & $-0.51^{+0.19}_{-0.16}$             & \textemdash                        & $1180_{-158}^{+210}$                         & 24-3000                          & 0.51                                                  & 96                             & 1.17                        \\[3pt]
021205                         & 03:17:24.3                      & CPL                              & $-1.79^{+0.14}_{-0.13}$             & \textemdash                        & $219_{-66}^{+76}$                            & 24-3000                          & 76.3                                                  & 93                             & 0.76                        \\[3pt]
030922B                        & 18:30:50.2                      & Band                             & $-1.100^{+0.094}_{-0.090}$          & $-2.47^{+0.16}_{-0.21}$            & $424 \pm 45$                                 & 24-3000                          & 11.0                                                  & 92                             & 0.86                        \\[3pt]
031027$^{1,2}$                 & 17:07:11.6                      & CPL                              & $-1.476^{+0.045}_{-0.043}$          & \textemdash                        & $323_{-16}^{+18}$                            & 24-3000                          & 39.2                                                  & 93                             & 1.22                        \\[3pt]
031111$^{1,2}$                 & 16:45:19.3                      & Band                             & $-1.099 \pm 0.052$                  & $-2.19^{+0.09}_{-0.11}$            & $724_{-80}^{+81}$                            & 24-3000                          & 4.08                                                  & 92                             & 1.50                        \\[3pt]
050717$^{3,4}$                 & 10:30:54.9                      & CPL                              & $-1.151 \pm 0.048$                  & \textemdash                        & $1930_{-324}^{+442}$                         & 24-5000                          & 9.50                                                  & 97                             & 1.14                        \\[3pt]
061121$^{3}$                   & 15:23:32.0                      & Band                             & $-0.28^{+0.13}_{-0.14}$             & $-2.11^{+0.10}_{-0.12}$            & $379_{-40}^{+43}$                            & 34-2600                          & 14.7                                                  & 86                             & 0.91                        \\[3pt]
061126$^{3}$                   & 08:48:01.6                      & CPL                              & $-0.72^{+0.17}_{-0.14}$             & \textemdash                        & $722_{-87}^{+109}$                           & 24-3000                          & 16.6                                                  & 93                             & 0.94                        \\[3pt]
\tableline
\end{tabular}
\tablenotetext{a}{The \textit{RHESSI} GRB number.}
\tablenotetext{b}{The starting time of the time-averaged spectrum.}
\tablenotetext{c}{The fitted spectral model.}
\tablenotetext{d}{$\alpha$ is the low-energy spectral index or only spectral index in case of CPL.}
\tablenotetext{e}{$\beta$ is the high-energy spectral index.}
\tablenotetext{f}{$E_\mathrm{peak}$ is the peak energy.}
\tablenotetext{g}{The energy range used for the fit.}
\tablenotetext{h}{The time interval during which the time-averaged spectrum was obtained.}
\tablenotetext{i}{The number of degrees of freedom.}
\tablenotetext{j}{The reduced $\chi^2$ of the fit.}
\tablecomments{All mentioned errors are $1\sigma$ statistical errors.
The following references already published spectral fits given by \textit{RHESSI}: $^{1}$\citet{bel10}, $^{2}$\citet{wig08}, $^{3}$\citet{bel08b}, $^{4}$\citet{wig06}.}
\end{table*}

\begin{table*}[t]
\caption{\label{tab:spec_fluences}
Fluences, total counts and hardness ratios of the \textit{RHESSI} GRBs used to check the conversion between
$H_{\frac{120-1500}{25-120}}$\,(erg\,cm$^{-2}$\,erg$^{-1}$\,cm$^{2}$) and $\widetilde{H}_{\frac{120-1500}{25-120}}$\,(cnt\,cnt$^{-1}$).
}
\small
\centering
\begin{tabular}{@{}clccccccccc@{}}
\tableline \\[-9pt]
Ref.\tablenotemark{a}    & GRB\tablenotemark{b} & $S_{25-120}$\tablenotemark{c}                  & $S_{120-1500}$\tablenotemark{d}                 & $C_{25-120}$\tablenotemark{e}  & $C_{120-1500}$\tablenotemark{f} & $H_{\frac{120-1500}{25-120}}$\tablenotemark{g}          & $\widetilde{H}_{\frac{120-1500}{25-120}}$\tablenotemark{h}   & $\theta$\tablenotemark{i} \\
                         &                      & \scriptsize{($10^{-6} \mathrm{erg\,cm}^{-2}$)} & \scriptsize{($10^{-6} \mathrm{erg\,cm}^{-2}$)}  & (cnt)                          & (cnt)                           & $(\frac{\mathrm{erg\,cm}^{-2}}{\mathrm{erg\,cm}^{-2}})$ & $(\frac{\mathrm{cnt}}{\mathrm{cnt}})$                        & (deg)   \\[3pt]
\tableline \\[-9pt]
1                        &   020418             & $1.386 \pm 0.053$                              & $12.91 \pm 0.29$                                & $3\,013 \pm  103$              & $4\,649 \pm  102$               & $9.32 \pm 0.41$                                          & $1.54 \pm 0.06$                                             & 100.5 \\
2                        &   021020             & $6.61 \pm 0.23$                                & $19.30 \pm 0.47$                                & $7\,210 \pm  181$              & $7\,372 \pm  219$               & $2.92 \pm 0.13$                                          & $1.02 \pm 0.04$                                             & 114.8 \\
3                        &   021201             & $0.204 \pm 0.031$                              & $3.70 \pm  0.15$                                & $481 \pm    34$                & $1\,074 \pm   43$               & $18.16 \pm 2.86$                                         & $2.23 \pm 0.18$                                             & 129.3 \\
4                        &   021205             & $11.70 \pm 0.66$                               & $18.6 \pm   1.2$                                & $9\,978 \pm  357$              & $7\,367 \pm  381$               & $1.59 \pm 0.14$                                          & $0.74 \pm 0.05$                                             & 116.9 \\
5                        &   030922B            & $5.75 \pm 0.22$                                & $21.25 \pm 0.42$                                & $7\,245 \pm  154$              & $8\,310 \pm  165$               & $3.70 \pm 0.16$                                          & $1.15 \pm 0.03$                                             & 144.3 \\
6                        &   031027             & $18.52 \pm 0.35$                               & $40.23 \pm 0.74$                                & $23\,785 \pm 314$              & $19\,462 \pm 341$               & $2.17 \pm 0.06$                                          & $0.82 \pm 0.02$                                             & 101.5 \\
7                        &   031111             & $7.61 \pm 0.17$                                & $41.15 \pm 0.58$                                & $10\,164 \pm 127$              & $14\,224 \pm 144$               & $5.41 \pm 0.14$                                          & $1.40 \pm 0.02$                                             & 155.6 \\
8                        &   050717             & $2.45 \pm 0.12$                                & $18.71 \pm 0.46$                                & $3\,510 \pm  125$              & $5\,902 \pm  147$               & $7.64 \pm 0.41$                                          & $1.68 \pm 0.07$                                             & 110.8 \\
9                        &   061121             & $2.69 \pm 0.10$                                & $19.75 \pm 0.40$                                & $5\,607 \pm  139$              & $7\,768 \pm  190$               & $7.35 \pm 0.30$                                          & $1.39 \pm 0.05$                                             & 85.3  \\
10                       &   061126             & $1.63 \pm 0.16$                                & $13.99 \pm 0.45$                                & $2\,863 \pm  133$              & $5\,048 \pm  188$               & $8.57 \pm 0.87$                                          & $1.76 \pm 0.11$                                             & 133.7 \\                     
\tableline
\end{tabular}
\tablenotetext{a}{The reference number of the measurements (see Fig.~\ref{fig:rhessi_Hcnt_T90} and Fig.~\ref{fig:rhessi_Hcnt_Herg}).}
\tablenotetext{b}{The \textit{RHESSI} GRB number.}
\tablenotetext{c}{The fluence at the energy range $25-120$\,keV obtained from the spectral fit.}
\tablenotetext{d}{The fluence at the energy range $120-1500$\,keV obtained from the spectral fit.}
\tablenotetext{e}{The total detected counts in the range $25-120$\,keV in the time interval $\Delta t_{\mathrm{spec}}$ from a GRB (background subtracted).}
\tablenotetext{f}{The total detected counts in the range $120-1500$\,keV in the time interval $\Delta t_{\mathrm{spec}}$ from a GRB (background subtracted).}
\tablenotetext{g}{The hardness calculated as a ratio of two fluences $S_{120-1500}/S_{25-120}$.}
\tablenotetext{h}{The ``pseudo-hardness'' calculated as a ratio of the detected counts $C_{120-1500}/C_{25-120}$.}
\tablenotetext{i}{The off-axis angle, i.e., the angle between the axis of the detector and the direction to the given GRB.}
\tablecomments{All mentioned errors are $1\sigma$ statistical errors.}
\end{table*}

\section{Discussion of the instrumental effects}
\label{sec:instrum_effects}

\subsection{Types of the biases}
\label{sec:biases_types}

In the previous two sections we obtained the results that for the BATSE database not all of the intermediate-duration
GRBs can be identified as XRFs; for the \textit{RHESSI}
database we obtained that either none or only a small minority of the intermediate-duration GRBs can be related to XRFs.
These two results seem to be in a contradiction with the result of \citet{ver10}, which claims a close relation of the \textit{Swift}'s 
intermediate-duration GRBs with XRFs. In fact, there can be two - essentially different - explanations for this discrepancy.
First, it can either happen that  something is wrong in these analyses, or, second, it can be that all three results are correct and the different
conclusions simply follow from the instrumental effects of the satellites. The purpose of this section is to debut the second alternative.

The instrumental effects indeed play an important role in the 
classification of GRBs based on the durations, fluences, spectral
properties and on the hardnesses. For example, the ratio of numbers of detected short to long GRBs depends 
on the instrument's energy range sensitivity, because the shorter GRBs tend to be harder \citep{qin13}. 
Different trigger criteria applied for different missions 
might also affect the results in the measured GRB distributions.
It is also known that the $T_{90}$ of a burst depends on the energy range of the detector used, because
lower detector's energy sensitivity leads to the
longer measured $T_{90}$ \citep{rich96,bis11,qin13}.
Another instrumental effect which can play a role is that a detector with a rather small effective area
in a given energy range will pick up the peaks of the time profiles of GRBs. This would lead to the underestimation
of $T_{90}$ durations (tip-of-the-iceberg effect) as well as to the overestimation of the hardness ratio due to
the hardness-intensity correlation \citep{gol83,lia96,bor01,koc03,koc13}. This can be the case for the \textit{RHESSI} satellite,
because the maximum effective area, as shown in Fig.~\ref{fig:rhessi_eff_area}, is only $142$\,cm$^{2}$.
The effective area of the \textit{Swift}/BAT is $\approx1400$\,cm$^{2}$
(on-axis maximum)\footnote{http://heasarc.gsfc.nasa.gov/docs/heasarc/caldb/swift/docs/ bat/SWIFT-BAT-CALDB-BTI-V6.pdf}
and the maximal effective area of \textit{CGRO}/BATSE (LAD) is $\approx1900$\,cm$^{2}$ \citep{fish85}.
Compared to these the \textit{RHESSI}'s effective area is about an order of magnitude smaller.

Any discussion of the instrumental biases can be a highly complicated task - see, 
for example, a discussion for the BATSE database in \citet{ho06}. Roughly, 
there are two types of biases, which can be important for the purpose of this
article, and hence they should be discussed - at least briefly.
The first type of the bias is given by the errors of the given quantities of a given GRB measured by a given instrument.
The second type of the bias is given by the fact that different satellites have different instrumentations.
We used the measured $T_{90}$, measured fluences, and the measured spectral parameters.
Hence, the first type can cause that these quantities are uncertain due to standard empirical errors following from the
instrumentation of a given satellite. The second type can even cause a given GRB to be detectable by a particular satellite,
but not detectable by another one.

For the sake of completeness it must also be noted that other types of biases can be present.
For example, it is possible to discuss two hypothetical datasets for a given satellite containing both
the GRBs that are actually detected and the GRBs that would be detected by an ideal detector - see the Subsection 4.3. of \citet{ho06}.
Since for our purpose the empirically determined hardnesses are sufficient, this type of discussion can be omitted in this work.

Concerning the first type of biases here an eventual change of a $T_{90}$ value
for a given GRB due to an error would cause a horizontal shift in the $T_{90}$ vs. hardness figure, which would not change its
position with respect to the horizontal XRF limiting values, however it would effect the classification into the
short-, intermediate-, and long-duration classes and thus effect the numbers of XRFs distributed among them.
The durations with their errors are usually accurately determined in the datasets used.
This allowed us to study such an effect in Subsections~\ref{sec:batse_uncertainties_in_fractions} and
\ref{sec:rhessi_uncertainties_in_fractions}.
The hardnesses of GRBs are calculated from the fluences, which are either calculated from the spectra or are given in detected counts.
All these effects constitute sources of uncertainty. Any change in the hardness value causes a vertical shift in the $T_{90}$ vs. hardness figure. Hence, the biases in the hardness values 
are very important in this study. This was the reason of the detailed discussions of the 
uncertainties in the previous two sections. After these detailed discussions we can say that the first type of biases
gave large uncertainties, but the results collected in Table~\ref{tab:batse_H_50-100/25-50} and
Table~\ref{tab:rhessi_pseudo-hardness_T90}, respectively, hold. On the other hand,
because the hardnesses are differently defined for the BATSE, \textit{RHESSI} and \textit{Swift} databases, a comparison of these instruments
is still needed to discuss the second type of biases. This is briefly done in the following subsection.

\subsection{Comparison of the \textit{Swift}, BATSE and \textit{RHESSI} data}
\label{sec:swift_biases}

For the \textit{Swift} database \citet{ver10} ``found evidence that the intermediate population is closely related to XRFs''.
This evidence was obtained from a \textit{Swift} sample containing 408 GRBs.
In this sample 24 soft GRBs were identified as XRFs by applying the XRF definition introduced by \citet{sak08}.
Then it was shown that all these 24 GRBs were the members of the intermediate
group with high probabilities (Table~7 of \citet{ver10}).
This procedure doubtlessly implies a strong relation between the intermediate group and XRFs.
However, 24 objects define only a 5.9\,\% fraction of the whole sample.
On the other hand, the intermediate group itself gives a higher fraction
(12\,\%, see Table~1 of \citet{ver10}). From Figure~2 of \citet{ver10} it also follows immediately that - roughly -
only the half of the intermediate group is below the XRF limit.
Hence, no strict conclusion can be said about the relation of
the intermediate-duration GRBs to XRFs being above the XRF limit from their Figure~2.
This implies that even for the \textit{Swift} database it can be argued that roughly only half of the
intermediate-duration GRBs are certainly XRFs
and a strong one-to-one identification of the intermediate-duration bursts to XRFs cannot be claimed yet.

A comparison of the intermediate-duration bursts of the BATSE and \textit{Swift} datasets, respectively, shows immediately that these two sets of
GRBs are different. For the \textit{Swift} database it is clear from Figure~2 and Figure~8 of \citet{ver10} that the intermediate
group contains the softest GRBs. In addition, there is no anti-correlation between the hardness and duration $T_{90}$.
Contrary to this, for the BATSE's intermediate group there is a strong anti-correlation between the hardness and $T_{90}$
(see Figure~1 of \citet{ho06}), because there the intermediate group contains also hard GRBs
having hardnesses comparable with the hardnesses of the short GRBs. In any case, this anti-correlation for
the BATSE's intermediate group remains further a curious phenomenon.

This behaviour of the two groups could be an instrumental effect, because the \textit{Swift} satellite is detecting photons only
below $150$\,keV. In other words, it is possible that \textit{Swift} simply cannot detect some hard intermediate-duration
GRBs, which are still detectable by BATSE. This different instrumental behaviour means
that the short hard bursts give only 8\,\% of all GRBs in the
\textit{Swift} database (Table~1 of \citet{ver10}), but in the BATSE database the short hard bursts give 25\,\% of all GRBs
(Table~2 of \citet{ho06}). For the short hard GRBs this difference
can be partially given by that difference in the energy sensitivity.
This can also occur for the spectrally hard part of intermediate-duration bursts
- BATSE could detect them, but \textit{Swift} cannot.
Another instrumental effect responsible for this difference is that in the BATSE case,
if the detectors see excess in the light curve, this means detection. 
However, in the \textit{Swift}/BAT case, not only the excess in the light curve is needed,
but also a new source needs to be identified in the image to get the position.
\textit{Swift}/BAT requires more photons to successfully image short GRBs.

Concerning the \textit{RHESSI}-\textit{Swift} relation the \textit{RHE\-SSI}'s intermediate-duration GRBs are on average as hard as the short GRBs.
\textit{RHESSI} is more sensitive for the higher photon energies and thus spectrally harder
GRBs should be more populated in the \textit{RHESSI} database than in the \textit{Swift} one.

Different fractions of XRFs - concerning the in\-ter\-me\-dia\-te-du\-ra\-tion GRBs
in the BATSE, \textit{RHESSI} and \textit{Swift} databases - might be caused by different satellite instrumentations.
A more precise work employing the instruments' response functions and trigger criteria would be required to confirm that claim.

\subsection{Comparison of \textit{RHESSI}, \textit{BeppoSAX}, \textit{HETE-2}, and \textit{Swift}}
\label{sec:GRB030528}

\begin{figure}[t]
\centering
\includegraphics[width=0.48\textwidth]{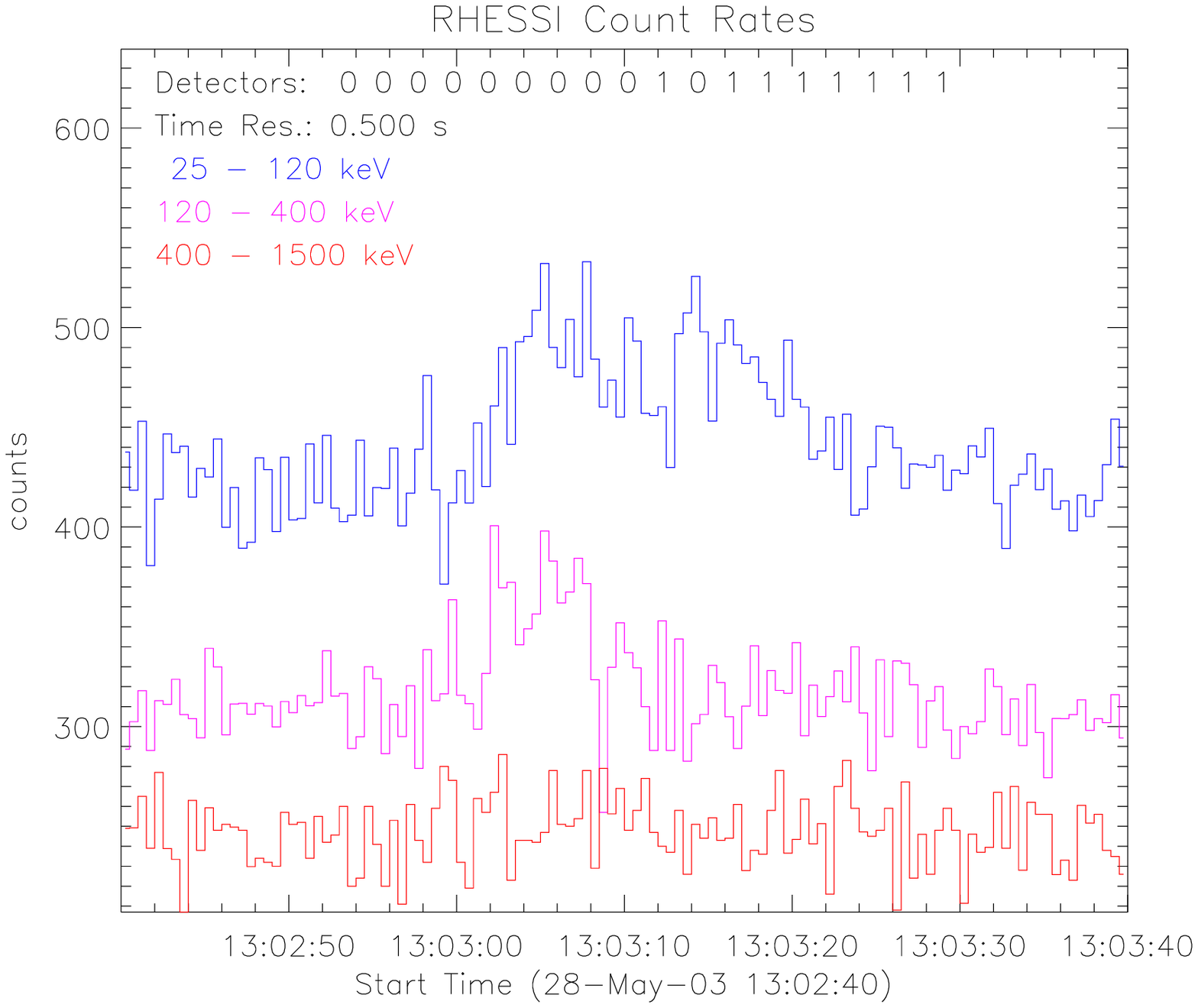}
\caption{The count-curve of GRB 030528 as observed by \textit{RHESSI} by all rear detector segments
(except for the malfunctioning No.2). This event was also detected by \textit{HETE-2} and classified as XRF.}
\label{fig:RHESSI_GRB030528}
\end{figure}

We checked the published lists of XRFs \break \citep{DAle06,sak05,sak08} detected by \textit{BeppoSAX}, \textit{HETE-2}, and \textit{Swift}
and searched if there was also a detection by \textit{RHESSI}.
We found that GRB 030528 detected by \textit{HETE-2} and classified as XRF by \citet{sak05} was also observed by the \textit{RHESSI} satellite
(see Fig.～\ref{fig:RHESSI_GRB030528}) and is presented in our sample. \citet{sak05} reported that this XRF had $T_{90} = 49.2\pm1.2$\,s,
$E_\mathrm{peak}=32\pm5$ KeV, the amplitude of the spectral fit normalized at 15\,KeV
$K_{15}=(14\pm2)$ $\times10^{-2}$ ph\,cm$^{-2}$\,s$^{-1}$ keV$^{-1}$, and the energy fluence
at the range $2-400$ keV $S=(119\pm8)$ $\times10^{-7}$ erg cm$^{-2}$. This means it had the
highest fluence of all the 16 detected \textit{HETE-2} XRFs with a typical $E_\mathrm{peak}$ and long duration.
\textit{RHESSI} detected this XRF with $T_{90} = 21.5\pm1.7$\,s and $\widetilde{H}_{\frac{120-1500}{25-120}}=0.59\pm0.1$.
Although, this is only one confirmed detection of an XRF by \textit{RHESSI} it demonstrates that \textit{RHESSI} can detect
long-duration XRFs if they are bright enough. We note that \textit{RHESSI} also detected flashes from
Soft-Gamma-Repeaters \citep{smi03,hur05} which are sources of soft gamma radiation ($\lesssim150$\,keV) \citep{mer08}.

Moreover, we performed simulations testing the detectability of the \textit{HETE-2} and \textit{Swift} XRFs by \textit{RHESSI}
using its detector response function and the real measured background.

We took the spectral parameters of the time averaged spectra of XRFs detected by \textit{HETE-2}
(Table~3 of \citet{sak05}) and \textit{Swift} (Table~1 of \citet{sak08}).

Then using these spectral parameters we simulated spectra of XRFs and folded them with the \textit{RHESSI}'s response function
summed over all rear segments of all detectors (except the malfunctioning No. 2) and averaged over all azimuth angles.

We used the response function for the off-axis angles $\theta_{\rm{sim}}$ nearest to the actual off-axis angles
$\theta$ of the simulated XRFs according to their celestial coordinates, the orientation of \textit{RHESSI},
and the trigger times of the XRFs. When the actual off-axis angle $\theta$ was not available then we used
response function for $\theta_{\rm{sim}}=90^{\circ}$ as approximation. 

Next, since the spectral parameters provided in \citet{sak05} and \citet{sak08} are time averaged, we assumed constant light curves
of durations $T$ equal to the durations of the actual XRFs (see Table~4 of \citet{sak05} and Table~1 of \citet{sak08}).

Having these simulated photon fluxes of XRFs folded with the \textit{RHESSI}'s response function one obtains the total number
of detector counts $C_{\rm{XRF}}$ which would be registered by \textit{RHESSI} over the duration $T$ and
energy range $25-1500$\,keV. 

The next step was to compare $C_{\rm{XRF}}$ with the real background.
We fitted the measured \textit{RHESSI} count rates by a linear function in a $\pm 50$\,s
interval around the trigger times of the detected \textit{HETE-2} and \textit{Swift} XRFs.
Then we calculated the average background count rates $n_{\rm{bkg}}$ at the times of the triggers.
When the \textit{RHESSI} data were not available then we used mean value $n_{\rm{bkg}}=2000$\,cnt\,s$^{-1}$
of all other 18 measured background count rates.
The total number of detector counts by \textit{RHESSI} due to the background summed over the duration $T$ and in the
range $25-1500$\,keV was $C_{\rm{bkg}}=T n_{\rm{bkg}}$.

The last step is to calculate S/N. Following the way the S/N was calculated in the \textit{RHESSI} database in
\citet{rip09} one has $\rm{S/N}=C_{\rm{XRF}}/$ \break $\sqrt{C_{\rm{XRF}}+2C_{\rm{bkg}}}$ because $C_{\rm{XRF}}=$
$C_{\rm{tot}}-C_{\rm{bkg}}$, where $C_{\rm{tot}}$ is the total number of counts in the detector.
In the \textit{RHESSI} database used by \citep{rip09} and in this work only the events with $\rm{S/N}>6$ were used.

The results of these simulations are shown in Table~\ref{tab:rhessi_hete2_swift_xrf_sim}.
We found that only one XRF out of 26 investigated ones is detectable by \textit{RHESSI}. It is GRB 030528 with
$\rm{S/N}=8.5$ obtained from the simulation. This XRF was actually detected by \textit{RHESSI} as described above.

We also tested the detectability of XRFs at ``best'' conditions, i.e. for $\theta_{\rm{sim}}=90^{\circ}$ and
$n_{\rm{bkg}}=1150$\,cnt\,s$^{-1}$ which is the minimal background rate from measured ones in
Table~\ref{tab:rhessi_hete2_swift_xrf_sim}. $\theta_{\rm{sim}}=90^{\circ}$ provides the highest affective area
especially at low energies (see Fig.\ref{fig:rhessi_eff_area}). The resultant S/N is above 6 only for GRB 030528.
For other five XRFs the S/N is above 3: GRB 021104 (S/N = 3.9), GRB 030823 (S/N = 5.8), GRB 030824 (S/N = 4.4), GRB 050416A
(S/N = 3.6), GRB 060923B (S/N = 3.5).

From these simulations we conclude that the vast majority of \textit{HETE-2} and \textit{Swift} XRFs is not detectable by
\textit{RHESSI}. Only sufficiently bright XRFs with significant flux above 25\,keV, like GRB 030528, can be detected
assuming the average background.

\begin{table*}[h]
\caption{\label{tab:rhessi_hete2_swift_xrf_sim}
Results of the detectability of the simulated \textit{HETE-2} and \textit{Swift} XRFs by \textit{RHESSI} using its detector response function
and the measured background.
}
\centering
\begin{tabular}{@{}lcccccccc@{}}
\tableline \\[-9pt]
GRB	&	Model\tablenotemark{a}	&	$\theta$\tablenotemark{b}	&	$\theta_{\rm{sim}}$\tablenotemark{c}	&	$T$\tablenotemark{d}	&	$n_{\rm{bkg}}$\tablenotemark{e}		&	$C_{\rm{bkg}}$\tablenotemark{f}	&	$C_{\rm{XRF}}$\tablenotemark{g}	&	S/N\tablenotemark{h}	\\
        &				&	(deg)				&	(deg)					&	(s)			&	($10^3$\,cnt\,s$^{-1}$)			&	($10^5$\,cnt)			&	($10^2$\,cnt)			&				\\
\tableline \\[-9pt]
010213	&	Band	&	N/A	&	90	&	34.4	&	N/A	&	6.88	&	1.20	&	0.3	\\
010225	&	CPL	&	N/A	&	90	&	9.8	&	N/A	&	1.96	&	4.03	&	2.0	\\
011019	&	CPL	&	N/A	&	90	&	24.6	&	N/A	&	4.92	&	1.92	&	0.6	\\
011130	&	PL	&	N/A	&	90	&	50.0	&	N/A	&	10.00	&	1.43	&	0.3	\\
011212	&	PL	&	N/A	&	90	&	57.7	&	N/A	&	11.54	&	4.16	&	0.9	\\
020317	&	CPL	&	155.9	&	150	&	10.0	&	N/A	&	2.00	&	0.72	&	0.4	\\
020625	&	CPL	&	132.7	&	135	&	41.9	&	1.86	&	7.79	&	0.04	&	0.0	\\
020903	&	PL	&	166.6	&	165	&	13.0	&	1.15	&	1.50	&	0.14	&	0.1	\\
021021	&	CPL	&	155.0	&	150	&	49.2	&	2.97	&	14.61	&	0.29	&	0.1	\\
021104	&	CPL	&	151.7	&	150	&	31.4	&	1.94	&	6.09	&	4.21	&	1.2	\\
030416	&	PL	&	142.1	&	135	&	78.6	&	2.21	&	17.37	&	2.78	&	0.5	\\
030429	&	CPL	&	147.9	&	150	&	24.6	&	2.03	&	4.99	&	2.73	&	0.9	\\
030528	&	Band	&	169.7	&	165	&	83.6	&	1.96	&	16.39	&	48.90	&	8.5	\\
030723	&	PL	&	155.8	&	150	&	31.3	&	2.17	&	6.79	&	3.39	&	0.9	\\
030823	&	CPL	&	145.3	&	150	&	55.6	&	1.63	&	9.06	&	8.69	&	2.0	\\
030824	&	PL	&	138.4	&	135	&	15.7	&	2.38	&	3.74	&	5.09	&	1.9	\\
050406	&	PL	&	59.1	&	60	&	6.4	&	N/A	&	1.28	&	0.57	&	0.4	\\
050416A	&	PL	&	144.9	&	150	&	3.0	&	2.32	&	0.70	&	1.20	&	1.0	\\
050714B	&	PL	&	65.7	&	60	&	50.3	&	2.12	&	10.66	&	4.14	&	0.9	\\
050819	&	PL	&	132.6	&	135	&	47.3	&	2.29	&	10.83	&	1.51	&	0.3	\\
050824	&	PL	&	129.7	&	135	&	26.6	&	2.05	&	5.45	&	1.02	&	0.3	\\
060219	&	PL	&	97.0	&	90	&	65.3	&	1.79	&	11.69	&	4.20	&	0.9	\\
060428B	&	CPL	&	102.3	&	105	&	65.7	&	1.71	&	11.24	&	0.54	&	0.1	\\
060512	&	PL	&	112.7	&	105	&	9.7	&	N/A	&	1.94	&	2.00	&	1.0	\\
060923B	&	PL	&	62.8	&	60	&	9.9	&	1.86	&	1.84	&	3.17	&	1.6	\\
060926	&	PL	&	81.3	&	75	&	8.7	&	1.64	&	1.43	&	1.87	&	1.1	\\
\tableline
\end{tabular}
\tablenotetext{a}{The used spectral model of the simulated XRFs. The spectral models and the best fit spectral parameters
were taken from Table~3 of \citet{sak05} and Table~1 of \citet{sak08}.}
\tablenotetext{b}{The actual off-axis angle of the simulated XRF according to its celestial coordinates, the orientation of the satellite,
and the trigger time of the XRF. ``N/A'' means that the value is not available because the date is before the launch of \textit{RHESSI}.}
\tablenotetext{c}{The off-axis angle of the used \textit{RHESSI} detector response function. When the actual off-axis angle $\theta$
was not available then we used $\theta_{\rm{sim}}=90^{\circ}$.}
\tablenotetext{d}{The duration of the XRF. The values were taken from Table~4 of \citet{sak05} and Table~1 of \citet{sak08}.}
\tablenotetext{e}{The measured background count rate by \textit{RHESSI} at the time of the given XRF in the range $25-1500$\,keV.
``N/A'' means that data were not available.}
\tablenotetext{f}{The number of the detector counts by \textit{RHESSI} due to the background summed over the duration $T$
and in the range $25-1500$\,keV. When the measured $n_{\rm{bkg}}$ was not available then we assumed mean background count rate
$n_{\rm{bkg}}=2000$\,cnt\,s$^{-1}$ of all other 18 measured count rates.}
\tablenotetext{g}{The calculated number of counts due to the simulated XRF summed over the duration $T$, in the range
$25-1500$\,keV, and assuming the \textit{RHESSI} detector response function.}
\tablenotetext{h}{The obtained signal-to-noise ratio of the simulated XRF.}

\end{table*}

\section{Conclusions}
\label{sec:conclusions}

The astrophysical meaning of the intermediate-duration GRBs remains an open question.
After \citet{ver10} it seemed that the question was answered,
because it was claimed that the \textit{Swift} intermediate-duration bursts can be related to XRFs. 
The results of this article show that this point of view can hold for the BATSE database (at least partially),
but it most likely does not hold for the \textit{RHESSI} dataset. 

Summing up we conclude:
\begin{enumerate}
\item
For the BATSE databases we used different spectral models and different GRB samples.
No essential difference followed from the different spectral models and from the different samples.
Only the fractions of XRFs were determined with large scatters.
After these detailed studies we deduced that there was a
$1.3-4.2$\,\% fraction of events classified as XRFs in the BATSE dataset.
For the three groups separately we obtained that the vast majority of the BATSE short bursts are not XRFs:
specifically, only $0.7-5.7$\,\% of the short bursts can be given by XRFs;
a $1-85$\,\% fraction of the BATSE intermediate-duration bursts,
and a $1.0-3.4$\,\% fraction of the long bursts
as identified by \citet{ho06} can be given by XRFs. The lower limit for intermediate-duration bursts
(1\,\%) is so small that one cannot exclude that this small fraction is given by outliers and
the BATSE database does not contain any XRFs from the intermediate-duration group whatsoever.
On the other hand, even at the other marginal case, 15\,\% of intermediate GRBs cannot
be identical to XRFs - hence not all bursts of the entire intermediate-duration group can be identical to XRFs.
\newline
\item
For the \textit{RHESSI} dataset a detailed analysis of the hardness of GRBs was provided. The connection of the
hardness and the ``pseudo-hardness'' was intensively studied - both by using numerical simulations and by analysing the actual data.
The short and the intermediate-duration GRBs, as identified in works \citet{rip09} and \citet{rip12},
were found most likely not to be associated with XRFs.
For the sake of precision it should be added that there are three short GRBs, which can potentially be XRFs.
However, even taking these into account more than $79$\,\% of short GRBs should not be XRFs,
and at least $53$\,\% of \textit{RHESSI} intermediate-duration bursts should not be XRFs.
At least 45\,\% of the \textit{RHESSI} long bursts are not given by XRFs.
A simulation of XRFs observed by \textit{HETE-2} and \textit{Swift} 
has shown that \textit{RHESSI} would detect, and in fact detected, only one long-duration XRF
out of 26 ones observed by those two satellites. Concerning the \textit{RHESSI} intermediate-duration
bursts, the conclusion that they are most likely not given by XRFs, was expected from 
the fact that their hardnesses are too high and actually they are comparable with the hardnesses of the short bursts.
\newline
\item
In the hardness vs. $T_{90}$ duration plots there is not seen any apparent separation between XRFs and GRBs
at the hardnesses given by the XRF limits used in this work and in the intermediate-to-long durations.
This suggests that XRFs could constitute a soft tail of the long GRB population
and could arise from the same phenomenon as stated already by \citet{kip03} and \citet{sak05}.
\newline
\item
A close relation of the intermediate-duration bursts and XRFs, suggested by \citet{ver10} from the \textit{Swift} database,
does not hold for the other two databases. The intermediate-duration bursts in the BATSE database can be partly populated by XRFs,
but the \textit{RHESSI} intermediate-duration bursts are most likely not given by XRFs.
In fact, this is the key result of this paper.
\end{enumerate}

\acknowledgments{\small
We wish to thank S. Appleby, Z. Bagoly, L.G. Bal\'azs, G.J. Fishman, I. Horv\'ath, J. K\'obori, P. M\'esz\'aros, D. Sz\'ecsi and P. Veres 
for the useful discussions and comments. We kindly thank to the anonymous referee for the useful suggestions
which improved this manuscript.
We gratefully acknowledge the use of the online BATSE Current
Catalog and Spectral Catalog released by the BATSE team and National Space, Science, and Technology Center.
Next, we acknowledge the use of the online data from the \textit{RHESSI} instrument and gratefully thank to C. Wigger and W. Hajdas
for providing us the \textit{RHESSI} GRB analysis software. We also kindly thank to E. Bellm and D. Smith for providing us
the off-axis \textit{RHESSI} response matrices. This study was supported by Taiwan's Ministry of Science and Technology (MOST)
funding number 104-2811-M-002-160.
}

\bibliographystyle{spr-mp-nameyear-cnd}
\bibliography{ripa_meszaros}

\begin{thebibliography}{93}
\ifx \bisbn   \undefined \def \bisbn  #1{ISBN #1}\fi
\ifx \binits  \undefined \def \binits#1{#1} \fi
\ifx \bauthor  \undefined \def \bauthor#1{#1} \fi
\ifx \batitle  \undefined \def \batitle#1{#1} \fi
\ifx \bjtitle  \undefined \def \bjtitle#1{#1}\fi
\ifx \bvolume  \undefined \def \bvolume#1{\textbf{#1}}\fi
\ifx \byear  \undefined \def \byear#1{#1} \fi
\ifx \bissue  \undefined \def \bissue#1{#1} \fi
\ifx \bfpage  \undefined \def \bfpage#1{#1} \fi
\ifx \blpage  \undefined \def \blpage #1{#1} \fi
\ifx \burl  \undefined \def \burl#1{\textsf{#1}} \fi
\ifx \doiurl  \undefined \def \doiurl#1{\textsf{#1}} \fi
\ifx \betal  \undefined \def \betal{\textit{et al.}} \fi
\ifx \binstitute  \undefined \def \binstitute#1{#1} \fi
\ifx \binstitutionaled  \undefined \def \binstitutionaled#1{#1} \fi
\ifx \bctitle  \undefined \def \bctitle#1{#1} \fi
\ifx \beditor  \undefined \def \beditor#1{#1} \fi
\ifx \bpublisher  \undefined \def \bpublisher#1{#1} \fi
\ifx \bbtitle  \undefined \def \bbtitle#1{#1} \fi
\ifx \bedition  \undefined \def \bedition#1{#1} \fi
\ifx \bseriesno  \undefined \def \bseriesno#1{#1} \fi
\ifx \blocation  \undefined \def \blocation#1{#1} \fi
\ifx \bsertitle  \undefined \def \bsertitle#1{#1} \fi
\ifx \bsnm \undefined \def \bsnm#1{#1} \fi
\ifx \bsuffix \undefined \def \bsuffix#1{#1} \fi
\ifx \bparticle \undefined \def \bparticle#1{#1} \fi
\ifx \barticle \undefined \def \barticle#1{#1} \fi
\ifx \bconfdate \undefined \def \bconfdate #1{#1} \fi
\ifx \botherref \undefined \def \botherref #1{#1} \fi
\ifx \url \undefined \def \url#1{\textsf{#1}} \fi
\ifx \bchapter \undefined \def \bchapter#1{#1} \fi
\ifx \bbook \undefined \def \bbook#1{#1} \fi
\ifx \bcomment \undefined \def \bcomment#1{#1} \fi
\ifx \oauthor \undefined \def \oauthor#1{#1} \fi
\ifx \citeauthoryear \undefined \def \citeauthoryear#1{#1} \fi
\ifx \endbibitem  \undefined \def \endbibitem {}\fi
\ifx \bconflocation  \undefined \def \bconflocation#1{#1} \fi
\ifx \arxivurl  \undefined \def \arxivurl#1{\textsf{#1}} \fi

\bibitem[\protect\citeauthoryear{{Axel\-sson} and {Borgonovo}}{2015}]{axe15}
\begin{barticle}
\bauthor{\bsnm{{Axel\-sson}}, \binits{M.}},
\bauthor{\bsnm{{Borgonovo}}, \binits{L.}}:
\bjtitle{\mnras}
\bvolume{447},
\bfpage{3150}
(\byear{2015})
\end{barticle}
\endbibitem

\bibitem[\protect\citeauthoryear{{Bagoly} et~al.}{1998}]{bag98}
\begin{barticle}
\bauthor{\bsnm{{Bagoly}}, \binits{Z.}},
\bauthor{\bsnm{{M{\'e}sz{\'a}ros}}, \binits{A.}},
\bauthor{\bsnm{{Horv{\'a}th}}, \binits{I.}},
\bauthor{\bsnm{{Bal{\'a}zs}}, \binits{L.G.}},
\bauthor{\bsnm{{M{\'e}sz{\'a}ros}}, \binits{P.}}:
\bjtitle{\apj}
\bvolume{498},
\bfpage{342}
(\byear{1998})
\end{barticle}
\endbibitem

\bibitem[\protect\citeauthoryear{{Bagoly} et~al.}{2009}]{bag09}
\begin{barticle}
\bauthor{\bsnm{{Bagoly}}, \binits{Z.}},
\bauthor{\bsnm{{Borgonovo}}, \binits{L.}},
\bauthor{\bsnm{{M{\'e}sz{\'a}ros}}, \binits{A.}},
\bauthor{\bsnm{{Bal{\'a}zs}}, \binits{L.G.}},
\bauthor{\bsnm{{Horv{\'a}th}}, \binits{I.}}:
\bjtitle{\aap}
\bvolume{493},
\bfpage{51}
(\byear{2009})
\end{barticle}
\endbibitem

\bibitem[\protect\citeauthoryear{{Balastegui} et~al.}{2001}]{balas01}
\begin{barticle}
\bauthor{\bsnm{{Balastegui}}, \binits{A.}},
\bauthor{\bsnm{{Ruiz-Lapuente}}, \binits{P.}},
\bauthor{\bsnm{{Canal}}, \binits{R.}}:
\bjtitle{\mnras}
\bvolume{328},
\bfpage{283}
(\byear{2001})
\end{barticle}
\endbibitem

\bibitem[\protect\citeauthoryear{{Bal{\'a}zs} et~al.}{2003}]{bal03}
\begin{barticle}
\bauthor{\bsnm{{Bal{\'a}zs}}, \binits{L.G.}},
\bauthor{\bsnm{{Bagoly}}, \binits{Z.}},
\bauthor{\bsnm{{Horv{\'a}th}}, \binits{I.}},
\bauthor{\bsnm{{M{\'e}sz{\'a}ros}}, \binits{A.}},
\bauthor{\bsnm{{M{\'e}sz{\'a}ros}}, \binits{P.}}:
\bjtitle{\aap}
\bvolume{401},
\bfpage{129}
(\byear{2003})
\end{barticle}
\endbibitem

\bibitem[\protect\citeauthoryear{{Band} et~al.}{1993}]{band93}
\begin{barticle}
\bauthor{\bsnm{{Band}}, \binits{D.}},
\bauthor{\bsnm{{Matteson}}, \binits{J.}},
\bauthor{\bsnm{{Ford}}, \binits{L.}},
\bauthor{\bsnm{{Schaefer}}, \binits{B.}},
\bauthor{\bsnm{{Palmer}}, \binits{D.}},
\bauthor{\bsnm{{Teegarden}}, \binits{B.}},
\bauthor{\bsnm{{Cline}}, \binits{T.}},
\bauthor{\bsnm{{Briggs}}, \binits{M.}},
\bauthor{\bsnm{{Paciesas}}, \binits{W.}},
\bauthor{\bsnm{{Pendleton}}, \binits{G.}},
\bauthor{\bsnm{{Fishman}}, \binits{G.}},
\bauthor{\bsnm{{Kouveliotou}}, \binits{C.}},
\bauthor{\bsnm{{Meegan}}, \binits{C.}},
\bauthor{\bsnm{{Wilson}}, \binits{R.}},
\bauthor{\bsnm{{Lestrade}}, \binits{P.}}:
\bjtitle{\apj}
\bvolume{413},
\bfpage{281}
(\byear{1993})
\end{barticle}
\endbibitem

\bibitem[\protect\citeauthoryear{{Bellm}}{2010}]{bel10}
\begin{barticle}
\bauthor{\bsnm{{Bellm}}, \binits{E.C.}}:
\bjtitle{\apj}
\bvolume{714},
\bfpage{881}
(\byear{2010})
\end{barticle}
\endbibitem

\bibitem[\protect\citeauthoryear{{Bellm}}{2011}]{bel11}
\begin{botherref}
\oauthor{\bsnm{{Bellm}}, \binits{E.C.}}:
{Studies of Gamma-Ray Burst Prompt Emission with RHESSI and NCT}.
PhD thesis,
University of California, Berkeley
(2011)
\end{botherref}
\endbibitem

\bibitem[\protect\citeauthoryear{{Bellm} et~al.}{2008a}]{bel08a}
\begin{barticle}
\bauthor{\bsnm{{Bellm}}, \binits{E.C.}},
\bauthor{\bsnm{{Hurley}}, \binits{K.}},
\bauthor{\bsnm{{Pal'shin}}, \binits{V.}},
\bauthor{\bsnm{{Yamaoka}}, \binits{K.}},
\bauthor{\bsnm{{Bandstra}}, \binits{M.S.}},
\bauthor{\bsnm{{Boggs}}, \binits{S.E.}},
\bauthor{\bsnm{{Hong}}, \binits{S.}},
\bauthor{\bsnm{{Kodaka}}, \binits{N.}},
\bauthor{\bsnm{{Kozyrev}}, \binits{A.S.}},
\bauthor{\bsnm{{Litvak}}, \binits{M.L.}},
\bauthor{\bsnm{{Mitrofanov}}, \binits{I.G.}},
\bauthor{\bsnm{{Nakagawa}}, \binits{Y.E.}},
\bauthor{\bsnm{{Ohno}}, \binits{M.}},
\bauthor{\bsnm{{Onda}}, \binits{K.}},
\bauthor{\bsnm{{Sanin}}, \binits{A.B.}},
\bauthor{\bsnm{{Sugita}}, \binits{S.}},
\bauthor{\bsnm{{Tashiro}}, \binits{M.}},
\bauthor{\bsnm{{Tretyakov}}, \binits{V.I.}},
\bauthor{\bsnm{{Urata}}, \binits{Y.}},
\bauthor{\bsnm{{Wigger}}, \binits{C.}}:
\bjtitle{\apj}
\bvolume{688},
\bfpage{491}
(\byear{2008}a)
\end{barticle}
\endbibitem

\bibitem[\protect\citeauthoryear{{Bellm} et~al.}{2008b}]{bel08b}
\begin{bchapter}
\bauthor{\bsnm{{Bellm}}, \binits{E.C.}},
\bauthor{\bsnm{{Bandstra}}, \binits{M.E.}},
\bauthor{\bsnm{{Boggs}}, \binits{S.E.}},
\bauthor{\bsnm{{Hajdas}}, \binits{W.}},
\bauthor{\bsnm{{Hurley}}, \binits{K.}},
\bauthor{\bsnm{{Smith}}, \binits{D.M.}},
\bauthor{\bsnm{{Wigger}}, \binits{C.}}:
In: \beditor{\bsnm{{Galassi}}, \binits{M.}},
\beditor{\bsnm{{Palmer}}, \binits{D.}},
\beditor{\bsnm{{Fenimore}}, \binits{E.}} (eds.)
\bbtitle{Gamma-Ray Bursts 2007: Proceedings of the Santa Fe Conference}.
\bsertitle{AIP Conf. Proc.},
vol. \bseriesno{1000},
p. \bfpage{154}
(\byear{2008}b)
\end{bchapter}
\endbibitem

\bibitem[\protect\citeauthoryear{{Bissaldi} et~al.}{2011}]{bis11}
\begin{barticle}
\bauthor{\bsnm{{Bissaldi}}, \binits{E.}},
\bauthor{\bsnm{{von Kienlin}}, \binits{A.}},
\bauthor{\bsnm{{Kouveliotou}}, \binits{C.}},
\bauthor{\bsnm{{Briggs}}, \binits{M.S.}},
\bauthor{\bsnm{{Connaughton}}, \binits{V.}},
\bauthor{\bsnm{{Greiner}}, \binits{J.}},
\bauthor{\bsnm{{Gruber}}, \binits{D.}},
\bauthor{\bsnm{{Lichti}}, \binits{G.}},
\bauthor{\bsnm{{Bhat}}, \binits{P.N.}},
\bauthor{\bsnm{{Burgess}}, \binits{M.}},
\bauthor{\bsnm{{Chaplin}}, \binits{V.}},
\bauthor{\bsnm{{Diehl}}, \binits{R.}},
\bauthor{\bsnm{{Fishman}}, \binits{G.J.}},
\bauthor{\bsnm{{Fitzpatrick}}, \binits{G.}},
\bauthor{\bsnm{{Foley}}, \binits{S.}},
\bauthor{\bsnm{{Gibby}}, \binits{M.H.}},
\bauthor{\bsnm{{Giles}}, \binits{M.M.}},
\bauthor{\bsnm{{Goldstein}}, \binits{A.}},
\bauthor{\bsnm{{Guiriec}}, \binits{S.}},
\bauthor{\bsnm{{van der Horst}}, \binits{A.J.}},
\bauthor{\bsnm{{Kippen}}, \binits{R.M.}},
\bauthor{\bsnm{{Lin}}, \binits{L.}},
\bauthor{\bsnm{{McBreen}}, \binits{S.}},
\bauthor{\bsnm{{Meegan}}, \binits{C.A.}},
\bauthor{\bsnm{{Paciesas}}, \binits{W.S.}},
\bauthor{\bsnm{{Preece}}, \binits{R.D.}},
\bauthor{\bsnm{{Rau}}, \binits{A.}},
\bauthor{\bsnm{{Tierney}}, \binits{D.}},
\bauthor{\bsnm{{Wilson-Hodge}}, \binits{C.}}:
\bjtitle{\apj}
\bvolume{733},
\bfpage{97}
(\byear{2011})
\end{barticle}
\endbibitem

\bibitem[\protect\citeauthoryear{{Borgonovo}}{2004}]{bor04}
\begin{barticle}
\bauthor{\bsnm{{Borgonovo}}, \binits{L.}}:
\bjtitle{\aap}
\bvolume{418},
\bfpage{487}
(\byear{2004})
\end{barticle}
\endbibitem

\bibitem[\protect\citeauthoryear{{Borgonovo} and {Bj{\"o}rnsson}}{2006}]{bor06}
\begin{barticle}
\bauthor{\bsnm{{Borgonovo}}, \binits{L.}},
\bauthor{\bsnm{{Bj{\"o}rnsson}}, \binits{C.-I.}}:
\bjtitle{\apj}
\bvolume{652},
\bfpage{1423}
(\byear{2006})
\end{barticle}
\endbibitem

\bibitem[\protect\citeauthoryear{{Borgonovo} and {Ryde}}{2001}]{bor01}
\begin{barticle}
\bauthor{\bsnm{{Borgonovo}}, \binits{L.}},
\bauthor{\bsnm{{Ryde}}, \binits{F.}}:
\bjtitle{\apj}
\bvolume{548},
\bfpage{770}
(\byear{2001})
\end{barticle}
\endbibitem

\bibitem[\protect\citeauthoryear{{Bromberg} et~al.}{2013}]{brom13}
\begin{barticle}
\bauthor{\bsnm{{Bromberg}}, \binits{O.}},
\bauthor{\bsnm{{Nakar}}, \binits{E.}},
\bauthor{\bsnm{{Piran}}, \binits{T.}},
\bauthor{\bsnm{{Sari}}, \binits{R.}}:
\bjtitle{\apj}
\bvolume{764},
\bfpage{179}
(\byear{2013})
\end{barticle}
\endbibitem

\bibitem[\protect\citeauthoryear{{Dado} et~al.}{2004}]{dad04}
\begin{barticle}
\bauthor{\bsnm{{Dado}}, \binits{S.}},
\bauthor{\bsnm{{Dar}}, \binits{A.}},
\bauthor{\bsnm{{De R{\'u}jula}}, \binits{A.}}:
\bjtitle{\aap}
\bvolume{422},
\bfpage{381}
(\byear{2004})
\end{barticle}
\endbibitem

\bibitem[\protect\citeauthoryear{{D'Alessio} et~al.}{2006}]{DAle06}
\begin{barticle}
\bauthor{\bsnm{{D'Alessio}}, \binits{V.}},
\bauthor{\bsnm{{Piro}}, \binits{L.}},
\bauthor{\bsnm{{Rossi}}, \binits{E.M.}}:
\bjtitle{\aap}
\bvolume{460},
\bfpage{653}
(\byear{2006})
\end{barticle}
\endbibitem

\bibitem[\protect\citeauthoryear{{Dermer} et~al.}{1999}]{der99}
\begin{barticle}
\bauthor{\bsnm{{Dermer}}, \binits{C.D.}},
\bauthor{\bsnm{{Chiang}}, \binits{J.}},
\bauthor{\bsnm{{B{\"o}ttcher}}, \binits{M.}}:
\bjtitle{\apj}
\bvolume{513},
\bfpage{656}
(\byear{1999})
\end{barticle}
\endbibitem

\bibitem[\protect\citeauthoryear{{Eichler} and {Levinson}}{2004}]{eich04}
\begin{barticle}
\bauthor{\bsnm{{Eichler}}, \binits{D.}},
\bauthor{\bsnm{{Levinson}}, \binits{A.}}:
\bjtitle{\apjl}
\bvolume{614},
\bfpage{13}
(\byear{2004})
\end{barticle}
\endbibitem

\bibitem[\protect\citeauthoryear{{Fishman} et~al.}{1985}]{fish85}
\begin{botherref}
\oauthor{\bsnm{{Fishman}}, \binits{G.J.}},
\oauthor{\bsnm{{Meegan}}, \binits{C.A.}},
\oauthor{\bsnm{{Parnell}}, \binits{T.A.}},
\oauthor{\bsnm{{Wilson}}, \binits{R.B.}},
\oauthor{\bsnm{{Paciesas}}, \binits{W.}},
\oauthor{\bsnm{{Mateson}}, \binits{J.L.}},
\oauthor{\bsnm{{Cline}}, \binits{T.L.}},
\oauthor{\bsnm{{Teegarden}}, \binits{B.J.}}:
International Cosmic Ray Conference
\textbf{3}
(1985)
\end{botherref}
\endbibitem

\bibitem[\protect\citeauthoryear{{Fishman} et~al.}{1994}]{fish94}
\begin{barticle}
\bauthor{\bsnm{{Fishman}}, \binits{G.J.}},
\bauthor{\bsnm{{Meegan}}, \binits{C.A.}},
\bauthor{\bsnm{{Wilson}}, \binits{R.B.}},
\bauthor{\bsnm{{Brock}}, \binits{M.N.}},
\bauthor{\bsnm{{Horack}}, \binits{J.M.}},
\bauthor{\bsnm{{Kouveliotou}}, \binits{C.}},
\bauthor{\bsnm{{Howard}}, \binits{S.}},
\bauthor{\bsnm{{Paciesas}}, \binits{W.S.}},
\bauthor{\bsnm{{Briggs}}, \binits{M.S.}},
\bauthor{\bsnm{{Pendleton}}, \binits{G.N.}},
\bauthor{\bsnm{{Koshut}}, \binits{T.M.}},
\bauthor{\bsnm{{Mallozzi}}, \binits{R.S.}},
\bauthor{\bsnm{{Stollberg}}, \binits{M.}},
\bauthor{\bsnm{{Lestrade}}, \binits{J.P.}}:
\bjtitle{\apjs}
\bvolume{92},
\bfpage{229}
(\byear{1994})
\end{barticle}
\endbibitem

\bibitem[\protect\citeauthoryear{{Gehrels} et~al.}{2004}]{geh04}
\begin{barticle}
\bauthor{\bsnm{{Gehrels}}, \binits{N.}},
\bauthor{\bsnm{{Chincarini}}, \binits{G.}},
\bauthor{\bsnm{{Giommi}}, \binits{P.}},
\bauthor{\bsnm{{Mason}}, \binits{K.O.}},
\bauthor{\bsnm{{Nousek}}, \binits{J.A.}},
\bauthor{\bsnm{{Wells}}, \binits{A.A.}},
\bauthor{\bsnm{{White}}, \binits{N.E.}},
\bauthor{\bsnm{{Barthelmy}}, \binits{S.D.}},
\bauthor{\bsnm{{Burrows}}, \binits{D.N.}},
\bauthor{\bsnm{{Cominsky}}, \binits{L.R.}},
\bauthor{\bsnm{{Hurley}}, \binits{K.C.}},
\bauthor{\bsnm{{Marshall}}, \binits{F.E.}},
\bauthor{\bsnm{{M{\'e}sz{\'a}ros}}, \binits{P.}},
\bauthor{\bsnm{{Roming}}, \binits{P.W.A.}},
\bauthor{\bsnm{{Angelini}}, \binits{L.}},
\bauthor{\bsnm{{Barbier}}, \binits{L.M.}},
\bauthor{\bsnm{{Belloni}}, \binits{T.}},
\bauthor{\bsnm{{Campana}}, \binits{S.}},
\bauthor{\bsnm{{Caraveo}}, \binits{P.A.}},
\bauthor{\bsnm{{Chester}}, \binits{M.M.}},
\bauthor{\bsnm{{Citterio}}, \binits{O.}},
\bauthor{\bsnm{{Cline}}, \binits{T.L.}},
\bauthor{\bsnm{{Cropper}}, \binits{M.S.}},
\bauthor{\bsnm{{Cummings}}, \binits{J.R.}},
\bauthor{\bsnm{{Dean}}, \binits{A.J.}},
\bauthor{\bsnm{{Feigelson}}, \binits{E.D.}},
\bauthor{\bsnm{{Fenimore}}, \binits{E.E.}},
\bauthor{\bsnm{{Frail}}, \binits{D.A.}},
\bauthor{\bsnm{{Fruchter}}, \binits{A.S.}},
\bauthor{\bsnm{{Garmire}}, \binits{G.P.}},
\bauthor{\bsnm{{Gendreau}}, \binits{K.}},
\bauthor{\bsnm{{Ghisellini}}, \binits{G.}},
\bauthor{\bsnm{{Greiner}}, \binits{J.}},
\bauthor{\bsnm{{Hill}}, \binits{J.E.}},
\bauthor{\bsnm{{Hunsberger}}, \binits{S.D.}},
\bauthor{\bsnm{{Krimm}}, \binits{H.A.}},
\bauthor{\bsnm{{Kulkarni}}, \binits{S.R.}},
\bauthor{\bsnm{{Kumar}}, \binits{P.}},
\bauthor{\bsnm{{Lebrun}}, \binits{F.}},
\bauthor{\bsnm{{Lloyd-Ronning}}, \binits{N.M.}},
\bauthor{\bsnm{{Markwardt}}, \binits{C.B.}},
\bauthor{\bsnm{{Mattson}}, \binits{B.J.}},
\bauthor{\bsnm{{Mushotzky}}, \binits{R.F.}},
\bauthor{\bsnm{{Norris}}, \binits{J.P.}},
\bauthor{\bsnm{{Osborne}}, \binits{J.}},
\bauthor{\bsnm{{Paczynski}}, \binits{B.}},
\bauthor{\bsnm{{Palmer}}, \binits{D.M.}},
\bauthor{\bsnm{{Park}}, \binits{H.-S.}},
\bauthor{\bsnm{{Parsons}}, \binits{A.M.}},
\bauthor{\bsnm{{Paul}}, \binits{J.}},
\bauthor{\bsnm{{Rees}}, \binits{M.J.}},
\bauthor{\bsnm{{Reynolds}}, \binits{C.S.}},
\bauthor{\bsnm{{Rhoads}}, \binits{J.E.}},
\bauthor{\bsnm{{Sasseen}}, \binits{T.P.}},
\bauthor{\bsnm{{Schaefer}}, \binits{B.E.}},
\bauthor{\bsnm{{Short}}, \binits{A.T.}},
\bauthor{\bsnm{{Smale}}, \binits{A.P.}},
\bauthor{\bsnm{{Smith}}, \binits{I.A.}},
\bauthor{\bsnm{{Stella}}, \binits{L.}},
\bauthor{\bsnm{{Tagliaferri}}, \binits{G.}},
\bauthor{\bsnm{{Takahashi}}, \binits{T.}},
\bauthor{\bsnm{{Tashiro}}, \binits{M.}},
\bauthor{\bsnm{{Townsley}}, \binits{L.K.}},
\bauthor{\bsnm{{Tueller}}, \binits{J.}},
\bauthor{\bsnm{{Turner}}, \binits{M.J.L.}},
\bauthor{\bsnm{{Vietri}}, \binits{M.}},
\bauthor{\bsnm{{Voges}}, \binits{W.}},
\bauthor{\bsnm{{Ward}}, \binits{M.J.}},
\bauthor{\bsnm{{Willingale}}, \binits{R.}},
\bauthor{\bsnm{{Zerbi}}, \binits{F.M.}},
\bauthor{\bsnm{{Zhang}}, \binits{W.W.}}:
\bjtitle{\apj}
\bvolume{611},
\bfpage{1005}
(\byear{2004})
\end{barticle}
\endbibitem

\bibitem[\protect\citeauthoryear{{Goldstein} et~al.}{2013}]{gold13}
\begin{barticle}
\bauthor{\bsnm{{Goldstein}}, \binits{A.}},
\bauthor{\bsnm{{Preece}}, \binits{R.D.}},
\bauthor{\bsnm{{Mallozzi}}, \binits{R.S.}},
\bauthor{\bsnm{{Briggs}}, \binits{M.S.}},
\bauthor{\bsnm{{Fishman}}, \binits{G.J.}},
\bauthor{\bsnm{{Kouveliotou}}, \binits{C.}},
\bauthor{\bsnm{{Paciesas}}, \binits{W.S.}},
\bauthor{\bsnm{{Burgess}}, \binits{J.M.}}:
\bjtitle{\apjs}
\bvolume{208},
\bfpage{21}
(\byear{2013})
\end{barticle}
\endbibitem

\bibitem[\protect\citeauthoryear{{Golenetskii} et~al.}{1983}]{gol83}
\begin{barticle}
\bauthor{\bsnm{{Golenetskii}}, \binits{S.V.}},
\bauthor{\bsnm{{Mazets}}, \binits{E.P.}},
\bauthor{\bsnm{{Aptekar}}, \binits{R.L.}},
\bauthor{\bsnm{{Ilinskii}}, \binits{V.N.}}:
\bjtitle{\nat}
\bvolume{306},
\bfpage{451}
(\byear{1983})
\end{barticle}
\endbibitem

\bibitem[\protect\citeauthoryear{{Hajdas} et~al.}{2004}]{haj04}
\begin{bchapter}
\bauthor{\bsnm{{Hajdas}}, \binits{W.}},
\bauthor{\bsnm{{Wigger}}, \binits{C.}},
\bauthor{\bsnm{{Arzner}}, \binits{K.}},
\bauthor{\bsnm{{Eggel}}, \binits{C.}},
\bauthor{\bsnm{{Guedel}}, \binits{M.}},
\bauthor{\bsnm{{Zehnder}}, \binits{A.}},
\bauthor{\bsnm{{Smith}}, \binits{D.}}:
\bctitle{{RHESSI Satellite as Efficient Gamma Ray Burst Detector}}.
\bsertitle{ESA Special Publication},
vol. \bseriesno{552},
p. \bfpage{805}
(\byear{2004})
\end{bchapter}
\endbibitem

\bibitem[\protect\citeauthoryear{{Hakkila} et~al.}{2000}]{hak00}
\begin{barticle}
\bauthor{\bsnm{{Hakkila}}, \binits{J.}},
\bauthor{\bsnm{{Haglin}}, \binits{D.J.}},
\bauthor{\bsnm{{Pendleton}}, \binits{G.N.}},
\bauthor{\bsnm{{Mallozzi}}, \binits{R.S.}},
\bauthor{\bsnm{{Meegan}}, \binits{C.A.}},
\bauthor{\bsnm{{Roiger}}, \binits{R.J.}}:
\bjtitle{\apj}
\bvolume{538},
\bfpage{165}
(\byear{2000})
\end{barticle}
\endbibitem

\bibitem[\protect\citeauthoryear{{Hakkila} et~al.}{2003}]{hak03}
\begin{barticle}
\bauthor{\bsnm{{Hakkila}}, \binits{J.}},
\bauthor{\bsnm{{Giblin}}, \binits{T.W.}},
\bauthor{\bsnm{{Roiger}}, \binits{R.J.}},
\bauthor{\bsnm{{Haglin}}, \binits{D.J.}},
\bauthor{\bsnm{{Paciesas}}, \binits{W.S.}},
\bauthor{\bsnm{{Meegan}}, \binits{C.A.}}:
\bjtitle{\apj}
\bvolume{582},
\bfpage{320}
(\byear{2003})
\end{barticle}
\endbibitem

\bibitem[\protect\citeauthoryear{{Heise}}{2003}]{hei03}
\begin{bchapter}
\bauthor{\bsnm{{Heise}}, \binits{J.}}:
In: \beditor{\bsnm{{Ricker}}, \binits{G.R.}},
\beditor{\bsnm{{Vanderspek}}, \binits{R.K.}} (eds.)
\bbtitle{Gamma-Ray Burst and Afterglow Astronomy 2001: A Workshop Celebrating
  the First Year of the HETE Mission}.
\bsertitle{AIP Conf. Proc.},
vol. \bseriesno{662},
p. \bfpage{229}
(\byear{2003})
\end{bchapter}
\endbibitem

\bibitem[\protect\citeauthoryear{{Heise} et~al.}{2001}]{hei01}
\begin{bchapter}
\bauthor{\bsnm{{Heise}}, \binits{J.}},
\bauthor{\bsnm{{Zand}}, \binits{J.I.}},
\bauthor{\bsnm{{Kippen}}, \binits{R.M.}},
\bauthor{\bsnm{{Woods}}, \binits{P.M.}}:
In: \beditor{\bsnm{{Costa}}, \binits{E.}},
\beditor{\bsnm{{Frontera}}, \binits{F.}},
\beditor{\bsnm{{Hjorth}}, \binits{J.}} (eds.)
\bbtitle{Gamma-ray Bursts in the Afterglow Era: Proceedings of the
  International Workshop Held in Rome, Italy, 17-20 October 2000},
p. \bfpage{16}
(\byear{2001})
\end{bchapter}
\endbibitem

\bibitem[\protect\citeauthoryear{{Horv{\'a}th}}{1998}]{ho98}
\begin{barticle}
\bauthor{\bsnm{{Horv{\'a}th}}, \binits{I.}}:
\bjtitle{\apj}
\bvolume{508},
\bfpage{757}
(\byear{1998})
\end{barticle}
\endbibitem

\bibitem[\protect\citeauthoryear{{Horv{\'a}th}}{2002}]{ho02}
\begin{barticle}
\bauthor{\bsnm{{Horv{\'a}th}}, \binits{I.}}:
\bjtitle{\aap}
\bvolume{392},
\bfpage{791}
(\byear{2002})
\end{barticle}
\endbibitem

\bibitem[\protect\citeauthoryear{{Horv{\'a}th}}{2009}]{hoi09}
\begin{barticle}
\bauthor{\bsnm{{Horv{\'a}th}}, \binits{I.}}:
\bjtitle{\apss}
\bvolume{323},
\bfpage{83}
(\byear{2009})
\end{barticle}
\endbibitem

\bibitem[\protect\citeauthoryear{{Horv{\'a}th} et~al.}{2006}]{ho06}
\begin{barticle}
\bauthor{\bsnm{{Horv{\'a}th}}, \binits{I.}},
\bauthor{\bsnm{{Bal{\'a}zs}}, \binits{L.G.}},
\bauthor{\bsnm{{Bagoly}}, \binits{Z.}},
\bauthor{\bsnm{{Ryde}}, \binits{F.}},
\bauthor{\bsnm{{M{\'e}sz{\'a}ros}}, \binits{A.}}:
\bjtitle{\aap}
\bvolume{447},
\bfpage{23}
(\byear{2006})
\end{barticle}
\endbibitem

\bibitem[\protect\citeauthoryear{{Horv{\'a}th} et~al.}{2008}]{ho08}
\begin{barticle}
\bauthor{\bsnm{{Horv{\'a}th}}, \binits{I.}},
\bauthor{\bsnm{{Bal{\'a}zs}}, \binits{L.G.}},
\bauthor{\bsnm{{Bagoly}}, \binits{Z.}},
\bauthor{\bsnm{{Veres}}, \binits{P.}}:
\bjtitle{\aap}
\bvolume{489},
\bfpage{L1}
(\byear{2008})
\end{barticle}
\endbibitem

\bibitem[\protect\citeauthoryear{{Horv\'ath} et~al.}{(2009)}]{ho09}
\begin{bchapter}
\bauthor{\bsnm{{Horv\'ath}}, \binits{I.}},
\bauthor{\bsnm{{Bagoly}}, \binits{Z.}},
\bauthor{\bsnm{{Bal\'azs}}, \binits{L.G.}},
\bauthor{\bsnm{{Tusn\'ady}}, \binits{G.}},
\bauthor{\bsnm{{Veres}}, \binits{P.}}:
In: \bbtitle{Proceedings of 2009 Fermi Symposium, eConf Proc. C0911022},
\byear{(2009)}.
\arxivurl{astro-ph/0912.3724}
\end{bchapter}
\endbibitem

\bibitem[\protect\citeauthoryear{{Horv{\'a}th} et~al.}{2010}]{ho10}
\begin{barticle}
\bauthor{\bsnm{{Horv{\'a}th}}, \binits{I.}},
\bauthor{\bsnm{{Bagoly}}, \binits{Z.}},
\bauthor{\bsnm{{Bal{\'a}zs}}, \binits{L.G.}},
\bauthor{\bsnm{{de Ugarte Postigo}}, \binits{A.}},
\bauthor{\bsnm{{Veres}}, \binits{P.}},
\bauthor{\bsnm{{M{\'e}sz{\'a}ros}}, \binits{A.}}:
\bjtitle{\apj}
\bvolume{713},
\bfpage{552}
(\byear{2010})
\end{barticle}
\endbibitem

\bibitem[\protect\citeauthoryear{{Huja} et~al.}{2009}]{huj09}
\begin{barticle}
\bauthor{\bsnm{{Huja}}, \binits{D.}},
\bauthor{\bsnm{{M{\'e}sz{\'a}ros}}, \binits{A.}},
\bauthor{\bsnm{{{\v R}{\'{\i}}pa}}, \binits{J.}}:
\bjtitle{\aap}
\bvolume{504},
\bfpage{67}
(\byear{2009})
\end{barticle}
\endbibitem

\bibitem[\protect\citeauthoryear{{Hurley} et~al.}{2005}]{hur05}
\begin{barticle}
\bauthor{\bsnm{{Hurley}}, \binits{K.}},
\bauthor{\bsnm{{Boggs}}, \binits{S.E.}},
\bauthor{\bsnm{{Smith}}, \binits{D.M.}},
\bauthor{\bsnm{{Duncan}}, \binits{R.C.}},
\bauthor{\bsnm{{Lin}}, \binits{R.}},
\bauthor{\bsnm{{Zoglauer}}, \binits{A.}},
\bauthor{\bsnm{{Krucker}}, \binits{S.}},
\bauthor{\bsnm{{Hurford}}, \binits{G.}},
\bauthor{\bsnm{{Hudson}}, \binits{H.}},
\bauthor{\bsnm{{Wigger}}, \binits{C.}},
\bauthor{\bsnm{{Hajdas}}, \binits{W.}},
\bauthor{\bsnm{{Thompson}}, \binits{C.}},
\bauthor{\bsnm{{Mitrofanov}}, \binits{I.}},
\bauthor{\bsnm{{Sanin}}, \binits{A.}},
\bauthor{\bsnm{{Boynton}}, \binits{W.}},
\bauthor{\bsnm{{Fellows}}, \binits{C.}},
\bauthor{\bsnm{{von Kienlin}}, \binits{A.}},
\bauthor{\bsnm{{Lichti}}, \binits{G.}},
\bauthor{\bsnm{{Rau}}, \binits{A.}},
\bauthor{\bsnm{{Cline}}, \binits{T.}}:
\bjtitle{\nat}
\bvolume{434},
\bfpage{1098}
(\byear{2005})
\end{barticle}
\endbibitem

\bibitem[\protect\citeauthoryear{{Kaneko} et~al.}{2006}]{kane06}
\begin{barticle}
\bauthor{\bsnm{{Kaneko}}, \binits{Y.}},
\bauthor{\bsnm{{Preece}}, \binits{R.D.}},
\bauthor{\bsnm{{Briggs}}, \binits{M.S.}},
\bauthor{\bsnm{{Paciesas}}, \binits{W.S.}},
\bauthor{\bsnm{{Meegan}}, \binits{C.A.}},
\bauthor{\bsnm{{Band}}, \binits{D.L.}}:
\bjtitle{\apjs}
\bvolume{166},
\bfpage{298}
(\byear{2006})
\end{barticle}
\endbibitem

\bibitem[\protect\citeauthoryear{{Kann} et~al.}{2010}]{kan10}
\begin{barticle}
\bauthor{\bsnm{{Kann}}, \binits{D.A.}},
\bauthor{\bsnm{{Klose}}, \binits{S.}},
\bauthor{\bsnm{{Zhang}}, \binits{B.}},
\bauthor{\bsnm{{Malesani}}, \binits{D.}},
\bauthor{\bsnm{{Nakar}}, \binits{E.}},
\bauthor{\bsnm{{Pozanenko}}, \binits{A.}},
\bauthor{\bsnm{{Wilson}}, \binits{A.C.}},
\bauthor{\bsnm{{Butler}}, \binits{N.R.}},
\bauthor{\bsnm{{Jakobsson}}, \binits{P.}},
\bauthor{\bsnm{{Schulze}}, \binits{S.}},
\bauthor{\bsnm{{Andreev}}, \binits{M.}},
\bauthor{\bsnm{{Antonelli}}, \binits{L.A.}},
\bauthor{\bsnm{{Bikmaev}}, \binits{I.F.}},
\bauthor{\bsnm{{Biryukov}}, \binits{V.}},
\bauthor{\bsnm{{B{\"o}ttcher}}, \binits{M.}},
\bauthor{\bsnm{{Burenin}}, \binits{R.A.}},
\bauthor{\bsnm{{Castro Cer{\'o}n}}, \binits{J.M.}},
\bauthor{\bsnm{{Castro-Tirado}}, \binits{A.J.}},
\bauthor{\bsnm{{Chincarini}}, \binits{G.}},
\bauthor{\bsnm{{Cobb}}, \binits{B.E.}},
\bauthor{\bsnm{{Covino}}, \binits{S.}},
\bauthor{\bsnm{{D'Avanzo}}, \binits{P.}},
\bauthor{\bsnm{{D'Elia}}, \binits{V.}},
\bauthor{\bsnm{{Della Valle}}, \binits{M.}},
\bauthor{\bsnm{{de Ugarte Postigo}}, \binits{A.}},
\bauthor{\bsnm{{Efimov}}, \binits{Y.}},
\bauthor{\bsnm{{Ferrero}}, \binits{P.}},
\bauthor{\bsnm{{Fugazza}}, \binits{D.}},
\bauthor{\bsnm{{Fynbo}}, \binits{J.P.U.}},
\bauthor{\bsnm{{G{\aa}lfalk}}, \binits{M.}},
\bauthor{\bsnm{{Grundahl}}, \binits{F.}},
\bauthor{\bsnm{{Gorosabel}}, \binits{J.}},
\bauthor{\bsnm{{Gupta}}, \binits{S.}},
\bauthor{\bsnm{{Guziy}}, \binits{S.}},
\bauthor{\bsnm{{Hafizov}}, \binits{B.}},
\bauthor{\bsnm{{Hjorth}}, \binits{J.}},
\bauthor{\bsnm{{Holhjem}}, \binits{K.}},
\bauthor{\bsnm{{Ibrahimov}}, \binits{M.}},
\bauthor{\bsnm{{Im}}, \binits{M.}},
\bauthor{\bsnm{{Israel}}, \binits{G.L.}},
\bauthor{\bsnm{{Jel{\'i}nek}}, \binits{M.}},
\bauthor{\bsnm{{Jensen}}, \binits{B.L.}},
\bauthor{\bsnm{{Karimov}}, \binits{R.}},
\bauthor{\bsnm{{Khamitov}}, \binits{I.M.}},
\bauthor{\bsnm{{Kizilo{\v g}lu}}, \binits{{\"U}.}},
\bauthor{\bsnm{{Klunko}}, \binits{E.}},
\bauthor{\bsnm{{Kub{\'a}nek}}, \binits{P.}},
\bauthor{\bsnm{{Kutyrev}}, \binits{A.S.}},
\bauthor{\bsnm{{Laursen}}, \binits{P.}},
\bauthor{\bsnm{{Levan}}, \binits{A.J.}},
\bauthor{\bsnm{{Mannucci}}, \binits{F.}},
\bauthor{\bsnm{{Martin}}, \binits{C.M.}},
\bauthor{\bsnm{{Mescheryakov}}, \binits{A.}},
\bauthor{\bsnm{{Mirabal}}, \binits{N.}},
\bauthor{\bsnm{{Norris}}, \binits{J.P.}},
\bauthor{\bsnm{{Ovaldsen}}, \binits{J.-E.}},
\bauthor{\bsnm{{Paraficz}}, \binits{D.}},
\bauthor{\bsnm{{Pavlenko}}, \binits{E.}},
\bauthor{\bsnm{{Piranomonte}}, \binits{S.}},
\bauthor{\bsnm{{Rossi}}, \binits{A.}},
\bauthor{\bsnm{{Rumyantsev}}, \binits{V.}},
\bauthor{\bsnm{{Salinas}}, \binits{R.}},
\bauthor{\bsnm{{Sergeev}}, \binits{A.}},
\bauthor{\bsnm{{Sharapov}}, \binits{D.}},
\bauthor{\bsnm{{Sollerman}}, \binits{J.}},
\bauthor{\bsnm{{Stecklum}}, \binits{B.}},
\bauthor{\bsnm{{Stella}}, \binits{L.}},
\bauthor{\bsnm{{Tagliaferri}}, \binits{G.}},
\bauthor{\bsnm{{Tanvir}}, \binits{N.R.}},
\bauthor{\bsnm{{Telting}}, \binits{J.}},
\bauthor{\bsnm{{Testa}}, \binits{V.}},
\bauthor{\bsnm{{Updike}}, \binits{A.C.}},
\bauthor{\bsnm{{Volnova}}, \binits{A.}},
\bauthor{\bsnm{{Watson}}, \binits{D.}},
\bauthor{\bsnm{{Wiersema}}, \binits{K.}},
\bauthor{\bsnm{{Xu}}, \binits{D.}}:
\bjtitle{\apj}
\bvolume{720},
\bfpage{1513}
(\byear{2010})
\end{barticle}
\endbibitem

\bibitem[\protect\citeauthoryear{{Kann} et~al.}{2011}]{kan11}
\begin{barticle}
\bauthor{\bsnm{{Kann}}, \binits{D.A.}},
\bauthor{\bsnm{{Klose}}, \binits{S.}},
\bauthor{\bsnm{{Zhang}}, \binits{B.}},
\bauthor{\bsnm{{Covino}}, \binits{S.}},
\bauthor{\bsnm{{Butler}}, \binits{N.R.}},
\bauthor{\bsnm{{Malesani}}, \binits{D.}},
\bauthor{\bsnm{{Nakar}}, \binits{E.}},
\bauthor{\bsnm{{Wilson}}, \binits{A.C.}},
\bauthor{\bsnm{{Antonelli}}, \binits{L.A.}},
\bauthor{\bsnm{{Chincarini}}, \binits{G.}},
\bauthor{\bsnm{{Cobb}}, \binits{B.E.}},
\bauthor{\bsnm{{D'Avanzo}}, \binits{P.}},
\bauthor{\bsnm{{D'Elia}}, \binits{V.}},
\bauthor{\bsnm{{Della Valle}}, \binits{M.}},
\bauthor{\bsnm{{Ferrero}}, \binits{P.}},
\bauthor{\bsnm{{Fugazza}}, \binits{D.}},
\bauthor{\bsnm{{Gorosabel}}, \binits{J.}},
\bauthor{\bsnm{{Israel}}, \binits{G.L.}},
\bauthor{\bsnm{{Mannucci}}, \binits{F.}},
\bauthor{\bsnm{{Piranomonte}}, \binits{S.}},
\bauthor{\bsnm{{Schulze}}, \binits{S.}},
\bauthor{\bsnm{{Stella}}, \binits{L.}},
\bauthor{\bsnm{{Tagliaferri}}, \binits{G.}},
\bauthor{\bsnm{{Wiersema}}, \binits{K.}}:
\bjtitle{\apj}
\bvolume{734},
\bfpage{96}
(\byear{2011})
\end{barticle}
\endbibitem

\bibitem[\protect\citeauthoryear{{Kippen} et~al.}{2003}]{kip03}
\begin{bchapter}
\bauthor{\bsnm{{Kippen}}, \binits{R.M.}},
\bauthor{\bsnm{{Woods}}, \binits{P.M.}},
\bauthor{\bsnm{{Heise}}, \binits{J.}},
\bauthor{\bsnm{{in't Zand}}, \binits{J.J.M.}},
\bauthor{\bsnm{{Briggs}}, \binits{M.S.}},
\bauthor{\bsnm{{Preece}}, \binits{R.D.}}:
In: \beditor{\bsnm{{Ricker}}, \binits{G.R.}},
\beditor{\bsnm{{Vanderspek}}, \binits{R.K.}} (eds.)
\bbtitle{Gamma-Ray Burst and Afterglow Astronomy 2001: A Workshop Celebrating
  the First Year of the HETE Mission}.
\bsertitle{AIP Conf. Proc.},
vol. \bseriesno{662},
p. \bfpage{244}
(\byear{2003})
\end{bchapter}
\endbibitem

\bibitem[\protect\citeauthoryear{{K{\'o}bori} et~al.}{2013}]{kob13}
\begin{barticle}
\bauthor{\bsnm{{K{\'o}bori}}, \binits{J.}},
\bauthor{\bsnm{{Bagoly}}, \binits{Z.}},
\bauthor{\bsnm{{Bal{\'a}zs}}, \binits{L.G.}},
\bauthor{\bsnm{{Horv{\'a}th}}, \binits{I.}}:
\bjtitle{Astronomische Nachrichten}
\bvolume{334},
\bfpage{1028}
(\byear{2013})
\end{barticle}
\endbibitem

\bibitem[\protect\citeauthoryear{{Kocevski} and {Petrosian}}{2013}]{koc13}
\begin{barticle}
\bauthor{\bsnm{{Kocevski}}, \binits{D.}},
\bauthor{\bsnm{{Petrosian}}, \binits{V.}}:
\bjtitle{\apj}
\bvolume{765},
\bfpage{116}
(\byear{2013})
\end{barticle}
\endbibitem

\bibitem[\protect\citeauthoryear{{Kocevski} et~al.}{2003}]{koc03}
\begin{barticle}
\bauthor{\bsnm{{Kocevski}}, \binits{D.}},
\bauthor{\bsnm{{Ryde}}, \binits{F.}},
\bauthor{\bsnm{{Liang}}, \binits{E.}}:
\bjtitle{\apj}
\bvolume{596},
\bfpage{389}
(\byear{2003})
\end{barticle}
\endbibitem

\bibitem[\protect\citeauthoryear{{Koen} and {Bere}}{2012}]{koen12}
\begin{barticle}
\bauthor{\bsnm{{Koen}}, \binits{C.}},
\bauthor{\bsnm{{Bere}}, \binits{A.}}:
\bjtitle{\mnras}
\bvolume{420},
\bfpage{405}
(\byear{2012})
\end{barticle}
\endbibitem

\bibitem[\protect\citeauthoryear{{Kouveliotou} et~al.}{2012}]{grb12}
\begin{bbook}
\bauthor{\bsnm{{Kouveliotou}}, \binits{C.}},
\bauthor{\bsnm{{Wijers}}, \binits{R.A.M.J.}},
\bauthor{\bsnm{{Woosley}}, \binits{S.}}:
\bbtitle{{Gamma-ray Bursts}}.
\bpublisher{Cambridge University Press},
\blocation{New York}
(\byear{2012})
\end{bbook}
\endbibitem

\bibitem[\protect\citeauthoryear{{Kouveliotou} et~al.}{1993}]{kou93}
\begin{barticle}
\bauthor{\bsnm{{Kouveliotou}}, \binits{C.}},
\bauthor{\bsnm{{Meegan}}, \binits{C.A.}},
\bauthor{\bsnm{{Fishman}}, \binits{G.J.}},
\bauthor{\bsnm{{Bhat}}, \binits{N.P.}},
\bauthor{\bsnm{{Briggs}}, \binits{M.S.}},
\bauthor{\bsnm{{Koshut}}, \binits{T.M.}},
\bauthor{\bsnm{{Paciesas}}, \binits{W.S.}},
\bauthor{\bsnm{{Pendleton}}, \binits{G.N.}}:
\bjtitle{\apjl}
\bvolume{413},
\bfpage{101}
(\byear{1993})
\end{barticle}
\endbibitem

\bibitem[\protect\citeauthoryear{{Lamb} and {Graziani}}{2003}]{lamb03}
\begin{bchapter}
\bauthor{\bsnm{{Lamb}}, \binits{D.Q.}},
\bauthor{\bsnm{{Graziani}}, \binits{C.}}:
In: \bbtitle{American Astronomical Society Meeting Abstracts \#202}.
\bsertitle{\baas},
vol. \bseriesno{35},
p. \bfpage{764}
(\byear{2003})
\end{bchapter}
\endbibitem

\bibitem[\protect\citeauthoryear{{Lamb} et~al.}{2004}]{lamb04}
\begin{bchapter}
\bauthor{\bsnm{{Lamb}}, \binits{D.Q.}},
\bauthor{\bsnm{{Donaghy}}, \binits{T.Q.}},
\bauthor{\bsnm{{Graziani}}, \binits{C.}}:
In: \beditor{\bsnm{{Fenimore}}, \binits{E.}},
\beditor{\bsnm{{Galassi}}, \binits{M.}} (eds.)
\bbtitle{Gamma-Ray Bursts: 30 Years of Discovery}.
\bsertitle{AIP Conf. Proc.},
vol. \bseriesno{727},
p. \bfpage{19}
(\byear{2004})
\end{bchapter}
\endbibitem

\bibitem[\protect\citeauthoryear{{Lamb} et~al.}{2005}]{la05}
\begin{barticle}
\bauthor{\bsnm{{Lamb}}, \binits{D.Q.}},
\bauthor{\bsnm{{Donaghy}}, \binits{T.Q.}},
\bauthor{\bsnm{{Graziani}}, \binits{C.}}:
\bjtitle{\apj}
\bvolume{620},
\bfpage{355}
(\byear{2005})
\end{barticle}
\endbibitem

\bibitem[\protect\citeauthoryear{{Liang} and {Kargatis}}{1996}]{lia96}
\begin{barticle}
\bauthor{\bsnm{{Liang}}, \binits{E.}},
\bauthor{\bsnm{{Kargatis}}, \binits{V.}}:
\bjtitle{\nat}
\bvolume{381},
\bfpage{49}
(\byear{1996})
\end{barticle}
\endbibitem

\bibitem[\protect\citeauthoryear{{Lin} et~al.}{2002}]{lin02}
\begin{barticle}
\bauthor{\bsnm{{Lin}}, \binits{R.P.}},
\bauthor{\bsnm{{Dennis}}, \binits{B.R.}},
\bauthor{\bsnm{{Hurford}}, \binits{G.J.}},
\bauthor{\bsnm{{Smith}}, \binits{D.M.}},
\bauthor{\bsnm{{Zehnder}}, \binits{A.}},
\bauthor{\bsnm{{Harvey}}, \binits{P.R.}},
\bauthor{\bsnm{{Curtis}}, \binits{D.W.}},
\bauthor{\bsnm{{Pankow}}, \binits{D.}},
\bauthor{\bsnm{{Turin}}, \binits{P.}},
\bauthor{\bsnm{{Bester}}, \binits{M.}},
\bauthor{\bsnm{{Csillaghy}}, \binits{A.}},
\bauthor{\bsnm{{Lewis}}, \binits{M.}},
\bauthor{\bsnm{{Madden}}, \binits{N.}},
\bauthor{\bsnm{{van Beek}}, \binits{H.F.}},
\bauthor{\bsnm{{Appleby}}, \binits{M.}},
\bauthor{\bsnm{{Raudorf}}, \binits{T.}},
\bauthor{\bsnm{{McTiernan}}, \binits{J.}},
\bauthor{\bsnm{{Ramaty}}, \binits{R.}},
\bauthor{\bsnm{{Schmahl}}, \binits{E.}},
\bauthor{\bsnm{{Schwartz}}, \binits{R.}},
\bauthor{\bsnm{{Krucker}}, \binits{S.}},
\bauthor{\bsnm{{Abiad}}, \binits{R.}},
\bauthor{\bsnm{{Quinn}}, \binits{T.}},
\bauthor{\bsnm{{Berg}}, \binits{P.}},
\bauthor{\bsnm{{Hashii}}, \binits{M.}},
\bauthor{\bsnm{{Sterling}}, \binits{R.}},
\bauthor{\bsnm{{Jackson}}, \binits{R.}},
\bauthor{\bsnm{{Pratt}}, \binits{R.}},
\bauthor{\bsnm{{Campbell}}, \binits{R.D.}},
\bauthor{\bsnm{{Malone}}, \binits{D.}},
\bauthor{\bsnm{{Landis}}, \binits{D.}},
\bauthor{\bsnm{{Barrington-Leigh}}, \binits{C.P.}},
\bauthor{\bsnm{{Slassi-Sennou}}, \binits{S.}},
\bauthor{\bsnm{{Cork}}, \binits{C.}},
\bauthor{\bsnm{{Clark}}, \binits{D.}},
\bauthor{\bsnm{{Amato}}, \binits{D.}},
\bauthor{\bsnm{{Orwig}}, \binits{L.}},
\bauthor{\bsnm{{Boyle}}, \binits{R.}},
\bauthor{\bsnm{{Banks}}, \binits{I.S.}},
\bauthor{\bsnm{{Shirey}}, \binits{K.}},
\bauthor{\bsnm{{Tolbert}}, \binits{A.K.}},
\bauthor{\bsnm{{Zarro}}, \binits{D.}},
\bauthor{\bsnm{{Snow}}, \binits{F.}},
\bauthor{\bsnm{{Thomsen}}, \binits{K.}},
\bauthor{\bsnm{{Henneck}}, \binits{R.}},
\bauthor{\bsnm{{McHedlishvili}}, \binits{A.}},
\bauthor{\bsnm{{Ming}}, \binits{P.}},
\bauthor{\bsnm{{Fivian}}, \binits{M.}},
\bauthor{\bsnm{{Jordan}}, \binits{J.}},
\bauthor{\bsnm{{Wanner}}, \binits{R.}},
\bauthor{\bsnm{{Crubb}}, \binits{J.}},
\bauthor{\bsnm{{Preble}}, \binits{J.}},
\bauthor{\bsnm{{Matranga}}, \binits{M.}},
\bauthor{\bsnm{{Benz}}, \binits{A.}},
\bauthor{\bsnm{{Hudson}}, \binits{H.}},
\bauthor{\bsnm{{Canfield}}, \binits{R.C.}},
\bauthor{\bsnm{{Holman}}, \binits{G.D.}},
\bauthor{\bsnm{{Crannell}}, \binits{C.}},
\bauthor{\bsnm{{Kosugi}}, \binits{T.}},
\bauthor{\bsnm{{Emslie}}, \binits{A.G.}},
\bauthor{\bsnm{{Vilmer}}, \binits{N.}},
\bauthor{\bsnm{{Brown}}, \binits{J.C.}},
\bauthor{\bsnm{{Johns-Krull}}, \binits{C.}},
\bauthor{\bsnm{{Aschwanden}}, \binits{M.}},
\bauthor{\bsnm{{Metcalf}}, \binits{T.}},
\bauthor{\bsnm{{Conway}}, \binits{A.}}:
\bjtitle{\solphys}
\bvolume{210},
\bfpage{3}
(\byear{2002})
\end{barticle}
\endbibitem

\bibitem[\protect\citeauthoryear{{Lloyd} and {Petrosian}}{2000}]{llo00}
\begin{barticle}
\bauthor{\bsnm{{Lloyd}}, \binits{N.M.}},
\bauthor{\bsnm{{Petrosian}}, \binits{V.}}:
\bjtitle{\apj}
\bvolume{543},
\bfpage{722}
(\byear{2000})
\end{barticle}
\endbibitem

\bibitem[\protect\citeauthoryear{{Mazets} et~al.}{1981}]{maz81}
\begin{barticle}
\bauthor{\bsnm{{Mazets}}, \binits{E.P.}},
\bauthor{\bsnm{{Golenetskii}}, \binits{S.V.}},
\bauthor{\bsnm{{Ilinskii}}, \binits{V.N.}},
\bauthor{\bsnm{{Panov}}, \binits{V.N.}},
\bauthor{\bsnm{{Aptekar}}, \binits{R.L.}},
\bauthor{\bsnm{{Gurian}}, \binits{I.A.}},
\bauthor{\bsnm{{Proskura}}, \binits{M.P.}},
\bauthor{\bsnm{{Sokolov}}, \binits{I.A.}},
\bauthor{\bsnm{{Sokolova}}, \binits{Z.I.}},
\bauthor{\bsnm{{Kharitonova}}, \binits{T.V.}}:
\bjtitle{\apss}
\bvolume{80},
\bfpage{3}
(\byear{1981})
\end{barticle}
\endbibitem

\bibitem[\protect\citeauthoryear{{Mereghetti}}{2008}]{mer08}
\begin{barticle}
\bauthor{\bsnm{{Mereghetti}}, \binits{S.}}:
\bjtitle{\aapr}
\bvolume{15},
\bfpage{225}
(\byear{2008})
\end{barticle}
\endbibitem

\bibitem[\protect\citeauthoryear{{M{\'e}sz{\'a}ros} et~al.}{2006}]{me06b}
\begin{barticle}
\bauthor{\bsnm{{M{\'e}sz{\'a}ros}}, \binits{A.}},
\bauthor{\bsnm{{Bagoly}}, \binits{Z.}},
\bauthor{\bsnm{{Bal{\'a}zs}}, \binits{L.G.}},
\bauthor{\bsnm{{Horv{\'a}th}}, \binits{I.}}:
\bjtitle{\aap}
\bvolume{455},
\bfpage{785}
(\byear{2006})
\end{barticle}
\endbibitem

\bibitem[\protect\citeauthoryear{{M{\'e}sz{\'a}ros}}{2006}]{meszp06}
\begin{barticle}
\bauthor{\bsnm{{M{\'e}sz{\'a}ros}}, \binits{P.}}:
\bjtitle{Reports on Progress in Physics}
\bvolume{69},
\bfpage{2259}
(\byear{2006})
\end{barticle}
\endbibitem

\bibitem[\protect\citeauthoryear{{Mochkovitch} et~al.}{2004}]{mo04}
\begin{bchapter}
\bauthor{\bsnm{{Mochkovitch}}, \binits{R.}},
\bauthor{\bsnm{{Daigne}}, \binits{F.}},
\bauthor{\bsnm{{Barraud}}, \binits{C.}},
\bauthor{\bsnm{{Atteia}}, \binits{J.L.}}:
In: \beditor{\bsnm{{Feroci}}, \binits{M.}},
\beditor{\bsnm{{Frontera}}, \binits{F.}},
\beditor{\bsnm{{Masetti}}, \binits{N.}},
\beditor{\bsnm{{Piro}}, \binits{L.}} (eds.)
\bbtitle{Gamma-Ray Bursts in the Afterglow Era}.
\bsertitle{Astronomical Society of the Pacific Conference Series},
vol. \bseriesno{312},
p. \bfpage{381}
(\byear{2004})
\end{bchapter}
\endbibitem

\bibitem[\protect\citeauthoryear{{Mukherjee} et~al.}{1998}]{muk98}
\begin{barticle}
\bauthor{\bsnm{{Mukherjee}}, \binits{S.}},
\bauthor{\bsnm{{Feigelson}}, \binits{E.D.}},
\bauthor{\bsnm{{Jogesh Babu}}, \binits{G.}},
\bauthor{\bsnm{{Murtagh}}, \binits{F.}},
\bauthor{\bsnm{{Fraley}}, \binits{C.}},
\bauthor{\bsnm{{Raftery}}, \binits{A.}}:
\bjtitle{\apj}
\bvolume{508},
\bfpage{314}
(\byear{1998})
\end{barticle}
\endbibitem

\bibitem[\protect\citeauthoryear{{Narayana Bhat} et~al.}{2016}]{bha16}
\begin{barticle}
\bauthor{\bsnm{{Narayana Bhat}}, \binits{P.}},
\bauthor{\bsnm{{Meegan}}, \binits{C.A.}},
\bauthor{\bsnm{{von Kienlin}}, \binits{A.}},
\bauthor{\bsnm{{Paciesas}}, \binits{W.S.}},
\bauthor{\bsnm{{Briggs}}, \binits{M.S.}},
\bauthor{\bsnm{{Burgess}}, \binits{J.M.}},
\bauthor{\bsnm{{Burns}}, \binits{E.}},
\bauthor{\bsnm{{Chaplin}}, \binits{V.}},
\bauthor{\bsnm{{Cleveland}}, \binits{W.H.}},
\bauthor{\bsnm{{Collazzi}}, \binits{A.C.}},
\bauthor{\bsnm{{Connaughton}}, \binits{V.}},
\bauthor{\bsnm{{Diekmann}}, \binits{A.M.}},
\bauthor{\bsnm{{Fitzpatrick}}, \binits{G.}},
\bauthor{\bsnm{{Gibby}}, \binits{M.H.}},
\bauthor{\bsnm{{Giles}}, \binits{M.M.}},
\bauthor{\bsnm{{Goldstein}}, \binits{A.M.}},
\bauthor{\bsnm{{Greiner}}, \binits{J.}},
\bauthor{\bsnm{{Jenke}}, \binits{P.A.}},
\bauthor{\bsnm{{Kippen}}, \binits{R.M.}},
\bauthor{\bsnm{{Kouveliotou}}, \binits{C.}},
\bauthor{\bsnm{{Mailyan}}, \binits{B.}},
\bauthor{\bsnm{{McBreen}}, \binits{S.}},
\bauthor{\bsnm{{Pelassa}}, \binits{V.}},
\bauthor{\bsnm{{Preece}}, \binits{R.D.}},
\bauthor{\bsnm{{Roberts}}, \binits{O.J.}},
\bauthor{\bsnm{{Sparke}}, \binits{L.S.}},
\bauthor{\bsnm{{Stanbro}}, \binits{M.}},
\bauthor{\bsnm{{Veres}}, \binits{P.}},
\bauthor{\bsnm{{Wilson-Hodge}}, \binits{C.A.}},
\bauthor{\bsnm{{Xiong}}, \binits{S.}},
\bauthor{\bsnm{{Younes}}, \binits{G.}},
\bauthor{\bsnm{{Yu}}, \binits{H.-F.}},
\bauthor{\bsnm{{Zhang}}, \binits{B.}}:
\bjtitle{\apjs}
\bvolume{223},
\bfpage{28}
(\byear{2016})
\end{barticle}
\endbibitem

\bibitem[\protect\citeauthoryear{{Norris} et~al.}{2001}]{nor01}
\begin{bchapter}
\bauthor{\bsnm{{Norris}}, \binits{J.P.}},
\bauthor{\bsnm{{Scargle}}, \binits{J.D.}},
\bauthor{\bsnm{{Bonnell}}, \binits{J.T.}}:
In: \beditor{\bsnm{{Costa}}, \binits{E.}},
\beditor{\bsnm{{Frontera}}, \binits{F.}},
\beditor{\bsnm{{Hjorth}}, \binits{J.}} (eds.)
\bbtitle{Gamma-Ray Bursts in the Afterglow Era: Proceedings of the
  International Workshop Held in Rome, Italy, 17-20 October 2000},
p. \bfpage{40}
(\byear{2001})
\end{bchapter}
\endbibitem

\bibitem[\protect\citeauthoryear{{Pendleton} et~al.}{1999}]{pen99}
\begin{barticle}
\bauthor{\bsnm{{Pendleton}}, \binits{G.N.}},
\bauthor{\bsnm{{Briggs}}, \binits{M.S.}},
\bauthor{\bsnm{{Kippen}}, \binits{R.M.}},
\bauthor{\bsnm{{Paciesas}}, \binits{W.S.}},
\bauthor{\bsnm{{Stollberg}}, \binits{M.}},
\bauthor{\bsnm{{Woods}}, \binits{P.}},
\bauthor{\bsnm{{Meegan}}, \binits{C.A.}},
\bauthor{\bsnm{{Fishman}}, \binits{G.J.}},
\bauthor{\bsnm{{McCollough}}, \binits{M.L.}},
\bauthor{\bsnm{{Connaughton}}, \binits{V.}}:
\bjtitle{\apj}
\bvolume{512},
\bfpage{362}
(\byear{1999})
\end{barticle}
\endbibitem

\bibitem[\protect\citeauthoryear{{Preece} et~al.}{2000}]{pre00}
\begin{barticle}
\bauthor{\bsnm{{Preece}}, \binits{R.D.}},
\bauthor{\bsnm{{Briggs}}, \binits{M.S.}},
\bauthor{\bsnm{{Mallozzi}}, \binits{R.S.}},
\bauthor{\bsnm{{Pendleton}}, \binits{G.N.}},
\bauthor{\bsnm{{Paciesas}}, \binits{W.S.}},
\bauthor{\bsnm{{Band}}, \binits{D.L.}}:
\bjtitle{\apjs}
\bvolume{126},
\bfpage{19}
(\byear{2000})
\end{barticle}
\endbibitem

\bibitem[\protect\citeauthoryear{{Press} et~al.}{2007}]{press07}
\begin{bbook}
\bauthor{\bsnm{{Press}}, \binits{W.H.}},
\bauthor{\bsnm{{Teukolsky}}, \binits{S.A.}},
\bauthor{\bsnm{{Vetterling}}, \binits{W.T.}},
\bauthor{\bsnm{{Flannery}}, \binits{B.P.}}:
\bbtitle{Numerical Recipes: The Art of Scientific Computing},
\bedition{3}rd edn.
\bpublisher{Cambridge University Press},
\blocation{Cambridge}
(\byear{2007})
\end{bbook}
\endbibitem

\bibitem[\protect\citeauthoryear{{Qin} et~al.}{2013}]{qin13}
\begin{barticle}
\bauthor{\bsnm{{Qin}}, \binits{Y.}},
\bauthor{\bsnm{{Liang}}, \binits{E.-W.}},
\bauthor{\bsnm{{Liang}}, \binits{Y.-F.}},
\bauthor{\bsnm{{Yi}}, \binits{S.-X.}},
\bauthor{\bsnm{{Lin}}, \binits{L.}},
\bauthor{\bsnm{{Zhang}}, \binits{B.-B.}},
\bauthor{\bsnm{{Zhang}}, \binits{J.}},
\bauthor{\bsnm{{L{\"u}}}, \binits{H.-J.}},
\bauthor{\bsnm{{Lu}}, \binits{R.-J.}},
\bauthor{\bsnm{{L{\"u}}}, \binits{L.-Z.}},
\bauthor{\bsnm{{Zhang}}, \binits{B.}}:
\bjtitle{\apj}
\bvolume{763},
\bfpage{15}
(\byear{2013})
\end{barticle}
\endbibitem

\bibitem[\protect\citeauthoryear{{Rajaniemi} and
  {M{\"a}h{\"o}nen}}{2002}]{raj02}
\begin{barticle}
\bauthor{\bsnm{{Rajaniemi}}, \binits{H.J.}},
\bauthor{\bsnm{{M{\"a}h{\"o}nen}}, \binits{P.}}:
\bjtitle{\apj}
\bvolume{566},
\bfpage{202}
(\byear{2002})
\end{barticle}
\endbibitem

\bibitem[\protect\citeauthoryear{{Ramirez-Ruiz} and
  {Lloyd-Ronning}}{2002}]{rr02}
\begin{barticle}
\bauthor{\bsnm{{Ramirez-Ruiz}}, \binits{E.}},
\bauthor{\bsnm{{Lloyd-Ronning}}, \binits{N.M.}}:
\bjtitle{\na}
\bvolume{7},
\bfpage{197}
(\byear{2002})
\end{barticle}
\endbibitem

\bibitem[\protect\citeauthoryear{{Richardson} et~al.}{1996}]{rich96}
\begin{bchapter}
\bauthor{\bsnm{{Richardson}}, \binits{G.}},
\bauthor{\bsnm{{Koshut}}, \binits{T.}},
\bauthor{\bsnm{{Paciesas}}, \binits{W.}},
\bauthor{\bsnm{{Kouveliotou}}, \binits{C.}}:
In: \beditor{\bsnm{{Kouveliotou}}, \binits{C.}},
\beditor{\bsnm{{Briggs}}, \binits{M.F.}},
\beditor{\bsnm{{Fishman}}, \binits{G.J.}} (eds.)
\bbtitle{Gamma-ray bursts: 3rd Huntsville symposium}.
\bsertitle{AIP Conf. Proc.},
vol. \bseriesno{384},
p. \bfpage{87}
(\byear{1996})
\end{bchapter}
\endbibitem

\bibitem[\protect\citeauthoryear{{Ricker} et~al.}{2003}]{ric03}
\begin{bchapter}
\bauthor{\bsnm{{Ricker}}, \binits{G.R.}},
\bauthor{\bsnm{{Atteia}}, \binits{J.-L.}},
\bauthor{\bsnm{{Crew}}, \binits{G.B.}},
\bauthor{\bsnm{{Doty}}, \binits{J.P.}},
\bauthor{\bsnm{{Fenimore}}, \binits{E.E.}},
\bauthor{\bsnm{{Galassi}}, \binits{M.}},
\bauthor{\bsnm{{Graziani}}, \binits{C.}},
\bauthor{\bsnm{{Hurley}}, \binits{K.}},
\bauthor{\bsnm{{Jernigan}}, \binits{J.G.}},
\bauthor{\bsnm{{Kawai}}, \binits{N.}},
\bauthor{\bsnm{{Lamb}}, \binits{D.Q.}},
\bauthor{\bsnm{{Matsuoka}}, \binits{M.}},
\bauthor{\bsnm{{Pizzichini}}, \binits{G.}},
\bauthor{\bsnm{{Shirasaki}}, \binits{Y.}},
\bauthor{\bsnm{{Tamagawa}}, \binits{T.}},
\bauthor{\bsnm{{Vanderspek}}, \binits{R.}},
\bauthor{\bsnm{{Vedrenne}}, \binits{G.}},
\bauthor{\bsnm{{Villasenor}}, \binits{J.}},
\bauthor{\bsnm{{Woosley}}, \binits{S.E.}},
\bauthor{\bsnm{{Yoshida}}, \binits{A.}}:
In: \beditor{\bsnm{{Ricker}}, \binits{G.R.}},
\beditor{\bsnm{{Vanderspek}}, \binits{R.K.}} (eds.)
\bbtitle{Gamma-Ray Burst and Afterglow Astronomy 2001: A Workshop Celebrating
  the First Year of the HETE Mission}.
\bsertitle{AIP Conf. Proc.},
vol. \bseriesno{662},
p. \bfpage{3}
(\byear{2003})
\end{bchapter}
\endbibitem

\bibitem[\protect\citeauthoryear{{\v{R}\'{\i}pa} et~al.}{2009}]{rip09}
\begin{barticle}
\bauthor{\bsnm{{\v{R}\'{\i}pa}}, \binits{J.}},
\bauthor{\bsnm{{M{\'e}sz{\'a}ros}}, \binits{A.}},
\bauthor{\bsnm{{Wigger}}, \binits{C.}},
\bauthor{\bsnm{{Huja}}, \binits{D.}},
\bauthor{\bsnm{{Hudec}}, \binits{R.}},
\bauthor{\bsnm{{Hajdas}}, \binits{W.}}:
\bjtitle{\aap}
\bvolume{498},
\bfpage{399}
(\byear{2009})
\end{barticle}
\endbibitem

\bibitem[\protect\citeauthoryear{{\v{R}\'{\i}pa} et~al.}{2012}]{rip12}
\begin{barticle}
\bauthor{\bsnm{{\v{R}\'{\i}pa}}, \binits{J.}},
\bauthor{\bsnm{{M{\'e}sz{\'a}ros}}, \binits{A.}},
\bauthor{\bsnm{{Veres}}, \binits{P.}},
\bauthor{\bsnm{{Park}}, \binits{I.H.}}:
\bjtitle{\apj}
\bvolume{756},
\bfpage{44}
(\byear{2012})
\end{barticle}
\endbibitem

\bibitem[\protect\citeauthoryear{{Ryde} et~al.}{2005}]{ryd05}
\begin{barticle}
\bauthor{\bsnm{{Ryde}}, \binits{F.}},
\bauthor{\bsnm{{Kocevski}}, \binits{D.}},
\bauthor{\bsnm{{Bagoly}}, \binits{Z.}},
\bauthor{\bsnm{{Ryde}}, \binits{N.}},
\bauthor{\bsnm{{M{\'e}sz{\'a}ros}}, \binits{A.}}:
\bjtitle{\aap}
\bvolume{432},
\bfpage{105}
(\byear{2005})
\end{barticle}
\endbibitem

\bibitem[\protect\citeauthoryear{{Sakamoto} et~al.}{2005}]{sak05}
\begin{barticle}
\bauthor{\bsnm{{Sakamoto}}, \binits{T.}},
\bauthor{\bsnm{{Lamb}}, \binits{D.Q.}},
\bauthor{\bsnm{{Kawai}}, \binits{N.}},
\bauthor{\bsnm{{Yoshida}}, \binits{A.}},
\bauthor{\bsnm{{Graziani}}, \binits{C.}},
\bauthor{\bsnm{{Fenimore}}, \binits{E.E.}},
\bauthor{\bsnm{{Donaghy}}, \binits{T.Q.}},
\bauthor{\bsnm{{Matsuoka}}, \binits{M.}},
\bauthor{\bsnm{{Suzuki}}, \binits{M.}},
\bauthor{\bsnm{{Ricker}}, \binits{G.}},
\bauthor{\bsnm{{Atteia}}, \binits{J.-L.}},
\bauthor{\bsnm{{Shirasaki}}, \binits{Y.}},
\bauthor{\bsnm{{Tamagawa}}, \binits{T.}},
\bauthor{\bsnm{{Torii}}, \binits{K.}},
\bauthor{\bsnm{{Galassi}}, \binits{M.}},
\bauthor{\bsnm{{Doty}}, \binits{J.}},
\bauthor{\bsnm{{Vanderspek}}, \binits{R.}},
\bauthor{\bsnm{{Crew}}, \binits{G.B.}},
\bauthor{\bsnm{{Villasenor}}, \binits{J.}},
\bauthor{\bsnm{{Butler}}, \binits{N.}},
\bauthor{\bsnm{{Prigozhin}}, \binits{G.}},
\bauthor{\bsnm{{Jernigan}}, \binits{J.G.}},
\bauthor{\bsnm{{Barraud}}, \binits{C.}},
\bauthor{\bsnm{{Boer}}, \binits{M.}},
\bauthor{\bsnm{{Dezalay}}, \binits{J.-P.}},
\bauthor{\bsnm{{Olive}}, \binits{J.-F.}},
\bauthor{\bsnm{{Hurley}}, \binits{K.}},
\bauthor{\bsnm{{Levine}}, \binits{A.}},
\bauthor{\bsnm{{Monnelly}}, \binits{G.}},
\bauthor{\bsnm{{Martel}}, \binits{F.}},
\bauthor{\bsnm{{Morgan}}, \binits{E.}},
\bauthor{\bsnm{{Woosley}}, \binits{S.E.}},
\bauthor{\bsnm{{Cline}}, \binits{T.}},
\bauthor{\bsnm{{Braga}}, \binits{J.}},
\bauthor{\bsnm{{Manchanda}}, \binits{R.}},
\bauthor{\bsnm{{Pizzichini}}, \binits{G.}},
\bauthor{\bsnm{{Takagishi}}, \binits{K.}},
\bauthor{\bsnm{{Yamauchi}}, \binits{M.}}:
\bjtitle{\apj}
\bvolume{629},
\bfpage{311}
(\byear{2005})
\end{barticle}
\endbibitem

\bibitem[\protect\citeauthoryear{{Sakamoto} et~al.}{2008}]{sak08}
\begin{barticle}
\bauthor{\bsnm{{Sakamoto}}, \binits{T.}},
\bauthor{\bsnm{{Hullinger}}, \binits{D.}},
\bauthor{\bsnm{{Sato}}, \binits{G.}},
\bauthor{\bsnm{{Yamazaki}}, \binits{R.}},
\bauthor{\bsnm{{Barbier}}, \binits{L.}},
\bauthor{\bsnm{{Barthelmy}}, \binits{S.D.}},
\bauthor{\bsnm{{Cummings}}, \binits{J.R.}},
\bauthor{\bsnm{{Fenimore}}, \binits{E.E.}},
\bauthor{\bsnm{{Gehrels}}, \binits{N.}},
\bauthor{\bsnm{{Krimm}}, \binits{H.A.}},
\bauthor{\bsnm{{Lamb}}, \binits{D.Q.}},
\bauthor{\bsnm{{Markwardt}}, \binits{C.B.}},
\bauthor{\bsnm{{Osborne}}, \binits{J.P.}},
\bauthor{\bsnm{{Palmer}}, \binits{D.M.}},
\bauthor{\bsnm{{Parsons}}, \binits{A.M.}},
\bauthor{\bsnm{{Stamatikos}}, \binits{M.}},
\bauthor{\bsnm{{Tueller}}, \binits{J.}}:
\bjtitle{\apj}
\bvolume{679},
\bfpage{570}
(\byear{2008})
\end{barticle}
\endbibitem

\bibitem[\protect\citeauthoryear{{Smith} et~al.}{2002}]{smi02}
\begin{barticle}
\bauthor{\bsnm{{Smith}}, \binits{D.M.}},
\bauthor{\bsnm{{Lin}}, \binits{R.P.}},
\bauthor{\bsnm{{Turin}}, \binits{P.}},
\bauthor{\bsnm{{Curtis}}, \binits{D.W.}},
\bauthor{\bsnm{{Primbsch}}, \binits{J.H.}},
\bauthor{\bsnm{{Campbell}}, \binits{R.D.}},
\bauthor{\bsnm{{Abiad}}, \binits{R.}},
\bauthor{\bsnm{{Schroeder}}, \binits{P.}},
\bauthor{\bsnm{{Cork}}, \binits{C.P.}},
\bauthor{\bsnm{{Hull}}, \binits{E.L.}},
\bauthor{\bsnm{{Landis}}, \binits{D.A.}},
\bauthor{\bsnm{{Madden}}, \binits{N.W.}},
\bauthor{\bsnm{{Malone}}, \binits{D.}},
\bauthor{\bsnm{{Pehl}}, \binits{R.H.}},
\bauthor{\bsnm{{Raudorf}}, \binits{T.}},
\bauthor{\bsnm{{Sangsingkeow}}, \binits{P.}},
\bauthor{\bsnm{{Boyle}}, \binits{R.}},
\bauthor{\bsnm{{Banks}}, \binits{I.S.}},
\bauthor{\bsnm{{Shirey}}, \binits{K.}},
\bauthor{\bsnm{{Schwartz}}, \binits{R.}}:
\bjtitle{\solphys}
\bvolume{210},
\bfpage{33}
(\byear{2002})
\end{barticle}
\endbibitem

\bibitem[\protect\citeauthoryear{{Smith} et~al.}{2003}]{smi03}
\begin{bchapter}
\bauthor{\bsnm{{Smith}}, \binits{D.M.}},
\bauthor{\bsnm{{Lin}}, \binits{R.P.}},
\bauthor{\bsnm{{Hurley}}, \binits{K.C.}},
\bauthor{\bsnm{{Coburn}}, \binits{W.}},
\bauthor{\bsnm{{Hurford}}, \binits{G.J.}},
\bauthor{\bsnm{{Wigger}}, \binits{C.}},
\bauthor{\bsnm{{Hajdas}}, \binits{W.}},
\bauthor{\bsnm{{Zehnder}}, \binits{A.}},
\bauthor{\bsnm{{McConnell}}, \binits{M.L.}}:
In: \beditor{\bsnm{{Truemper}}, \binits{J.E.}},
\beditor{\bsnm{{Tananbaum}}, \binits{H.D.}} (eds.)
\bbtitle{X-Ray and Gamma-Ray Telescopes and Instruments for Astronomy.}
\bsertitle{\procspie},
vol. \bseriesno{4851},
p. \bfpage{1163}
(\byear{2003})
\end{bchapter}
\endbibitem

\bibitem[\protect\citeauthoryear{{Tarnopolski}}{2015a}]{tar15a}
\begin{barticle}
\bauthor{\bsnm{{Tarnopolski}}, \binits{M.}}:
\bjtitle{\aap}
\bvolume{581},
\bfpage{A29}
(\byear{2015}a)
\end{barticle}
\endbibitem

\bibitem[\protect\citeauthoryear{{Tarnopolski}}{2015b}]{tar15b}
\begin{barticle}
\bauthor{\bsnm{{Tarnopolski}}, \binits{M.}}:
\bjtitle{\apss}
\bvolume{359},
\bfpage{20}
(\byear{2015}b)
\end{barticle}
\endbibitem

\bibitem[\protect\citeauthoryear{{Tarnopolski}}{2016}]{tar16}
\begin{barticle}
\bauthor{\bsnm{{Tarnopolski}}, \binits{M.}}:
\bjtitle{\mnras}
\bvolume{458},
\bfpage{2024}
(\byear{2016})
\end{barticle}
\endbibitem

\bibitem[\protect\citeauthoryear{{Toma} et~al.}{2005}]{to05}
\begin{barticle}
\bauthor{\bsnm{{Toma}}, \binits{K.}},
\bauthor{\bsnm{{Yamazaki}}, \binits{R.}},
\bauthor{\bsnm{{Nakamura}}, \binits{T.}}:
\bjtitle{\apj}
\bvolume{635},
\bfpage{481}
(\byear{2005})
\end{barticle}
\endbibitem

\bibitem[\protect\citeauthoryear{{Varga} et~al.}{2005}]{var05}
\begin{barticle}
\bauthor{\bsnm{{Varga}}, \binits{B.}},
\bauthor{\bsnm{{Horv{\'a}th}}, \binits{I.}},
\bauthor{\bsnm{{Bal{\'a}zs}}, \binits{L.G.}}:
\bjtitle{Nuovo Cimento C}
\bvolume{28},
\bfpage{861}
(\byear{2005})
\end{barticle}
\endbibitem

\bibitem[\protect\citeauthoryear{{Vavrek} et~al.}{2008}]{va08}
\begin{barticle}
\bauthor{\bsnm{{Vavrek}}, \binits{R.}},
\bauthor{\bsnm{{Bal{\'a}zs}}, \binits{L.G.}},
\bauthor{\bsnm{{M{\'e}sz{\'a}ros}}, \binits{A.}},
\bauthor{\bsnm{{Horv{\'a}th}}, \binits{I.}},
\bauthor{\bsnm{{Bagoly}}, \binits{Z.}}:
\bjtitle{\mnras}
\bvolume{391},
\bfpage{1741}
(\byear{2008})
\end{barticle}
\endbibitem

\bibitem[\protect\citeauthoryear{{Vedrenne} and {Atteia}}{2009}]{ved09}
\begin{bbook}
\bauthor{\bsnm{{Vedrenne}}, \binits{G.}},
\bauthor{\bsnm{{Atteia}}, \binits{J.-L.}}:
\bbtitle{{Gamma-Ray Bursts: The Brightest Explosions in the Universe}}.
\bpublisher{Springer},
\blocation{Berlin}
(\byear{2009})
\end{bbook}
\endbibitem

\bibitem[\protect\citeauthoryear{{Veres} et~al.}{2010}]{ver10}
\begin{barticle}
\bauthor{\bsnm{{Veres}}, \binits{P.}},
\bauthor{\bsnm{{Bagoly}}, \binits{Z.}},
\bauthor{\bsnm{{Horv{\'a}th}}, \binits{I.}},
\bauthor{\bsnm{{M{\'e}sz{\'a}ros}}, \binits{A.}},
\bauthor{\bsnm{{Bal{\'a}zs}}, \binits{L.G.}}:
\bjtitle{\apj}
\bvolume{725},
\bfpage{1955}
(\byear{2010})
\end{barticle}
\endbibitem

\bibitem[\protect\citeauthoryear{{Wigger} et~al.}{2004}]{wig04}
\begin{barticle}
\bauthor{\bsnm{{Wigger}}, \binits{C.}},
\bauthor{\bsnm{{Hajdas}}, \binits{W.}},
\bauthor{\bsnm{{Smith}}, \binits{D.M.}},
\bauthor{\bsnm{{G{\"u}del}}, \binits{M.}},
\bauthor{\bsnm{{Hurley}}, \binits{K.}},
\bauthor{\bsnm{{McHedlishvili}}, \binits{A.}},
\bauthor{\bsnm{{Zehnder}}, \binits{A.}}:
\bjtitle{Nuclear Physics B Proceedings Supplements}
\bvolume{132},
\bfpage{331}
(\byear{2004})
\end{barticle}
\endbibitem

\bibitem[\protect\citeauthoryear{{Wigger} et~al.}{2006}]{wig06}
\begin{bchapter}
\bauthor{\bsnm{{Wigger}}, \binits{C.}},
\bauthor{\bsnm{{Hajdas}}, \binits{W.}},
\bauthor{\bsnm{{Zehnder}}, \binits{A.}},
\bauthor{\bsnm{{Hurley}}, \binits{K.}},
\bauthor{\bsnm{{Bellm}}, \binits{E.}},
\bauthor{\bsnm{{Boggs}}, \binits{S.}},
\bauthor{\bsnm{{Bandstra}}, \binits{M.}},
\bauthor{\bsnm{{Smith}}, \binits{D.M.}}:
In: \bbtitle{Swift and GRBs: Unveiling the Relativistic Universe}.
\bsertitle{Il Nuovo Cimento B},
vol. \bseriesno{121},
p. \bfpage{1117}
(\byear{2006})
\end{bchapter}
\endbibitem

\bibitem[\protect\citeauthoryear{{Wigger} et~al.}{2008}]{wig08}
\begin{barticle}
\bauthor{\bsnm{{Wigger}}, \binits{C.}},
\bauthor{\bsnm{{Wigger}}, \binits{O.}},
\bauthor{\bsnm{{Bellm}}, \binits{E.}},
\bauthor{\bsnm{{Hajdas}}, \binits{W.}}:
\bjtitle{\apj}
\bvolume{675},
\bfpage{553}
(\byear{2008})
\end{barticle}
\endbibitem

\bibitem[\protect\citeauthoryear{{Yamazaki} et~al.}{2002}]{yam02}
\begin{barticle}
\bauthor{\bsnm{{Yamazaki}}, \binits{R.}},
\bauthor{\bsnm{{Ioka}}, \binits{K.}},
\bauthor{\bsnm{{Nakamura}}, \binits{T.}}:
\bjtitle{\apjl}
\bvolume{571},
\bfpage{31}
(\byear{2002})
\end{barticle}
\endbibitem

\bibitem[\protect\citeauthoryear{{Zhang} and {M{\'e}sz{\'a}ros}}{2002}]{zha02}
\begin{barticle}
\bauthor{\bsnm{{Zhang}}, \binits{B.}},
\bauthor{\bsnm{{M{\'e}sz{\'a}ros}}, \binits{P.}}:
\bjtitle{\apj}
\bvolume{581},
\bfpage{1236}
(\byear{2002})
\end{barticle}
\endbibitem

\bibitem[\protect\citeauthoryear{{Zhang} et~al.}{2004}]{zha04}
\begin{barticle}
\bauthor{\bsnm{{Zhang}}, \binits{B.}},
\bauthor{\bsnm{{Dai}}, \binits{X.}},
\bauthor{\bsnm{{Lloyd-Ronning}}, \binits{N.M.}},
\bauthor{\bsnm{{M{\'e}sz{\'a}ros}}, \binits{P.}}:
\bjtitle{\apjl}
\bvolume{601},
\bfpage{119}
(\byear{2004})
\end{barticle}
\endbibitem

\bibitem[\protect\citeauthoryear{{Zhang} et~al.}{2009}]{zha09}
\begin{barticle}
\bauthor{\bsnm{{Zhang}}, \binits{B.}},
\bauthor{\bsnm{{Zhang}}, \binits{B.-B.}},
\bauthor{\bsnm{{Virgili}}, \binits{F.J.}},
\bauthor{\bsnm{{Liang}}, \binits{E.-W.}},
\bauthor{\bsnm{{Kann}}, \binits{D.A.}},
\bauthor{\bsnm{{Wu}}, \binits{X.-F.}},
\bauthor{\bsnm{{Proga}}, \binits{D.}},
\bauthor{\bsnm{{Lv}}, \binits{H.-J.}},
\bauthor{\bsnm{{Toma}}, \binits{K.}},
\bauthor{\bsnm{{M{\'e}sz{\'a}ros}}, \binits{P.}},
\bauthor{\bsnm{{Burrows}}, \binits{D.N.}},
\bauthor{\bsnm{{Roming}}, \binits{P.W.A.}},
\bauthor{\bsnm{{Gehrels}}, \binits{N.}}:
\bjtitle{\apj}
\bvolume{703},
\bfpage{1696}
(\byear{2009})
\end{barticle}
\endbibitem

\bibitem[\protect\citeauthoryear{{Zitouni} et~al.}{2015}]{zit15}
\begin{barticle}
\bauthor{\bsnm{{Zitouni}}, \binits{H.}},
\bauthor{\bsnm{{Guessoum}}, \binits{N.}},
\bauthor{\bsnm{{Azzam}}, \binits{W.J.}},
\bauthor{\bsnm{{Mochkovitch}}, \binits{R.}}:
\bjtitle{\apss}
\bvolume{357},
\bfpage{7}
(\byear{2015})
\end{barticle}
\endbibitem

\end{thebibliography}

\end{document}